\documentclass[11pt, a4]{amsart}
 \usepackage{amsmath,amssymb,graphicx}
 \usepackage{wrapfig,pdfsync}
 \usepackage{subfig}

\newcommand{\e}{\epsilon}

\makeatletter \@addtoreset{equation}{section} \makeatother

\numberwithin{equation}{section}

\def\sech{{\rm sech}}

\newtheorem{theorem}{Theorem}[section]

\newtheorem{Definition}[theorem]{Definition}

\newtheorem{Remark}[theorem]{Remark}
\newenvironment{remark}{\begin{Remark}\rm}{\end{Remark}}
\newtheorem{Example}[theorem]{Example}

\newtheorem{Assumptions}[theorem]{Assumptions}

\begin{document}
\title[Small dispersion limit of KdV]{Numerical study of the small 
dispersion limit of the Korteweg-de Vries equation and asymptotic 
solutions}
\begin{abstract}
We study numerically the small dispersion limit for the Korteweg-de 
Vries (KdV) equation $u_t+6uu_x+\epsilon^{2}u_{xxx}=0$ for 
$\epsilon\ll1$ and give a quantitative comparison of the numerical 
solution with various 
asymptotic formulae for small $\epsilon$ in the whole $(x,t)$-plane. The matching of the 
asymptotic solutions is studied numerically.
\end{abstract}
\author[T. Grava]{T. Grava}
 \address{SISSA, Via Bonomea 265, I-34136 Trieste, Italy} 
 \email{grava@sissa.it}
\author[C. Klein]{C. Klein}
 \address{Institut de Math\'ematiques de Bourgogne,
		Universit\'e de Bourgogne, 9 avenue Alain Savary, 21078 Dijon
		Cedex, France} 
 \email{christian.klein@u-bourgogne.fr}
 \thanks{This work has been supported by 
 the project FroM-PDE funded by the European
 Research Council through the Advanced Investigator Grant Scheme.
 CK
 thanks for financial support by the Conseil R\'egional de Bourgogne
 via a FABER grant and the ANR via the program ANR-09-BLAN-0117-01.}
\maketitle
\section{Introduction}
%Recent works have studied the critical behavior of solutions to Hamiltonian perturbations
%of hyperbolic  and elliptic systems. In particular, the universal nature
%of this critical behavior is remarkable. For hyperbolic one- and two-component
%systems as well as for elliptic two-component systems, it has been conjectured by
%Dubrovin et al. [14, 15, 16] that solutions can be described in a certain critical
%regime by special solutions to the PainlevŽ I equation and its hierarchy.
%In the hyperbolic case, such critical regime has been rigorously proved for the Cauchy problem for the Korteweg de Vries equation with analytic initial data.
% For several hyperbolic systems, two other critical regimes have been observed. One of
%these regimes  is related to the second PainlevŽ equation, while the other critical is a solitonic regime  because the local asymtptotic behaviour is described by a train of solitons.
%The existence of such critical regimes has been rigorously proved for the Cauchy problem for the Korteweg de Vries equation.

The  behavior of solutions  to  Hamiltonian 
perturbations 
of hyperbolic  and elliptic systems  has seen a renewed interest in 
\cite{Dubrovin1,Dubrovin2, DubrovinGravaKlein}.  Specific integrable cases like the solution to  
the small dispersion limit of the Korteweg-de Vries (KdV) equation
or the semiclassical limit of the nonlinear Schr\"odinger equation 
have been studied  in  detail in some part of the  $(x,t)$ plane  in the seminal papers \cite{LL, DVZ, KMM}. 
However some detail description of the solution   in 
several critical regions of the $(x,t)$ plane can  be given  for non-integrable  Hamiltonian 
perturbations of hyperbolic or elliptic systems.     It has been conjectured by
Dubrovin  and Dubrovin  et al. \cite{Dubrovin1,DubrovinGravaKlein,Dubrovin2} that solutions can be approximated in one of the critical
regimes by special solutions to the Painlev\'e I  equation and its  hierarchy (see also \cite{BertolaTovbis}).  In particular the 
universal nature of the critical behavior is remarkable.
The conjecture  has been rigorously proved in one specific case, that is for the 
Cauchy problem for the KdV equation with analytic initial data.
 Further critical behaviours have been observed in solutions to Hamiltonian perturbations of  hyperbolic and elliptic equations (see e.g. \cite{Bertola,Miller, AGK}). 
  In particular in the  Hamiltonian perturbations of  hyperbolic systems, two other critical regimes have been observed:  one of
them  is related to the second Painlev\'e equation, while the other is  solitonic   because the local asymptotic behaviour is described by a train of solitons.
The existence of such critical regimes has been rigorously proved for 
the Cauchy problem for the KdV equation.
A review of these results as well as a new and improved   numerical 
comparison of all asymptotic formulae  with the numerical solution of KdV, is the subject of the present work.

\begin{figure}[thb!]
  \includegraphics[width=0.7\textwidth]{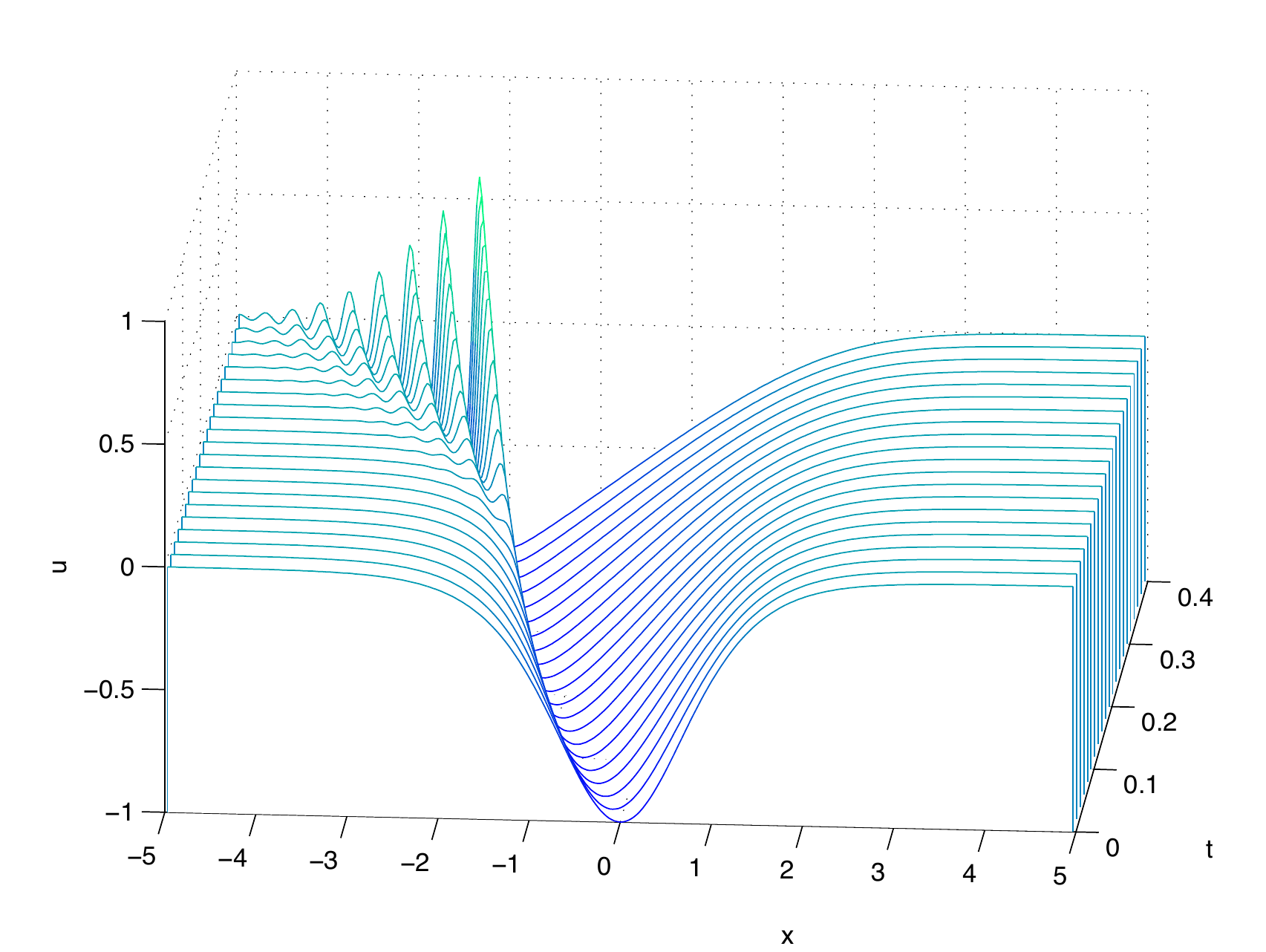}
 \caption{Solution to the KdV equation for the initial data 
 $u_{0}(x)=-\mbox{sech}^{2}x$ and $\epsilon=10^{-1}$.}
   \label{fig1}
\end{figure}

\begin{figure}[htb!]
  \includegraphics[width=0.8\textwidth]{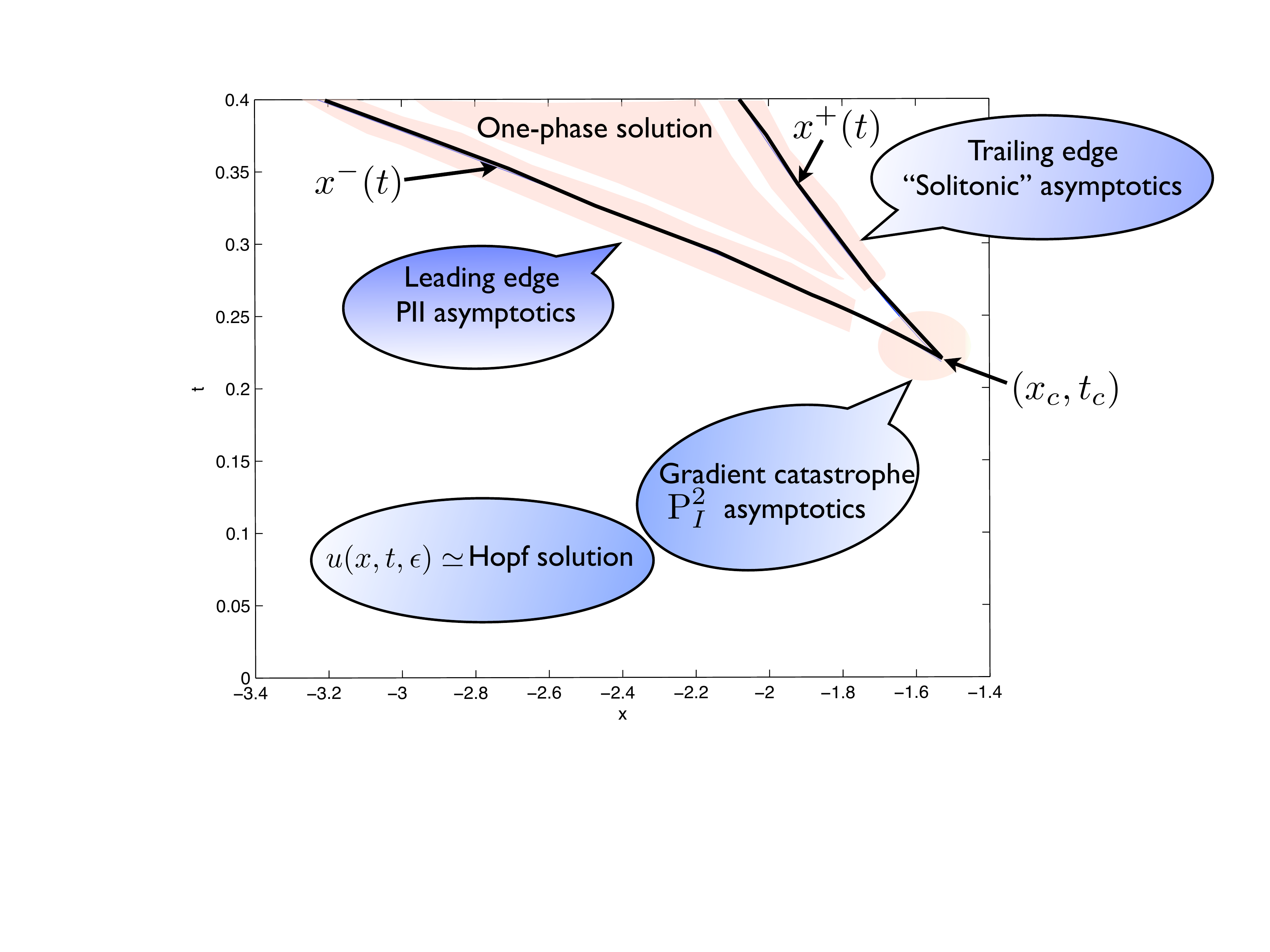}
 \caption{Whitham zone for the KdV equation for the initial data 
 $u_{0}(x)=-\mbox{sech}^{2}x$. The shadow regions indicate the various asymptotic approximations in a neighbourhood of the Whitham zone. }
   \label{fig2}
\end{figure}

  We consider the  Cauchy problem for the KdV equation with  small dispersion 
\begin{equation}\label{KdV}
u_t+6uu_x+\epsilon^{2}u_{xxx}=0, \qquad u(x,t=0,\e)=u_0(x),\qquad 
\e>0,\; x\in\mathbb{R},\;\;t\in\mathbb{R}^+;
\end{equation}
$u_0(x) $ is  real analytic negative initial data with sufficient decay at infinity and with a single negative hump (for detailed definition see \cite{CG1}).
For a much wider class of initial data than the one we consider, it 
is known that the KdV solution exists for all positive times $t$ (see for example \cite{Staffilani}). 
Up to a certain time $t_c$ the solution $u(x,t,\epsilon)$ as $\epsilon\rightarrow 0$   can be approximated \cite{LL} by the solution to the Cauchy problem for the (dispersionless) Hopf equation
\begin{equation}
u_t+6uu_x=0, \qquad u(x,0)=u_0(x),\qquad t>0,
\end{equation}
which can be solved by using the method of characteristics in the form
\begin{equation}
\label{Hopfsol}
u(x,t)=u_0(\xi),\quad x=6tu_0(\xi)+\xi.
\end{equation}
At time
\[
t_c=\dfrac{1}{\max_{\xi\in\mathbb{R}}[-6u'_0(\xi)]},
\]
the Hopf equation reaches a point  $(x_c,t_c)$ of gradient catastrophe where the derivative of the Hopf solution blows up.
Recently,  it has been proved \cite{MR}  that for initial data $u_0(x)$ in the Sobolev space $H^s(\mathbb{R})$, $s>\frac{3}{4}$,  and  for a larger class of equations than KdV
\begin{equation}
\label{Hopfasym}
u(x,t,\epsilon)=u(x,t)+O(\epsilon^2), \quad t<t_c,
\end{equation}
where $u(x,t)$ is the solution (\ref{Hopfsol}). The sub-leading term in the above expansion is determined explicitly.
For $t>t_c$, the Hopf solution (\ref{Hopfsol}) is multi-valued.
However, the KdV solution is well-defined for all positive $t$ and $\e>0$: the dispersive term $\epsilon^{2}u_{xxx}$ regularizes the gradient catastrophe.
For $t$ slightly smaller than $t_c$, the KdV solution 
$u(x,t,\epsilon)$ starts to oscillate. For $t>t_c$ a zone of  rapid modulated oscillations develops \cite{GP,LL}. In the $(x,t)$-plane, the oscillations take place in a cusp-shaped 
region (which depends on the initial data) as illustrated in 
Fig.~\ref{fig1}. These  oscillations in 
the limit $\epsilon\rightarrow 0$ are confined to a certain interval $[x^-(t),x^+(t)]$, see Figure \ref{fig2}. The interval $[x^-(t),x^+(t)]$ is usually called Whitham zone
because the oscillations are described  inside this interval through the Whtiham equations \cite{W} (see below (\ref{Whitham})).
Furthermore  the functions $x^+(t)$ and $x^-(t)$ are determined by 
the confluent form of the Whitham equations (see subsections \ref{leadingsection}, \ref{trailingsection}). 
%\begin{figure}[!htb]\label{fig1}
%\begin{center}
%\includegraphics[width=5in]{kdv1e2_04.eps}
%%\textwidth}
%\end{center}
%\caption{Solution of the KdV equation with $\e=10^{-2}$, initial data $u_0(x)=-\mbox{sech}^2(x)$, and $t=0.4$. }
%\end{figure}

Inside this cusp-shaped region, the exact one-phase  solution to the 
KdV equation in terms of elliptic functions gives   an asymptotic description of the 
oscillations, but on an elliptic surface where the branch points 
depend on $x$ and $t$ via the Whitham modulation equations as was proved by Lax and Levermore and Venakides  \cite{LL, Ve}.  
This averaging procedure works well inside the Whitham zone, but has 
to be amended near  the boundaries of the zone  as was found numerically in \cite{GK}. Near the  point of gradient catastrophe $(x_c,t_c)$ , Dubrovin 
\cite{Dubrovin1} conjectured that for a large class of equations 
containing KdV,  the corresponding solution is asymptotically given in terms of a 
special solution of the second equation in the Painlev\'e I 
hierarchy that we will call ${\rm P}_I^2$ equation. This was tested numerically in \cite{GK1} and  proven for KdV in \cite{CG}. Near the leading 
edge a multiscale expansion was presented for the oscillations in 
terms of a particular solution of the Painlev\'e II equation in 
\cite{GK2}. The validity of such an expansion has been proved rigorously in \cite{CG1}.  Near the trailing edge \cite{CG2} gave an asymptotic 
solution in terms of a series of pulses of the shape of KdV solitons (see Fig~\ref{fig2}).

It is remarkable that the KdV solutions can be approximated by Painlev\'e equations in the critical regimes described above. Those Painlev\'e equations have appeared in many branches of pure and applied mathematics during the past decades, for an overview we refer to \cite{FIKN}.

The universal critical behaviour  of KdV solutions  should be seen in relation
to the known universality results in random matrix theory. For large
unitary random matrix ensembles, local eigenvalue statistics turn
out to be, to some extent, independent of the choice of the ensemble
and independent of the reference point chosen \cite{Deift,DKMVZ1}. Critical break-up times occur when the eigenvalues move
from a one-cut regime to a multi-cut regime. These transitions 
take place in the presence of singular points, of which three
different types are distinguished \cite{DKMVZ1}. 
Such singular points have the same singularity type which appears in   the small dispersion limit of KdV. 
Singular interior points \cite{FIK, DK, BI, CK, CKV} show remarkable similarities with the leading edge of the
oscillatory region for the KdV equation.
The second possibility, which is related to the trailing edge for KdV, is that an interval in the spectrum shrinks and disappears afterwards. This case is also referred to as birth of a cut in unitary random matrix ensembles,  \cite{BertolaLee, C, Mo, Eynard}. 
 The point of gradient catastrophe, i.e.\ the
break-up point where the oscillations  set in is comparable to a singular edge point in unitary random
matrix ensembles. It was conjectured by Bowick and Br\'ezin, and by
Br\'ezin, Marinari, and Parisi \cite{BB, BMP} that local eigenvalue
statistics in this regime should be given in terms of the Painlev\'e
I hierarchy. In \cite{CV2}, it was proven that indeed double scaling
limits of the local eigenvalue correlation kernel are given in terms
of the Lax pair for the $P_I^2$ equation.
%In addition,an expansion similar to (\ref{univer0}) was obtained for the recurrence coefficients of orthogonal polynomials related to the relevant random matrix ensembles.

In this paper we implement numerically  all known  asymptotic approximations to the solution $u(x,t,\epsilon)$ of the KdV equation as $\epsilon\rightarrow 0$ 
and test them quantitatively for a concrete example. With 
respect to previous works, the numerical methods are overall improved 
which allows to study a wider range of values for the small 
dispersion parameter $\epsilon$. 
The asymptotic formulae for the trailing edge are implemented for the 
first time as well as the terms of order \( \epsilon^{4/7} \) at the 
breakup point.  We also address 
numerically the matching of  different approximations.

The paper is organized as follows: In sect.~2 we collect  various 
asymptotic formulae for KdV solutions in the small dispersion limit. 
In sect.~3 we give a brief overview of the used numerical methods. In 
sect.~4 we study numerically the asymptotic description given by the 
one-phase KdV and the Hopf solution. At the leading edge, the 
multiscale solution in terms of a special solution to the Painlev\'e 
II equation is studied in sect.~5. In sect.~6 the same is done for 
the asymptotic solution at the trailing edge, and in sect.~7 for the 
point of gradient catastrophe. In sect.~8 we add some concluding 
remarks and an outlook on the connection formulae for the various 
asymptotic regimes.
%\begin{figure}[t]\label{figcusp}
%\begin{center}
%\includegraphics[scale=0.35]{cusp.eps}
%\end{center}
%\caption{Cusp-shaped oscillatory region in the $(x,t)$-plane for $t>t_c$.}
%\end{figure}
\section{Asymptotic descriptions of the small dispersion limit}
In this section we summarize various asymptotic descriptions of the 
dispersive shocks which will be implemented numerically in the 
following. 

\subsection{One-phase solution in the Whitham zone}

Inside the cusp-shaped region in Fig.~\ref{fig2} for $t$ slightly bigger than $t_c$, the KdV solution $u(x,t,\epsilon)$ can be approximated for small $\epsilon$,  by the exact $1$-phase solution of KdV, where the branch points of the elliptic surface depends on $x$ and $t$ through Whitham equations. The one-phase solution of KdV can be written in terms of   the Jacobi elliptic theta function and the complete elliptic integrals of the first and second kind $E(s)$ and $K(s)$ \cite{LL, DVZ, GP}:
\begin{equation}
\label{elliptic}
u(x,t,\e)\simeq \beta_1+\beta_2+\beta_3+2\alpha+ 2\e^2\frac{\partial^2}{\partial
x^2}\log\vartheta(\Omega(x,t);\mathcal{T}),
\end{equation}
where $\beta_1>\beta_2>\beta_3$,  $\Omega$, $\alpha$, and $\mathcal{T}$ have the form
\begin{align}
&\label{Omega}
\Omega(x,t)=\dfrac{\sqrt{\beta_1-\beta_3}}{2\e K(s)}[x-2 t(\beta_1+\beta_2+\beta_3) -q],
\\
&\label{alpha}
\alpha(s)=-\beta_{1}+(\beta_{1}-\beta_{3})\frac{E(s)}{K(s)},\;\;\mathcal{T}=i\dfrac{K'(s)}{K(s)},
\;\; s^{2}=\frac{\beta_{2}-\beta_{3}}{\beta_{1}-\beta_{3}}.
\end{align}
Note that $K'(s)=K(\sqrt{1-s^{2}})$, and
 $\vartheta$ is defined by the
Fourier series
\[
\vartheta(z;\mathcal{T})=\sum_{n\in\mathbb{Z}}e^{\pi i n^2\mathcal{T}+2\pi i nz}.
\]
The formula for $q$ in the phase $\Omega$ in (\ref{Omega}) is equal to \cite{GK2, DVZ}
\begin{equation}
\label{q0}
    q(\beta_{1},\beta_{2},\beta_{3}) = \frac{1}{2\sqrt{2}\pi}
    \int_{-1}^{1}\int_{-1}^{1}d\mu d\nu \frac {f_L( \frac{1+\mu}{2}(\frac{1+\nu}{2}\beta_{1}
	+\frac{1-\nu}{2}\beta_{2})+\frac{1-\mu}{2}\beta_{3})}{\sqrt{1-\mu}
    \sqrt{1-\nu^{2}}},
\end{equation}
where $f_L(y)$ is the inverse function of the decreasing part of the 
initial data $u_0$.   The formula (\ref{q0}) for $q$ is valid as long as $\beta_3$ does 
 not reach the minimum  value of the initial data $u_0$. When 
 $\beta_3$ reaches and goes beyond the negative hump it is also necessary to take 
 into account the increasing part of the initial data $f_R$, \cite{GK}, \cite{FRT1}
 \begin{equation}
\label{qnm}
q(\beta_1,\beta_2,\beta_3)=\frac{1}{2\pi}\int_{\beta_2}^{\beta_1} \dfrac{d\lambda\left(\displaystyle\int_{\beta_{3}}^{-1}
    \frac{d\xi
    f_{R}(\xi)}{\sqrt{\lambda-\xi}}+\displaystyle\int_{-1}^{\lambda}
    \frac{d\xi f_{L}(\xi)}{\sqrt{\lambda-\xi}}\right)
}{
\sqrt{(\beta_1-\lambda)(\lambda-\beta_2)(\lambda-\beta_3)}}
.
\end{equation}
\begin{remark}
In the formula (\ref{elliptic}) the term $\beta_1+\beta_2+\beta_3+2\alpha$ is the weak limit of $u(x,t,\epsilon)$ as $\epsilon\rightarrow 0$  \cite{LL} and the term containing the $\theta$-function describes the oscillations \cite{DVZ, Ve}. The error term in the asymptotic expansion  (\ref{elliptic}) should be of order $O(\epsilon)$.
\end{remark}
Formula (\ref{elliptic}) can be written also in 
terms of the Jacobi elliptic function $\mbox{dn}$  in the form
\begin{equation}
\label{elliptic1}
u(x,t,\e)\simeq \beta_2+\beta_3-\beta_1+2(\beta_1-\beta_3)\mbox{dn}^2(2K(s)\Omega+K(s)).
\end{equation}
For constant values of  $\beta_1>\beta_2>\beta_3$, the right hand side of (\ref{elliptic1}) is an exact solution of KdV.
However in  the description
of the leading order asymptotics of $u(x,t,\epsilon)$ as $\epsilon\rightarrow 0$,
 the numbers $\beta_1>\beta_2>\beta_3$ depend on $x$ and $t$  and evolve
 according to the Whitham equations \cite{W}
\begin{equation}
\label{Whitham}
\dfrac{\partial}{\partial t}\beta_i+v_i\dfrac{\partial}{\partial x}\beta_i=0,\quad   v_{i}=4\frac{\prod_{k\neq
     i}^{}(\beta_{i}-\beta_{k})}{\beta_{i}+\alpha}+2(\beta_1+\beta_{2}+\beta_{3}),\;\;\; i=1,2,3,
\end{equation}
%where the speeds $v_i$ are given by the formula
%\begin{equation}
   % \label{eq:la0}
%\end{equation}
with $\alpha$ as in (\ref{alpha}).

The Whitham equations  (\ref{Whitham}) can be integrated through the
so-called hodograph transform, which generalizes the method of characteristics,
and which gives the solution in the implicit form \cite{Tsarev}
\begin{equation}
\label{hodograph}
x=v_it+w_i,\quad i=1,2,3,
\end{equation}
where  the $v_i$  are defined in (\ref{Whitham}), and where $w_i=w_i(\beta_1,\beta_2,\beta_3)$ for $i=1,2,3$
is obtained from an algebro-geometric procedure by the formula \cite{FRT}
\begin{equation}
    w_{i} =
    \frac{1}{2}\left(v_{i}-2\sum_{k=1}^{3}\beta_{k}\right)\frac{
    \partial q}{\partial\beta_{i}}+q,\quad i=1,2,3,
    \label{eq:w}
\end{equation}
with $q$ defined in (\ref{q0}) or (\ref{qnm}). Solvability of (\ref{hodograph})  for initial data with a single hump   was proved in \cite{FRT1}.

\medskip

Near the boundary of the oscillatory cusp-shaped region, neither the 
Hopf solution (\ref{Hopfsol})  nor the one-phase  solution (\ref{elliptic}) gives a satisfactory description of  the KdV solution  $u(x,t,\epsilon)$ as $\epsilon\rightarrow 0$. Three different transitional regimes can be distinguished: (1) the cusp point where the gradient catastrophe for the Hopf equation takes place and where $\beta_1=\beta_2=\beta_3=u_c$, (2) the leading edge of the oscillatory zone where $\beta_2=\beta_3$, and (3) the trailing edge of the oscillatory zone where $\beta_1=\beta_2$. We will illustrate in the next subsections  that in the above three cases
\begin{eqnarray*}
u(x,t,\epsilon)=\left\{
\begin{array}{lll}
u(x_c,t_c)+O(\epsilon^{\frac{2}{7}}),&\mbox{near  the point of gradient catastrophe  $(x_c,t_c)$ },&\\
u(x^-(t),t)+O(\epsilon^{\frac{1}{3}} ), &\mbox{near the leading edge},&\\
u(x^+(t),t)+O(1), &\mbox{near the trailing edge},&
\end{array}
\right.
\end{eqnarray*}
where $u(x,t)$ is the solution of the Hopf equation and $x^{\pm}(t)$ are the boundaries of the Whitham zone. 
The sub-leading terms are described respectively by a  ${\rm P_I^2}$ transcendent, a ${\rm P_{II} }$  transcendent, and a train of solitons.
Clearly the above asymptotic expansions are not uniform in $\epsilon$. Connections formula need to be developed.
\subsection{Point of gradient catastrophe}

It was conjectured in 
\cite{Dubrovin1} and proved afterwards in \cite{CG} that the KdV  solution  $u(x,t,\epsilon)$  as $\epsilon\rightarrow 0$  near  the point of gradient catastrophe $(x_c, t_c)$  for the Hopf solution (\ref{Hopfsol}),   is  given in terms of
a distinguished Painlev\'e transcendent, namely a special smooth solution $U(X,T)$ to the fourth order ODE
\begin{equation}
\label{PI2}
X=6T\, U -\left[ U^3  +\dfrac{1}{2} U_{X}^2 + U\, U_{XX} 
+\frac1{10} U_{XXXX}\right].
\end{equation}
This ODE is the second member of the Painlev\'e I hierarchy, and we
refer to it as the ${\rm P_I^2}$ equation. The relevant solution is real and has
the asymptotic behavior
     \begin{equation}
\label{PI2asym}
	U(X,T)=\mp (|X|)^{1/3}\mp 2T|X|^{-1/3}
	    +\mathcal{O}(|X|^{-1}),
	    \qquad\mbox{as $X\to\pm\infty$,}
\end{equation}
for any fixed $T\in\mathbb{R}$, and has no poles for real values of
$X$ and $T$ \cite{CV1, Dubrovin1}. It is remarkable that
$U(X,T)$ is an exact solution to the KdV equation 
\begin{equation}\label{UKdV}U_T+6UU_X+U_{XXX}=0.\end{equation}

 In a double scaling limit where $\e\to 0$ and simultaneously $x$ and $t$ approach the point and the  time of gradient catastrophe $x_c$ and $t_c$ in such a way that the limits
\[
\lim_{\begin{matrix}
\epsilon\rightarrow 0\\
x-6u_ct\rightarrow x_c-6u_ct_c
\end{matrix}}\left[ \dfrac{x-x_c-6u_c(t-t_c)}{\epsilon^{\frac{6}{7}}}\right],\quad 
\lim_{\begin{matrix}
\epsilon\rightarrow 0\\
t\rightarrow t_c
\end{matrix}}\left[ \dfrac{(t-t_c)}{\epsilon^{\frac{4}{7}} }\right], \;\;\;u_c=u(x_c,t_c)
\]
exist and are bounded,   the KdV solution has an expansion of the following form \cite{CG}.
Let 
\begin{equation}
\label{XT}
X=\dfrac{x-x_c-6u_c(t-t_c)}{(k)^{1/7}\epsilon^{\frac{6}{7}}},\quad T=\dfrac{(t-t_c)}{(k)^{3/7}\epsilon^{\frac{4}{7}} }
\end{equation}
with 
\begin{equation}
\label{k}
k=-f'''_L(u_c)/6
\end{equation}
and $f_L(u)$ the inverse of the decreasing part of the initial data.
Then the solution of KdV is approximated by
\begin{equation}
\label{ExpP12}
\begin{split}
u(x,t,\epsilon)&=u_c+ \left(\frac{\epsilon}{k}\right)^{2/7} U(X,T)
-\left(\frac{\epsilon}{k}\right)^{4/7}\dfrac{ f_L^{(IV)}(u_c)}{63 f_L'''(u_c)}\times\\
&\left[QU_X+2U_{XX}+4U^2+15T-90T^2U_X-3XUU_X-\dfrac{1}{2}XU_{XXX}\right]+O(\epsilon^{\frac{5}{7}})
\end{split}
\end{equation}
and  $Q(X,T)$ is the integral of $U(X,T)$, 
\[
Q=\dfrac{1}{10}U_XU_{XXX}-\dfrac{1}{20}U_{XX}^2+XU-3TU^2+\dfrac{1}{4}U^4+\dfrac{1}{2}UU_X^2, \;\; Q_X=U.
\] 
The correction term of order $\epsilon^{\frac{2}{7}}$ was  rigorously derived using steepest descent analysis for the Riemann-Hilbert problem for KdV in \cite{CG}. Such an approximation
was already discovered for the Gurevich-Pitaevskii solution of KdV in 
\cite{Su}. The correction of  order \( \epsilon^{4/7} \)  was  derived 
in \cite{CG3}.

\subsection{Leading edge\label{leadingsection}}

Near the leading edge at the left of the zone where the oscillations  
become small,  a multiscale analysis and numerical results \cite{GK2} showed that the envelope of the oscillations is asymptotically described by a particular solution to the second Painlev\'e equation (${\rm P}_{II}$)
 \begin{equation}\label{PII}
 q''(s)=sq+2q^{3}(s).
\end{equation}
The special solution we are interested in, is the Hastings-McLeod solution \cite{HastingsMcLeod}
which is uniquely determined by the boundary conditions
\begin{align}
&\label{HM1}q(s)=\sqrt{-s/2}(1+o(1)),&\mbox{ as $s\to -\infty$,}\\
&\label{HM2}q(s)=\mbox{Ai}(s)(1+o(1)), &\mbox{ as $s\to +\infty$,}
\end{align}
where $\mbox{Ai}(s)$ is the Airy function.
Although any nonzero Painlev\'e II solution has an infinite number of poles in the complex plane, it is known \cite{HastingsMcLeod} that the Hastings-McLeod solution $q(s)$ is smooth for all real values of $s$.
 
The leading edge corresponds   to the Whitham equations in the confluent case where
\[\beta_3(x,t)=\beta_2(x,t)=v(t), \qquad \beta_1(t)=u(t),\]
see also (\ref{elliptic}).
 There exists a time $\tilde{t}>t_c$ such that for $t_c<t<\tilde{t}$, the leading edge $x^-(t)$ is determined uniquely by the system of
equations  \cite{FRT, GT} 
\begin{align}
&\label{leading}x^-(t)=6tu(t)+f_L(u(t)),\\
&\label{leading2}6t+\theta(v(t);u(t))=0,\\
&\label{leading3}\partial_{v}\theta(v(t);u(t))=0,
\end{align}
with $u(t)>v(t)$ and with
\begin{equation}
\label{theta}
\theta(v;u)=\theta(v;u)=\dfrac{1}{2\sqrt{2}}\int_{-1}^1f'_L\left(\frac{1+m}{2}v+\frac{1-m}{2}u\right)\dfrac{dm}{\sqrt{1-m}}.
\end{equation}
Furthermore $x^-(t)$, $u(t)$, and $v(t)$ are smooth functions of $t$.
Throughout the rest of the subsection, whenever we refer to $u$, we mean by this the solution of the system (\ref{leading})-(\ref{leading3}) for a given time $t_c<t<\tilde{t}$, while we denote the solution of the KdV equation as $u(x,t,\e)$ and the solution of the Hopf equation as $u(x,t)$.

The behaviour of the solution of KdV near the leading edge as $\epsilon\rightarrow 0$ is described as follows.
Take a double scaling limit where  $\epsilon\to
0$ and at the same time  $x\to x^-(t)$  in such a way that
\begin{equation}
\lim _{\begin{matrix}
\epsilon\rightarrow 0\\
x\rightarrow x^-(t)
\end{matrix}}
\left[\dfrac{x- x^-(t)}{\e^{2/3}}\right]
\end{equation}
remains bounded. In this double scaling limit, the solution $u(x,t,\epsilon)$ of the KdV equation (\ref{KdV}) with initial data $u_0$ has the  asymptotic expansion \cite{CG1}
\begin{multline}
\label{expansionu}
u(x,t,\e) = u-\dfrac{4\e^{1/3}}{c^{1/3}}q\left(s(x,t,\e)\right)
\cos\left(\frac{\Theta(x,t)}{\epsilon}+\e^{1/3}\Theta_1(x,t,\e)\right)\\+\frac{x-x^-}{6t+f'_L(u)}
-\dfrac{4\e^{2/3}} {c^{2/3}(u-v)} q\left(s(x,t,\e)\right)^2
\sin^2\left(\frac{\Theta(x,t)}{\epsilon}\right)+O(\e).
\end{multline}
Here $x^-$ and $v<u$ (each of them depending on $t$) solve the system (\ref{leading}), and the phase $\Theta(x,t)$ is given by
\begin{equation}
\label{Theta}
\Theta(x,t)=2\sqrt{u-v}(x-x^-)+2\int_{v}^{u}(f_L'(\xi)+6t)\sqrt{\xi-v}d\xi.
\end{equation}
Furthermore
\begin{equation}\label{def c}
c=-\sqrt{u-v}\dfrac{\partial^2}{\partial v^2}\theta(v;u)>0,\qquad s(x,t,\e)=-\frac{x-x^-}{c^{1/3}\sqrt{u-v}\,\e^{2/3}},
\end{equation}
with $\theta$ defined by (\ref{theta}), and $q$ is the Hastings-McLeod solution to the Painlev\'e II equation.
The correction to the phase $\Theta_1(x,t,\e)$ takes the form
\begin{multline}
\label{Theta1}
\Theta_1(x,t,\e)= \frac{1}{c^{1/3}}\left[\left(\frac{q'}{q}+p\right)\frac{\partial^3_{v^3}\theta(v;u)}{6\partial^2_{v^2}\theta(v;u)}-\frac{5p+\frac{q'}{q}}{4(u-v)}\right.
\\ \left.+\dfrac{s(x,t,\e)^2}{4}\left(\frac{\partial^3_{v^3}\theta(v;u)}{3\partial^2_{v^2}\theta(v;u)}-\frac{3}{2(u-v)}
+\dfrac{2c\sqrt{u-v}}{6t+f'_L(u)}
\right)\right],
\end{multline} where we used the notations
\[q=q(s), \qquad q'=q'(s),\qquad p=p(s)=-q^4(s)-sq^2(s)+q'(s)^2,\] with $s=s(x,t,\e)$ and $p'(s)=-q^2$.
\begin{remark}
Note that the leading order term in the expansion (\ref{expansionu}) of $u(x,t,\e)$ is given by $u(t)$ the solution of the Hopf equation at the leading edge.
The second term in (\ref{expansionu}) is of order $\e^{1/3}$, while 
the remaining terms are of order $\e^{2/3}$.
From the $\mathcal{O}(\e^{1/3})$-term, we observe that $u(x,t,\e)$ develops oscillations of wavelength $\mathcal{O}(\e)$ at the leading edge, the envelope of the oscillations is proportional to
 the Hastings-McLeod solution $q$.  If we let $(x-x^-(t))/\epsilon^{\frac{2}{3}}\to -\infty$ (so that $x$ lies to the left of the leading edge), the terms with the oscillations disappear due to the exponential decay of $q$, see (\ref{HM2}). We are then left with only two terms in (\ref{expansionu}), which are the first two terms in the Taylor series of the Hopf solution $u(x,t)$ near $x^-$. 
\end{remark}
\begin{remark}
The formula (\ref{expansionu}) can be obtained from a multiple scale analysis \cite{GK2} from  (\ref{elliptic}), letting 
\[\beta_3(x,t)=v(t)-\dfrac{2\e^{1/3}}{c^{1/3}}q\left(s(x,t,\e)\right),\quad \beta_2(x,t)=v(t)+\dfrac{2\e^{1/3}}{c^{1/3}}q\left(s(x,t,\e)\right), 
\]
\[\qquad \beta_1(t)=u(t)+\frac{x-x^-}{6t+f'_L(u)}.\]
\end{remark}

\subsection{Trailing edge\label{trailingsection}}
The trailing edge $x^+(t)$ of the oscillatory interval  is uniquely 
determined by the confluent form $\beta_1=\beta_2=v$ and $\beta_3=u$, $v>u$,
of the equations  (\ref{hodograph}), namely \cite{FRT, GT}
\begin{align}
&\label{trailing1}x^+(t)=6tu(t)+f_L(u(t)),\\
&\label{trailing2}6t+\theta(v(t);u(t))=0,\\
&\label{trailing3}\int_{u(t)}^{v(t)}(6t+\theta(\lambda;u(t)))\sqrt{\lambda-u(t)}d\lambda=0,\\
\end{align}
where $\theta(v;u)$ has been defined in (\ref{theta}).

\medskip
Let $x^+=x^+(t)$, $u=u(t)$, and $v=v(t)$ solve the system (\ref{trailing1})-(\ref{trailing3}), and let
us take a double scaling limit where $\e\to 0$ simultaneously with $x\to x^+(t)$ in such a way that \[y:=2\sqrt{v-u}\, \frac{x-x^+}{\e\ln\e}\] remains bounded: there exists a real  $M>0$ such that $|y|<M$. Then there exists $\tilde{t}>t_c$ such that for $t_c<t<\tilde{t}$, we have the following expansion for the KdV solution $u(x,t,\e)$ in the double scaling limit \cite{CG2},
\begin{equation}\label{expansion u}
u\left(x,t,\e\right)=u+2(v-u)\sum_{j=0}^{\lceil M\rceil}\sech^2(X_j)+\mathcal{O}(\e\ln^2\e),
\end{equation}
where $\lceil M\rceil$ is the smallest integer $\geq M$,
\begin{equation}
\begin{split}
\label{Xj}
&X_j=\frac{1}{2}(\frac12-y+j)\ln\e-\ln(\sqrt{2\pi} h_j)-(j+\frac12)\log\gamma,\\
&h_j=\dfrac{2^{\frac{j}{2}} }{\pi^{\frac{1}{4}}\sqrt{j!}},\quad \gamma=4(v-u)^{\frac{5}{4}}\sqrt{-\partial_v\theta(v;u)},
\end{split}
\end{equation}
and $\theta$ is given by (\ref{theta}). Observe that $h_j$ are 
the normalization constants of the Hermite polynomials.
\begin{remark}
Observe that each term in the sum of (\ref{expansion u}) generates a pulse with amplitude $2(v-u)$ for $y$ near a half positive integer
 which can be seen as a soliton.  Indeed  the term $\sech^2(X_j)$ is 
 of the order $O(1)$ for $y=j+1/2$.  For $y=j$ or  $j+1$, it already 
 decreased to order $O(\epsilon^{\frac{1}{2}})$. For  
 $y=j-\frac{1}{2}$ or  $j+\frac{3}{2}$, the contribution of 
 $\sech^2(X_j)$ is absorbed by the error term $O(\e\ln^2\e)$. 
 Clearly, since $j$ is nonnegative, the solitons appear only for  $y$ positive,  
 that is for $x<x^+$, namely inside the Whitham zone.
\end{remark}
\begin{remark}
The phase defined in (\ref{Omega}) satisfies the formal limit 
\[
2K(s)\Omega|_{\beta_1\simeq\beta_2}=\dfrac{x-x^+}{\epsilon}\sqrt{v-u}+O((\beta_1-\beta_2)^2),
\]
and from the Whitham solution (\ref{hodograph}) one obtains in the limit $\beta_1\rightarrow \beta_2$
\[
x-x^+(t)\simeq -\dfrac{1}{4}\partial_v\theta(v;u)\left(\dfrac{\beta_1-\beta_2}{2}\right)^2\log\left(\dfrac{\beta_1-\beta_2}{2}\right)^2.
\]
  The Jacobi elliptic function $\mbox{dn}\rightarrow \sech$ as the modulus $s\rightarrow 1$ and
\[
K(s)\simeq\dfrac{1}{2}\log\left[\dfrac{8}{1-s}\right],\quad \mbox{as}\;\;s\rightarrow 1.
\]
Therefore  the formal limit  of the solution (\ref{elliptic1}) in 
terms of elliptic functions  as $s\rightarrow 1$, $\beta_1,\beta_2\rightarrow v$, $\beta_3\rightarrow u$,  gives
\begin{equation}\label{expansion formal}
u(x,t,\epsilon)\simeq u+2(v-u)\,\sech^2\left[  \dfrac{x-x^+}{\epsilon}\sqrt{v-u}+(j+\dfrac{1}{2})\log\left[\dfrac{8}{1-s}\right]\right],
\end{equation}
for any positive integer $j$, due to the periodicity of the elliptic function $\mbox{dn}$.
Choosing 
\[
\left(\dfrac{\beta_1-\beta_2}{2}\right)^2=\dfrac{4\epsilon}{\partial_v\theta(v,u)\sqrt{v-u}}
\]
and inserting it in (\ref{expansion formal}) one can partially 
reproduce the formula (\ref{expansion u}) in the sense that all the 
terms in the phase $X_j$  defined in (\ref{Xj}) can be reproduced  
except the one   containing the normalization constants of the Hemite polynomials $h_j$. 
We would like to remark that the formal  limit   (\ref{expansion formal})  has appeared several times in the literature, but such limit does not describe
 the small dispersion solution   of KdV near the trailing edge, since  the limiting value of the phase (\ref{Omega}) does not give the right result.
\end{remark}
\section{Numerical Methods}
The numerical task in treating the small dispersion limit of KdV and 
various asymptotic formulas consists in solving the KdV equation 
itself, certain ODEs of Painlev\'e type for a given asymptotic 
behavior, and of the Whitham equations for which the implicit 
solution (\ref{Whitham}) exists. We will summarize in this section 
how these different tasks are solved numerically, and how we control 
the numerical accuracy.
\subsection{KdV solution}
Since critical phenomena are generally believed to be independent of 
the chosen boundary conditions, we study a periodic 
setting in the following. This also includes rapidly decreasing functions which can be 
periodically continued as smooth functions within the finite numerical precision. 
This allows to approximate the spatial dependence 
via truncated Fourier series which leads for
the studied equations to large stiff systems of ordinary differential 
equations (ODEs), see below. 
The use of 
Fourier methods not only gives \emph{spectral accuracy} in the spatial 
coordinates (the numerical error in approximating smooth functions 
decreases faster than any power of the number $N$ of Fourier modes), but also minimizes the introduction of numerical 
dissipation which is important in the study of the purely dispersive 
effects we are interested in here.
In 
Fourier space, equation (\ref{KdV})  has the form
\begin{equation}
    v_{t}=\mathbf{L}v+\mathbf{N}(v,t)
    \label{utrans},
\end{equation}
where $v$ denotes the (discrete) Fourier transform of $u$, 
and where $\mathbf{L}$ and $\mathbf{N}$ denote linear and nonlinear 
operators, respectively. The resulting system of ODEs consists in 
this case of \emph{stiff} equations where the 
stiffness is related to the linear part $\mathbf{L}$ (it is 
a consequence of the distribution of the eigenvalues of 
$\mathbf{L}$), whereas the 
nonlinear part contains only low order derivatives. In the small dispersion 
limit, this stiffness is still present despite the small term 
$\epsilon^{2}$ in $\mathbf{L}$. This is due to the fact that the 
smaller $\epsilon$ is, the higher wavenumbers are needed to 
resolve the rapid oscillations. 

Loosely speaking a stiff system is a 
system for which explicit numerical schemes as explicit Runge-Kutta 
methods are inefficient, since prohibitively small time steps have to 
be chosen to control exponentially growing terms. 
The standard remedy for this is to use stable implicit schemes, which 
require, however, the iterative solution of a system of nonlinear equations at 
each time step which is computationally expensive. In addition the 
iteration often introduces numerical errors in the Fourier 
coefficients. Thus we used in \cite{GK} an integrating factor 
method, where the linear stiff part is explicitly integrated. This 
can be conveniently done here since the operator $\mathbf{L}$ corresponding to 
the third derivative with respect to $x$ is diagonal in 
Fourier space. As was shown in \cite{HO}, integrating factor methods 
can suffer from \emph{order reductions}, which means that the actual 
decrease of the numerical error with the numerical resolution is much 
lower than the classical order of the used method. This was confirmed for the small 
dispersion limit of KdV in \cite{etna}. There it was also shown that 
\emph{exponential time differencing} (ETD) schemes are very efficient 
for KdV. 
ETD schemes were developed originally by 
Certaine \cite{Cer} in the 60s,  see 
\cite{HO09} for a comprehensive 
review. The basic idea is to use equidistant time 
steps $h$ and to integrate equation (\ref{utrans}) exactly between 
the time steps $t_{n}$ and $t_{n+1}$ with respect to $t$. With $v(t_{n}) = v_{n}$ and 
$v(t_{n+1})=v_{n+1}$, we get 
$$
v_{n+1}=e^{\mathbf{L}h}v_{n}+\int_{0}^{h} e^{\mathbf{L}(h-\tau)}
    \mathbf{N}(v(t_{n}+\tau),t_{n}+\tau)d\tau.
$$
The integral will be computed in an approximate way for which 
different schemes exist. We use here a   
Runge-Kutta method of classical order 4 due to  Cox-Matthews 
\cite{CM}. As in \cite{AGK} for the Camassa-Holm equation, this 
approach could be amended by 
identifying two regimes $t\in[0,t_{1}]$ and $t\in[t_{1},t_{end}]$ 
with $t_{1}\ll t_{c}$. A much larger time step can be used in the 
first regime than in the second where the rapid modulated 
oscillations appear. We do not use this approach here since it was 
not necessary for the considered values of $\epsilon$. 

The accuracy of the numerical solution is controlled via the 
numerically computed conserved energy of the solution
\begin{equation}
    E[u] = \int_{\mathbb{T}}^{}(2u^3 - \epsilon^{2}u_{x}^2)dx
    \label{kdvE},
\end{equation}
which is an exactly conserved quantity for KdV. 
Numerically the energy $E$ will be a function of time. We define 
$\Delta E:=|(E(t)-E(0))/E(0)|$. It was shown in \cite{etna} that this quantity can 
be used as an indicator of the numerical accuracy if sufficient 
resolution in space is provided. The quantity $\Delta E$ typically 
overestimates the precision by two orders of magnitude. Since the 
numerical error has to be clearly smaller than the difference between 
KdV solution and the asymptotic descriptions we want to test (which give 
at best  descriptions of order \( \epsilon \)) we are 
interested in a numerical  value clearly below the smallest considered value of 
$\epsilon$. To ensure this  we will 
always ensure that the modulus of the Fourier coefficients of the final state 
decreases well below $10^{-5}$ (thus providing the needed resolution), and that the quantity $\Delta E$ is 
smaller than $10^{-6}$ (in general it is of the order of machine 
precision, i.e.\ $10^{-14}$).

\begin{figure}[htb!]
  \includegraphics[width=0.7\textwidth]{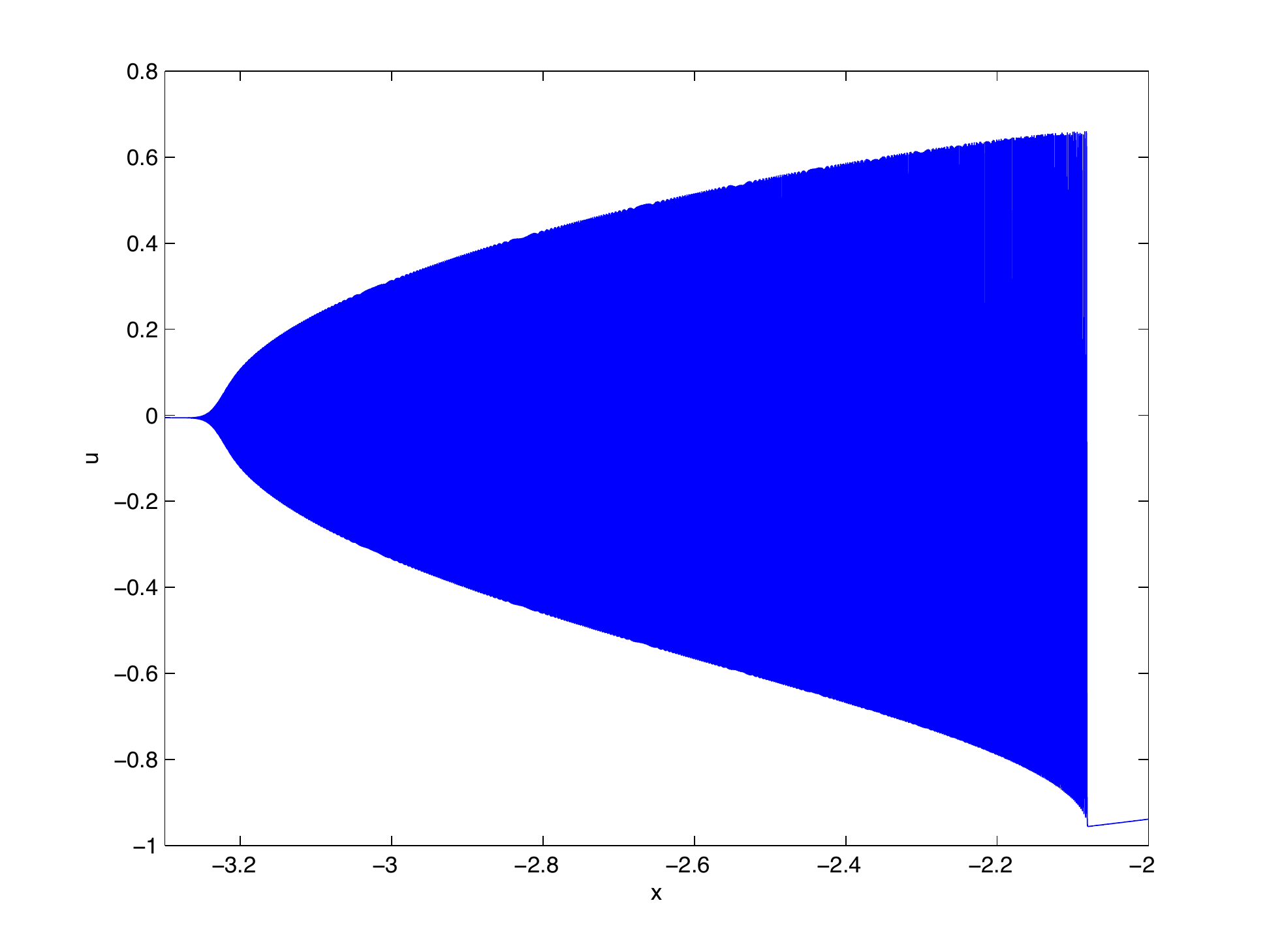}
 \caption{The oscillatory zone in the solution to the KdV equation for the initial data 
 $u_{0}(x)=-\mbox{sech}^{2}x$ and $\epsilon=10^{-3.5}$ for $t=0.4$. 
 The oscillations are so rapid, that they are graphically difficult to 
 represent though they are numerically well resolved.}
   \label{kdv1e7_t04}
\end{figure}

\begin{figure}[htb!]
  \includegraphics[width=0.6\textwidth]{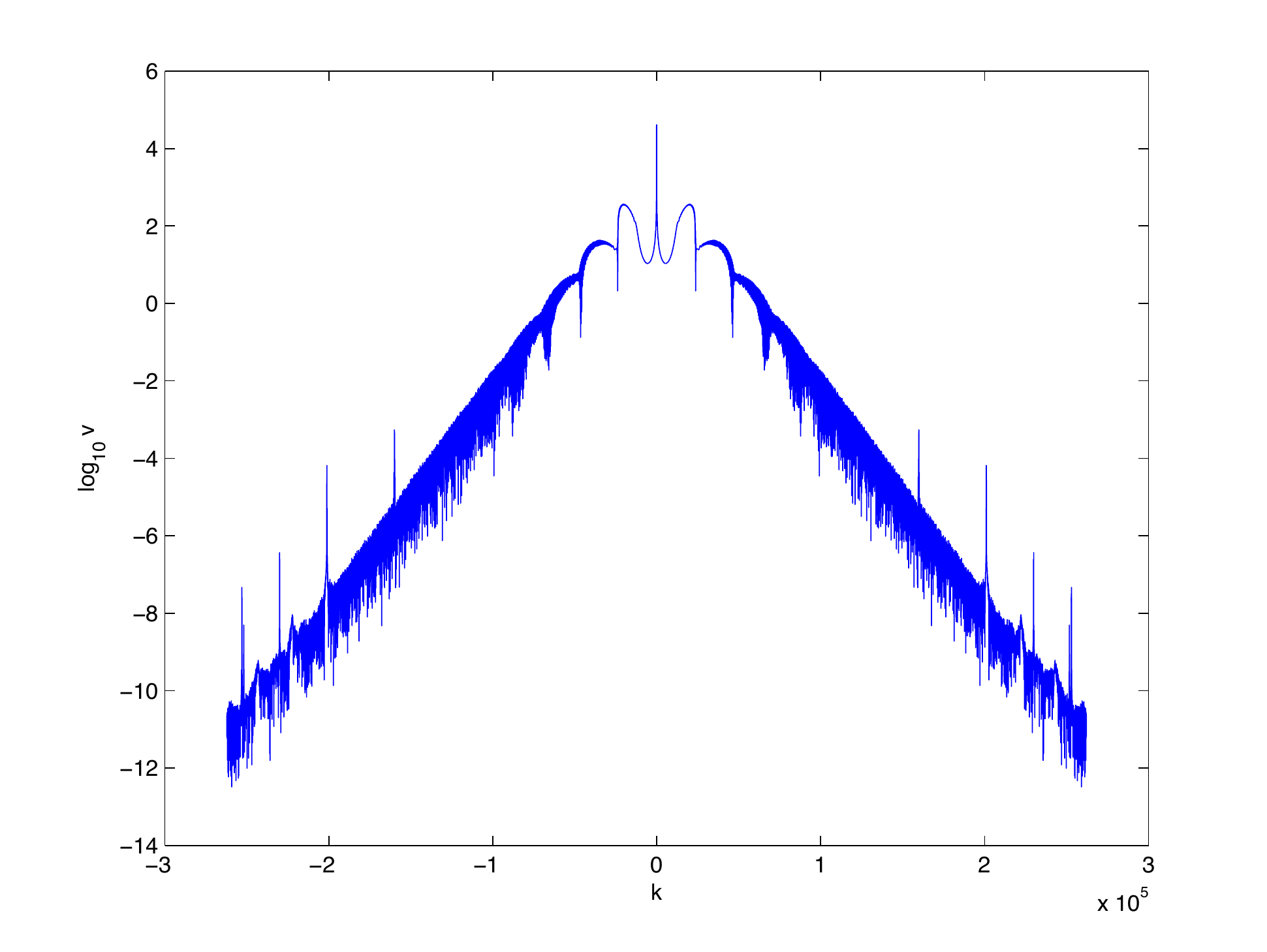}
 \caption{Fourier coefficients for the solution in Fig.~\ref{kdv1e7_t04}.}
   \label{kdv1e7_t04coeff}
\end{figure}

We consider in the following always the example 
$u_{0}=-\mbox{sech}^{2}x$ and values of 
$$\epsilon=10^{-1},10^{-1.25},\ldots 10^{-3.5}.$$ For the smallest 
values of $\epsilon$, we use $N=2^{19}$ Fourier modes and 
$N_{t}=4*10^{5}$ time steps; for larger values of $\epsilon$ between 
$2^{15}$ and $2^{17}$ Fourier modes, and between $10^{4}$ to $10^{5}$ 
time steps. The oscillatory zone for 
$\epsilon=10^{-3.5}$ can be seen in Fig.~\ref{kdv1e7_t04}. The 
Fourier coefficients for this solution are shown in 
Fig.~\ref{kdv1e7_t04coeff}.


\subsection{Numerical solution of the Whitham  equations and of the Hopf 
equation}
The Whitham equations (\ref{Whitham}) are solved
for given initial data by
inverting  the hodograph transform (\ref{hodograph}) 
to obtain $\beta_1>\beta_2>\beta_3$ as a function of $x$ and $t$, and 
similarly for the implicit solution of the Hopf equation 
(\ref{Hopfsol}).  
Since the hodograph transform becomes degenerate at the leading and 
trailing edge 
we solve the system (\ref{leading})--(\ref{leading3}) and 
(\ref{trailing1})--(\ref{trailing3}) instead of (\ref{hodograph}) 
to avoid convergence problems.

These equations are of the form 
\begin{equation}
    S_{i}(\{y_{i}\},x,t)=0,\quad i=1,\ldots,M
    \label{Si},
\end{equation}
where the $S_{i}$ denote some given real  function of the $y_{i}$ and $x$, 
$t$. The task is to determine the $y_{i}$ in dependence of $x$ and 
$t$. To this end we determine the $y_{i}$ for given $x$ and $t$ as 
the zeros of the function $S:=\sum_{i=1}^{M}S_{i}^{2}$. This is
done numerically by using the algorithm of \cite{optim} which is 
implemented as the function \emph{fminsearch} in Matlab. The 
algorithm provides an iterative approach which converges in our case 
rapidly if the starting values are close enough to the solution (see 
below how the starting values are chosen). 
We calculate the zeros to the order of machine precision. 

\begin{figure}[htb!]
  \includegraphics[width=0.5\textwidth]{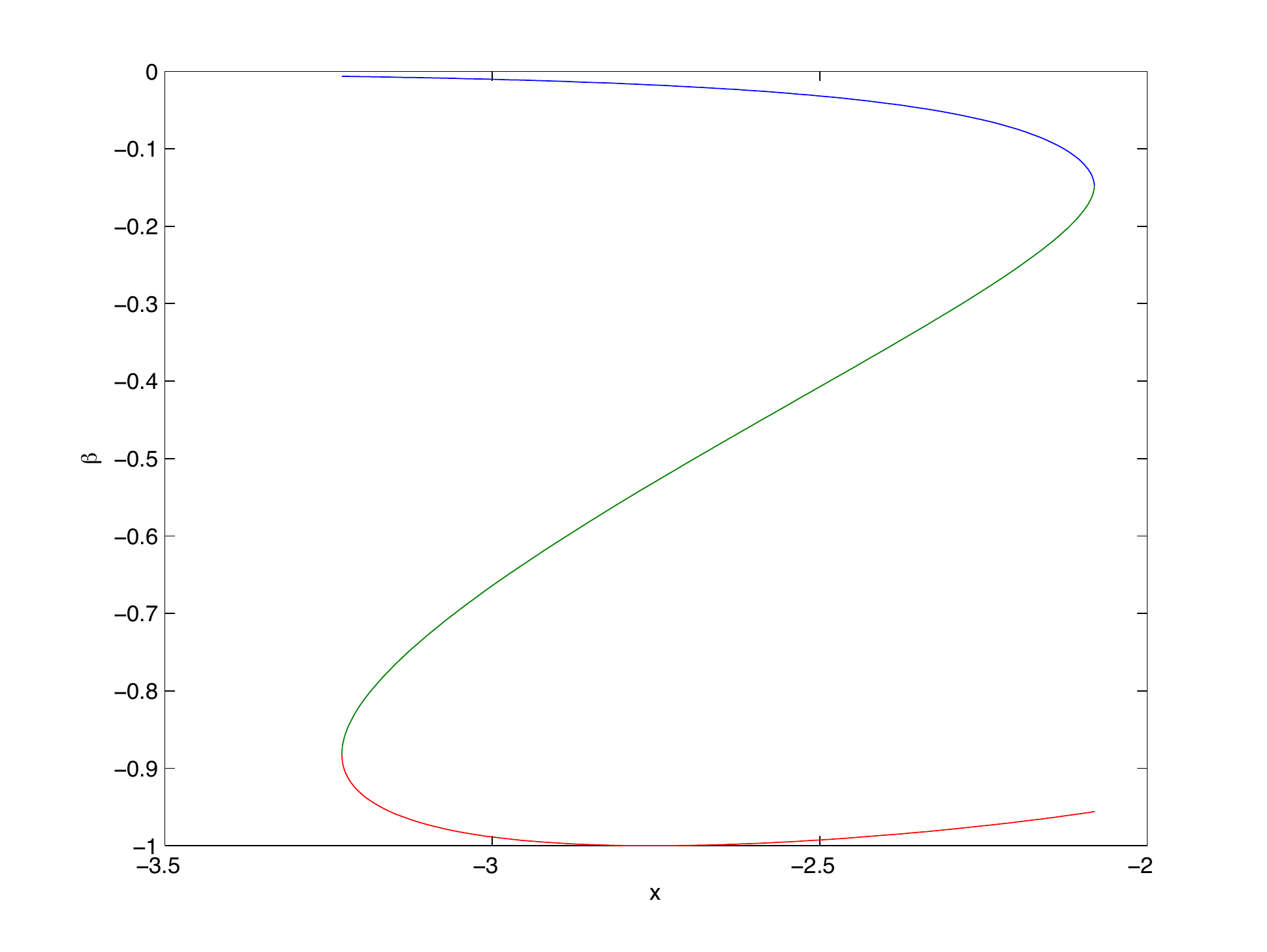}
 \caption{Solutions of the Whitham equations (\ref{hodograph}) for 
 the initial data $u_{0}=-\mbox{sech}^{2}x$ for $t=0.4$.}
   \label{beta}
\end{figure}

For a given $t>t_{c}$, we always first 
solve the system (\ref{leading})--(\ref{leading3}) to obtain 
the leading edge coordinate
$x^{-}(t)$ and \[
\beta_1^{-}(t)>\beta^{-}_2(t)=\beta^{-}_3(t).\]
 Similarly we solve the equations 
(\ref{expansion u}) for $x^{+}$ and $\beta_{1}^{+}=\beta_{2}^{+}$ and $\beta_{3}^{+}$ 
which fixes the interval $[x^{-},x^{+}]$. This interval is subdivided 
into a number of points $x_{n}$, $n=1,\ldots,N_{x}$. In contrast to 
\cite{GK}, we choose the $x_{n}$ to be related to Chebyshev 
collocation points $l_{j}=\cos(j\pi/N_{c})$, $j=0,1\ldots,N_{c}$, 
to allow for better interpolation formulas. Since the polynomial 
interpolation we will use works best for smooth functions, we use the 
analytic knowledge that $\beta_{2}\sim \beta_{3}\sim\sqrt{x-x^{-}(t)}$ 
for $x\sim x^{-}(t)$, and similarly $\beta_{1}\sim 
\beta_{2}\sim\sqrt{x^{+}(t)-x}$ 
for $x\sim x^{+}(t)$. Thus we put for $j=0,\ldots,N_{c}$
$$x_{j}=x^{-}(t)+\frac{x^{+}(t)-x^{-}(t)}{2}\frac{(1+l_{j})^{2}}{4}, \quad 
x\in[x^{-}(t),\frac{1}{2}(x^{-}(t)+x^{+}(t))]$$
and 
$$x_{j}=x^{+}(t)-\frac{x^{+}(t)-x^{-}(t)}{2}\frac{(1-l_{j})^{2}}{4}, \quad 
x\in[\frac{1}{2}(x^{-}(t)+x^{+}(t)),x^{+}(t)].$$
For given $x_{j}$ and $t$, the Whitham equations are solved as 
discussed in \cite{GK}. Thus the $\beta_{i}$ are sampled on 
Chebyshev collocation points which can be used to obtain an expansion of 
these functions in terms of Chebyshev polynomials, see for instance 
\cite{fornberg}. As for Fourier 
series, the order of magnitude of the modulus of the coefficient of the 
highest order polynomial gives for 
smooth functions an indication of the numerical resolution. For our 
example the Chebyshev coefficients decrease well below $10^{-6}$ 
with $N_{c}=64$ which is more than sufficient for our purposes. To obtain 
machine precision, the integrals in (\ref{hodograph}) would have to 
be computed  as described in \cite{GK} with higher precision for $x$ 
close to the boundaries of the Whitham zone. The $\beta_{i}$ for the 
initial data $u_{0}=-\mbox{sech}^{2}x$ for $t=0.4$ can be seen in 
Fig.~\ref{beta}.

At intermediate values of $x \in [x^{-}(t),x^{+}(t)]$, the 
$\beta_{i}$ are obtained from the values on the collocation points 
via numerically stable barycentric Lagrange interpolation, see \cite{barycentric}, which 
is essentially an efficient implementation of the Lagrange polynomial 
for Chebyshev collocation points.

\subsection{Painlev\'e transcendents}
The asymptotic solutions near the breakup point and the leading edge 
are given by pole-free solutions with a given asymptotic behaviour for 
$x\to\pm\infty$ to the P$^{2}_{I}$ equation and the 
Painlev\'e II equation respectively. A way to solve 
these equations is to give a series solution to the respective equation 
with the imposed asymptotics that is generally divergent. These 
divergent series are truncated at finite values of $x$, $x_{l}<x_{r}$ at the first 
term that is of the order of machine precision. The sum of this 
truncated series at these points is then used as boundary data, and 
similarly for derivatives at these points. Thus the problem is 
translated to a boundary value problem on the finite interval 
$[x_{l},x_{r}]$. 

In \cite{GK2} we used for the ${\rm P}_I^2$ solution a collocation method with 
cubic splines distributed as \emph{bvp4} with Matlab, in \cite{GK1} 
for the Hastings-McLeod solution of ${\rm P}_{II}$ a Chebyshev collocation 
method with a fixed point iteration. Here we use again a Chebyshev 
collocation method for both equations. As for the Whitham equations 
above, the solution of the ODEs is sampled on Chebyshev collocation 
points $x_{j}$, $j=0,\ldots,N_{c}$ which can be related to an expansion 
of the solution in terms  of Chebyshev 
polynomials. Since the derivative of a Chebyshev polynomial can be 
again expressed in terms of a linear combination of 
Chebyshev polynomials, the action of the 
derivative operator on the Hilbert space of Chebyshev polynomials is 
equivalent to the action of a matrix on this space. This leads to the 
well known Chebyshev differentiation matrices, see for instance 
\cite{trefethen}. Thus for the numerical solution, in an ODE of the 
form 
$F(u,\partial_{x}u,..)=0$, $u$ is replaced by the vector $u(x_{j})$, 
$j=0,\ldots,N_{c}$ and $\partial_{x}$ by the differentiation matrix. 
The ODE is in this setting replaced by $N_{c}+1$ algebraic 
equations. The boundary data are included via a so-called 
$\tau$-method: The equations for $j=0$ and for $j=N_{c}$ (for the 
fourth order equation $j=0,1,N_{c}-1,N_{c}$) are replaced by the 
boundary conditions. The resulting system of algebraic equations is 
solved with a standard Newton method.  The convergence of the 
solutions is in general very fast. We always stop the Newton 
iteration when machine precision is reached. Again the highest  
Chebyshev coefficients are taken as an indication of sufficient 
resolution of the solutions (they have to reach machine precision). A 
similar approach had been used in \cite{chebop} for the 
Hastings-McLeod solution. Solutions to the Painlev\'e II equation have
been computed as the solution of a 
Riemann-Hilbert problem in \cite{olver}. Certain Painlev\'e 
transcendents can be expressed in terms of Fredholm determinants 
which can be computed  with the methods of \cite{fredholm}. For the 
study of Painlev\'e solutions with poles in the complex plane, an 
approach based on Pad\'e approximants has been presented in 
\cite{FW}. 

\begin{figure}[htb!]
    \includegraphics[width=0.45\textwidth]{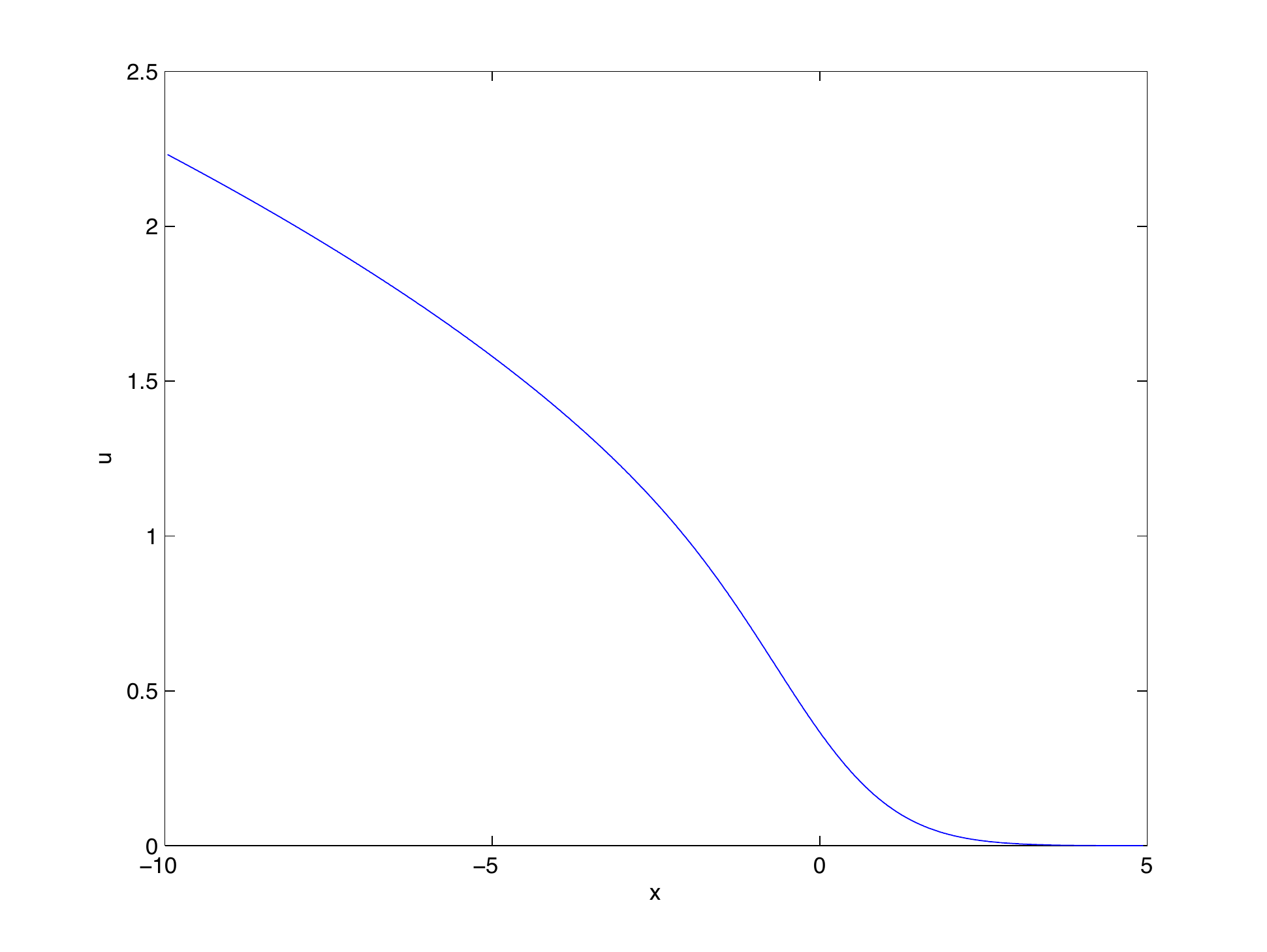}
    \includegraphics[width=0.45\textwidth]{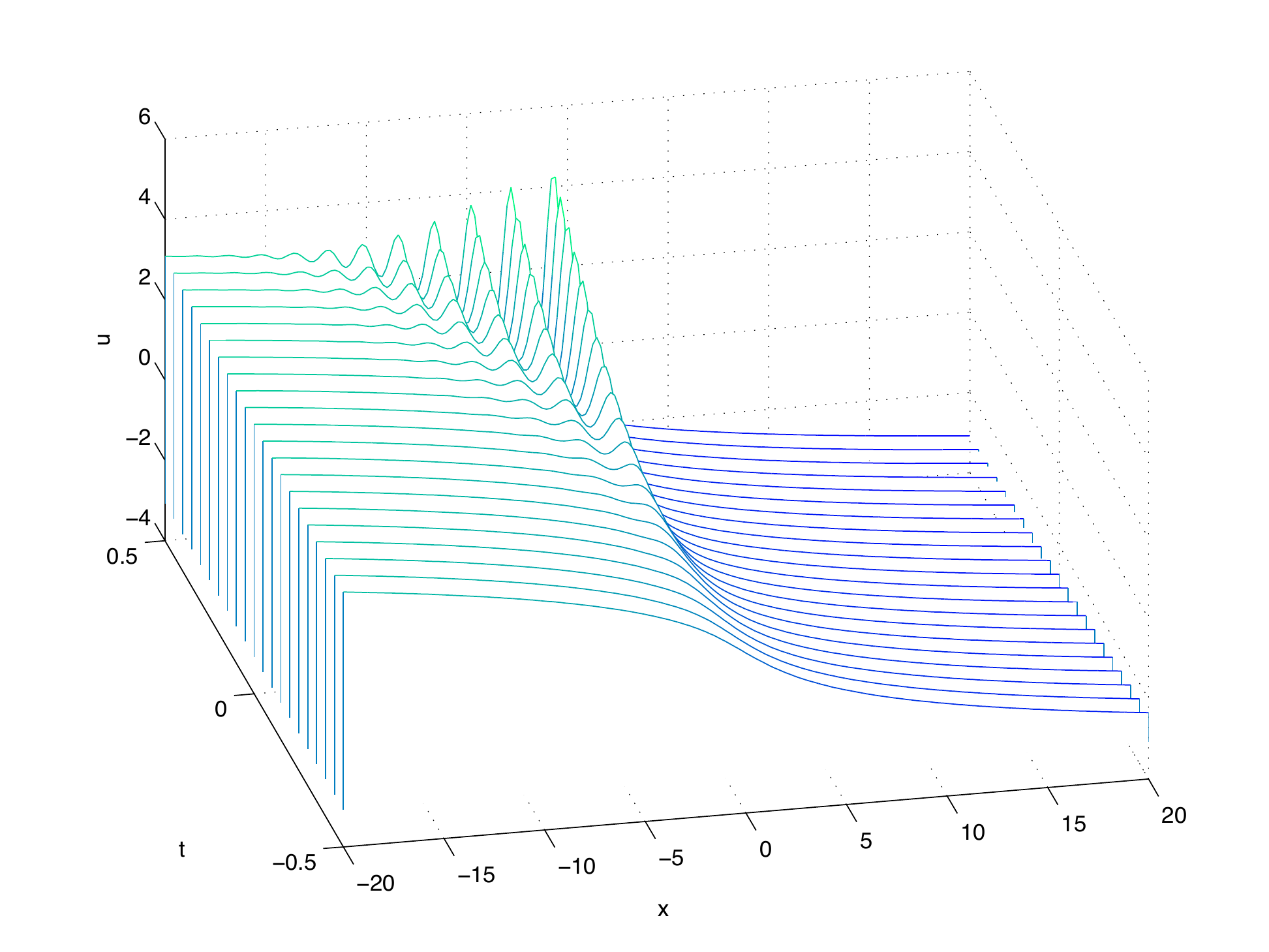}
 \caption{Hastings-McLeod solution of the Painlev\'e II equation on 
 the left, and the special solution to the Painlev\'e I2 equation for 
 several values of $t$ on the right.}
   \label{hasleod}
\end{figure}

The Hastings-McLeod solution and  the special
solution to the ${\rm P}_I^2$ equation for various values of $t$ can be seen in Fig.~\ref{hasleod}.

\section{Numerical solution of KdV, Whitham and Hopf equations}
In this and the following sections, we will numerically solve the KdV equation 
(\ref{KdV}) for the initial data $u_{0}(x)=-\mbox{sech}^{2}x$ for the 
values $\epsilon=10^{-1},10^{-1.25},\ldots,10^{-3.5}$, and compute 
the various asymptotic descriptions of the small dispersion limit 
from sect.~2. We will study the validity of these asymptotic 
descriptions  in various regions of the $(x, t)$-plane. To obtain the 
$\epsilon$-dependence of a certain quantity $A$, we perform a linear 
regression analysis for the dependence of the logarithms, $\ln A = a 
\ln \epsilon + b$.  This allows to obtain numerically the scaling of 
the difference between   numerical and asymptotic solutions also for the 
cases where no analytic behavior is yet known.  We first consider the 
asymptotic description based on the Hopf solution outside the Whitham 
zone and the one-phase KdV solution inside the zone with branch po
of the elliptic surface given by the Whitham equations. 

Outside the Whitham zone, the Hopf solution for the same initial data 
as the KdV solution gives an asymptotic description of the latter. 
Inside the Whitham zone, the one-phase KdV solution provides an 
asymptotic solution. 
\paragraph{\emph{Before breakup:}}
For times much smaller than the critical time, we find that the 
$L_{\infty}$ norm of the difference between Hopf and KdV solutions
decreases as $\epsilon^{2}$. More precisely we find by linear 
regression an exponent $a = 1.9987$ with correlation coefficient $r = 
0.999995$ and standard deviation $a=0.0051$.
\paragraph{\emph{At breakup, $t=t_{c}$:}}
For times close to the breakup time, the Hopf solution develops a 
gradient catastrophe. The largest difference between Hopf and KdV 
solution can be found close to the breakup point. We determine the 
scaling of the $L_{\infty}$ norm of the difference between Hopf and 
KdV solutions on the whole interval of computation. We find that its 
scaling is compatible with $\epsilon^{2/7}$ as conjectured in 
\cite{Dubrovin1} and proven in \cite{CG}. More precisely we find in a 
linear regression analysis $a = 0.2929$ ($2/7 = 0.2857\ldots$) with a 
correlation coefficient $r = 0.99996$ and standard deviation $a=0.0022$.

\paragraph{\emph{After breakup:}}

\begin{figure}[htb!]
  \includegraphics[width=\textwidth]{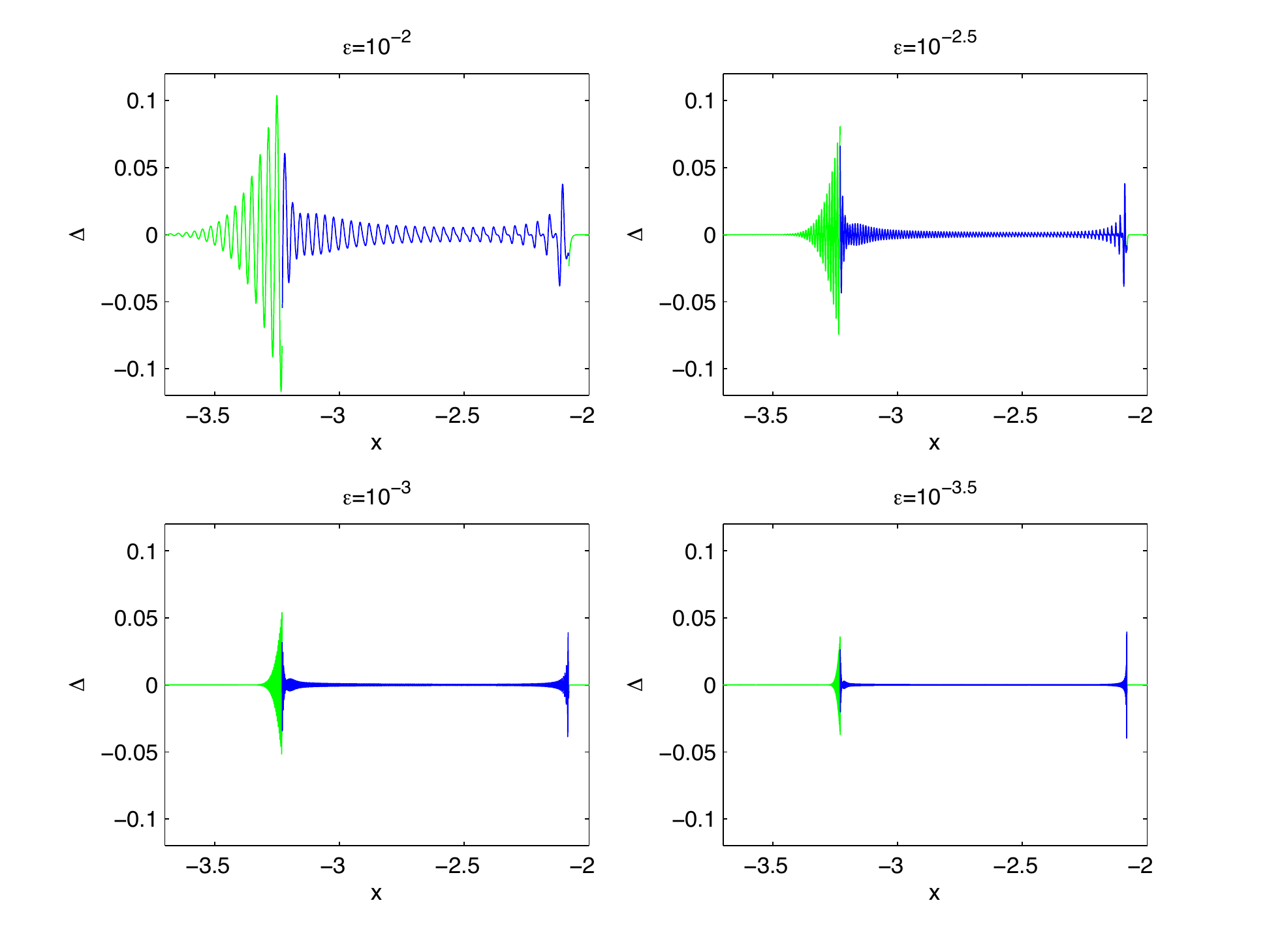}
 \caption{The blue line describes the  difference between the 
numerical solution of the KdV equation  and
the asymptotic formula (\ref{elliptic})
for the initial data $u_0(x)=-1/\cosh^2x$ and for $t=0.4$.
The green lines represent the  difference between the 
numerical solution of the KdV equation  and the Hopf solution 
(\ref{Hopfsol}).}
   \label{kdvwhitdiff4t}
\end{figure}

For times much greater than the critical time of the Hopf solution, 
we find that the asymptotic solution given by Hopf  solution and the one-phase 
KdV solution via the Whitham equations gives a very good description 
of the KdV solution. Thus it is necessary to plot the difference 
between these solutions as done in Fig.~\ref{kdvwhitdiff4t}

It can be seen that this difference is not uniform in $x$, and that this 
applies also for the decrease with $\epsilon$. The approximation is 
very good close to the centre of the Whitham zone, but much worse at 
the edges. We define the interior part of the zone tentatively as 
 the interval symmetric to the centre of half the length of the 
zone. The results do not depend on whether this zone is taken slightly 
smaller or bigger. We find that the $L_{\infty}$ norm of the 
difference of the 
numerical KdV solution and the one-phase KdV solution (\ref{elliptic}) 
decreases there roughly as $\epsilon$. More precisely we find in a 
linear regression analysis $a = 0.98$  with a 
correlation coefficient $r = 0.998$ and standard deviation $a=0.047$.

At the leading edge we find that the error is always biggest close to 
the boundary of the Whitham zone. In an interval symmetric to this 
boundary with the same length as the above interior zone, we find 
that the difference between KdV solution and asymptotic solutions via 
Hopf and the one-phase KdV solution (\ref{elliptic}) scale roughly as 
$\epsilon^{1/3}$. More precisely we find in a 
linear regression analysis $a = 0.33$  with a 
correlation coefficient $r = 0.999$ and standard deviation $a=0.012$.

The situation at the trailing edge is more complicated. It can be 
seen in Fig.~\ref{kdvwhitdiff4t} that the difference between the KdV  solution and 
one-phase KdV solution (\ref{elliptic}) is almost constant (roughly 0.04) 
there. Notice that this error of order $\mathcal{O}(1)$ which is supposed to appear 
in a zone of width $\epsilon\ln \epsilon$ close to the trailing edge 
of the Whitham zone was not seen in \cite{GK} because of a lack of 
resolution. This is one of the reasons why we redid the computations 
with a considerably higher resolution. It can also be seen in 
Fig.~\ref{kdvwhitdiff4t} that the $\mathcal{O}(1)$ oscillation is moving 
closer and closer to the edge with smaller $\epsilon$ as expected. 
The difference between KdV and Hopf solutions close to the trailing 
edge decreases, however, roughly as $\sqrt{\epsilon}$. More precisely we find in a 
linear regression analysis $a = 0.54$  with a 
correlation coefficient $r = 0.997$ and standard deviation $a=0.03$.

\section{Leading edge}
In this section we study numerically the asymptotic formula 
(\ref{expansionu}) via the Hastings-McLeod solution which approximates the KdV solution at the leading edge as $\epsilon\rightarrow 0$.  We will refer to this asymptotic solution as ${\rm P}_{II}$ asymptotics.
% It will be shown that the latter
%provides a better description of the asymptotic behavior near the
%leading edge of the Whitham zone than the Hopf or the elliptic
%KdV solution. 
We identify the
zone, where the   ${\rm P}_{II}$ asymptotics 
gives a better description of KdV than the
 Hopf  (\ref{Hopfasym}) or the one-phase KdV solution (\ref{elliptic})  and study the $\epsilon$-dependence of the errors. 

\begin{figure}[htb!]
  \includegraphics[width=.7\textwidth]{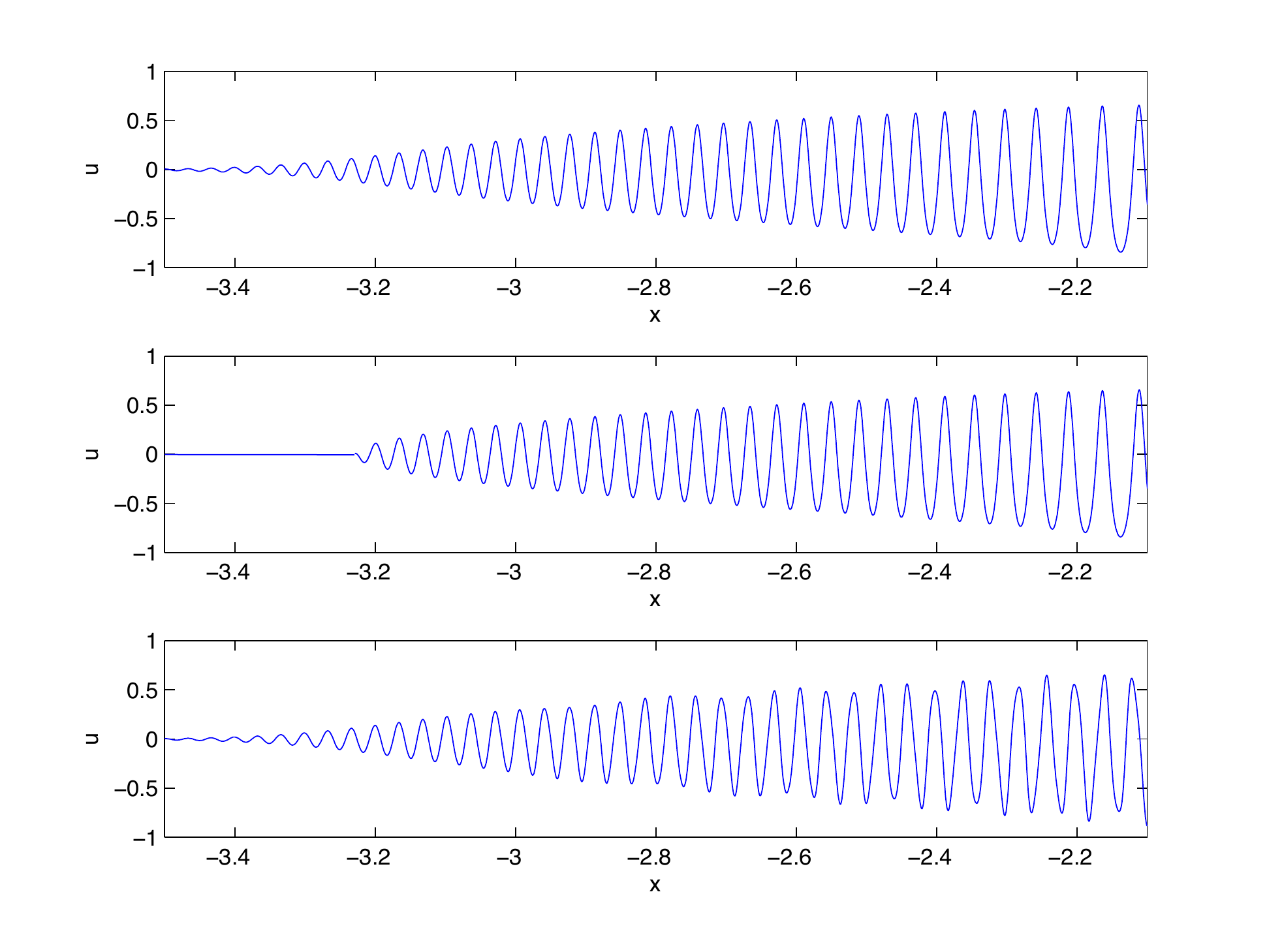}
 \caption{The
figure shows in the upper part the numerical solution to the KdV
equation for the initial datum $u_{0}=-\mbox{sech}^{2}x$ and 
$\epsilon=10^{-2}$ at $t=0.4$, in the middle the
corresponding asymptotic solution in terms of Hopf and one-phase KdV
solution, and in the lower part the ${\rm P}_{II}$ asymptotic  solution 
(\ref{expansionu}).}
   \label{kdv1e4t4lead3}
\end{figure}

In Fig.~\ref{kdv1e4t4lead3} we show the KdV solution, the asymptotic
solution via Whitham and Hopf and the ${\rm P}_{II}$- asymptotics  near
the leading edge of the Whitham zone. It can be seen that the
one-phase KdV solution gives a very good description in the interior
of the Whitham zone as discussed above, whereas the
${\rm P}_{II}$  asymptotics gives as expected a better description near
the leading edge.

\begin{figure}[htb!]
  \includegraphics[width=0.5\textwidth]{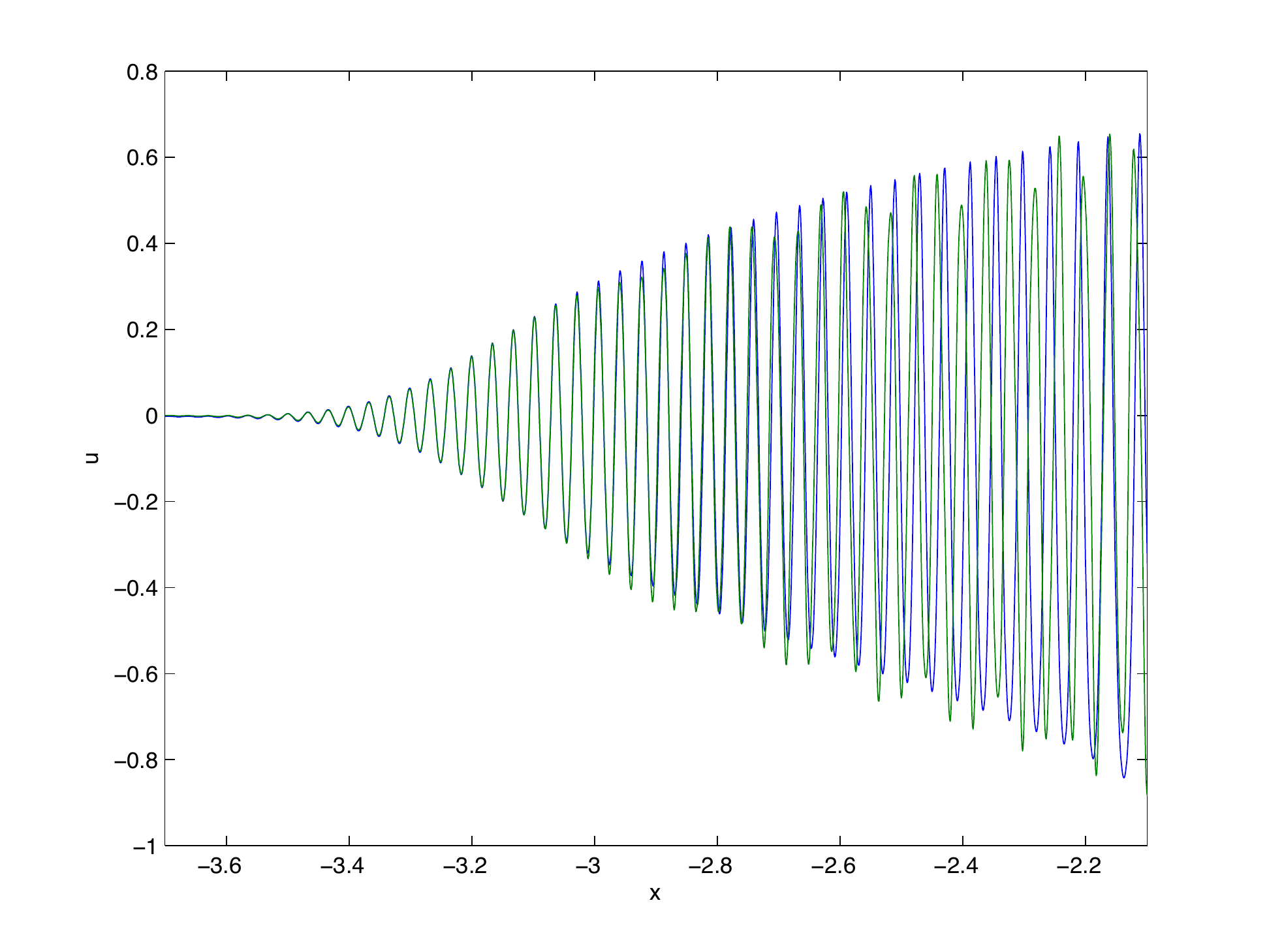}
 \caption{The
numerical solution to the KdV  equation for the initial datum
$u_{0}=-\mbox{sech}^{2}x$  and $\epsilon=10^{-2}$ at
$t=0.4$ in blue and the corresponding  ${\rm P}_{II}$ asymptotic  solution 
(\ref{expansionu}) in
green.}
   \label{kdv1e4t4lead}
\end{figure}
In Fig.~\ref{kdv1e4t4lead} the KdV  solution and the ${\rm P}_{II}$ asymptotics are
shown in one plot for $\epsilon=10^{-2}$. It can be seen that the
agreement near the edge of the Whitham zone is so good that one
has to study the difference of the solutions. The solution only
gives locally an asymptotic description and is quickly out of
phase for larger distances from the leading edge, whereas the 
amplitude is roughly of the right size.
\begin{figure}[htb!]
  \includegraphics[width=\textwidth]{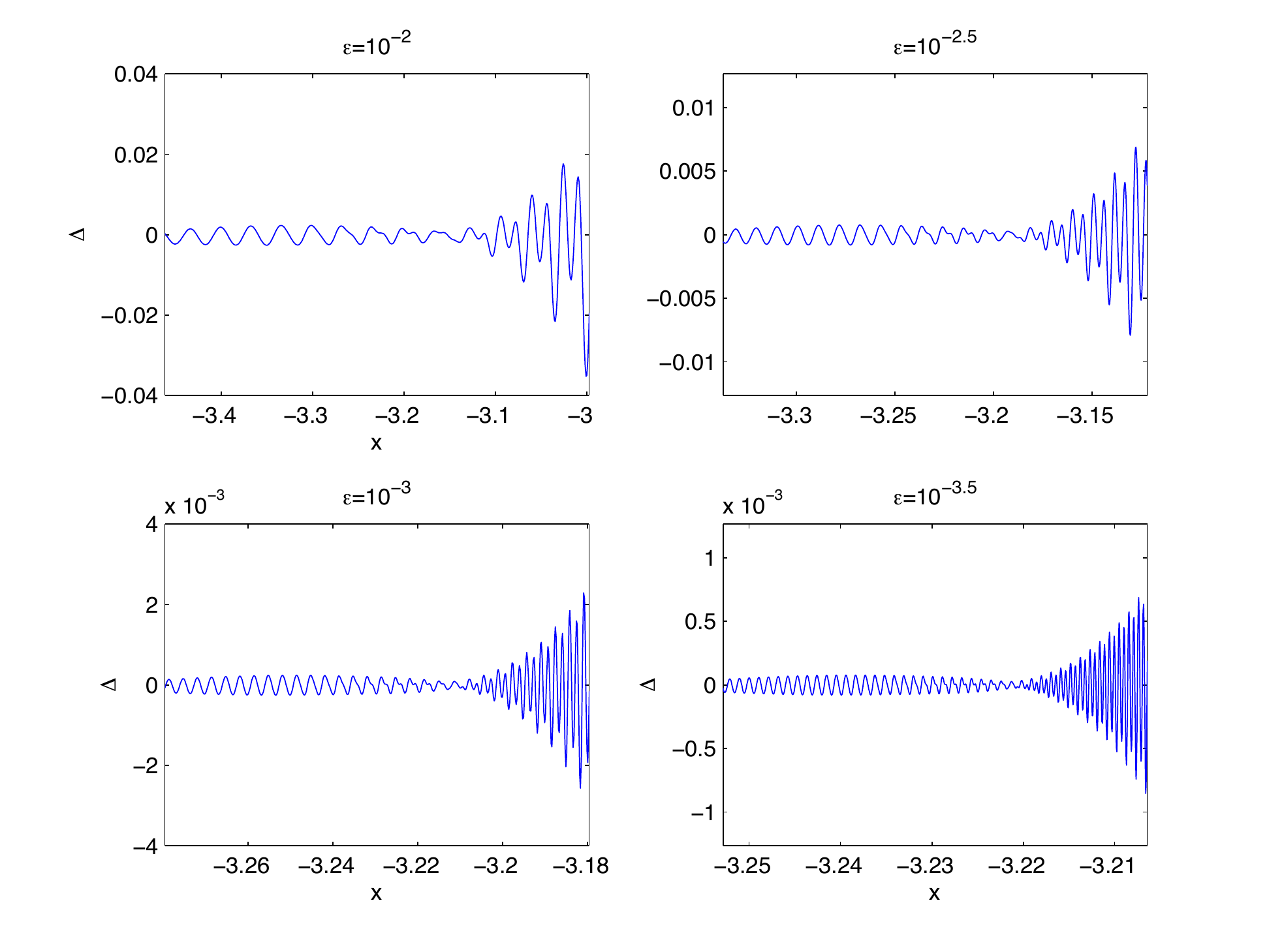}
 \caption{The difference between the numerical solution to the KdV
equation for the initial datum $u_{0}=-\mbox{sech}^{2}x$ at $t=0.4$ 
and the corresponding  multiscale solution (\ref{expansionu}) for
four values of $\epsilon$.  Notice the scaling of the $x$ and $\Delta$ axes with
a factor $\epsilon^{2/3}$ and $\epsilon$ respectively to take care of 
the expected scalings in $x$
and of the shown error next to the leading edge of the Whitham zone.}
   \label{kdvlead4e}
\end{figure}

The difference between KdV solution and the ${\rm P}_{II}$ asymptotics is shown for
several values of $\epsilon$ in Fig.~\ref{kdvlead4e}. It can
be seen that the  error  close to the Whitham
edge is almost constant. The scales in $x$ 
and $\Delta$ are rescaled by a factor $\epsilon^{2/3}$ and $\epsilon$ 
respectively which is the
expected scaling behavior of the zone, where the multiscale
solution should be applicable, and of the expected error. It can be seen
that with these rescalings the error is of the same order for 
different values of $\epsilon$.  A linear
regression analysis for the logarithm of the difference $\Delta$
between KdV and multiscale solution in the interval 
$[x^{-}-\epsilon^{2/3},x^{-}+\epsilon^{2/3}]$ gives a scaling
of the form $\Delta\propto \epsilon^{a}$ with $a=1.00$ with
standard deviation $\sigma_{a}=0.004$ and correlation coefficient 
$r=0.99999$. The result is almost the same in a larger interval, 
e.g., $[x^{-}-2\epsilon^{2/3},x^{-}+2\epsilon^{2/3}]$ with just a 
slightly worse correlation. The found scaling is thus as expected of 
order $\epsilon$.

\begin{figure}
  \centering
  \subfloat[]{\label{kdv1e4t42d} \includegraphics[width=0.5\textwidth]{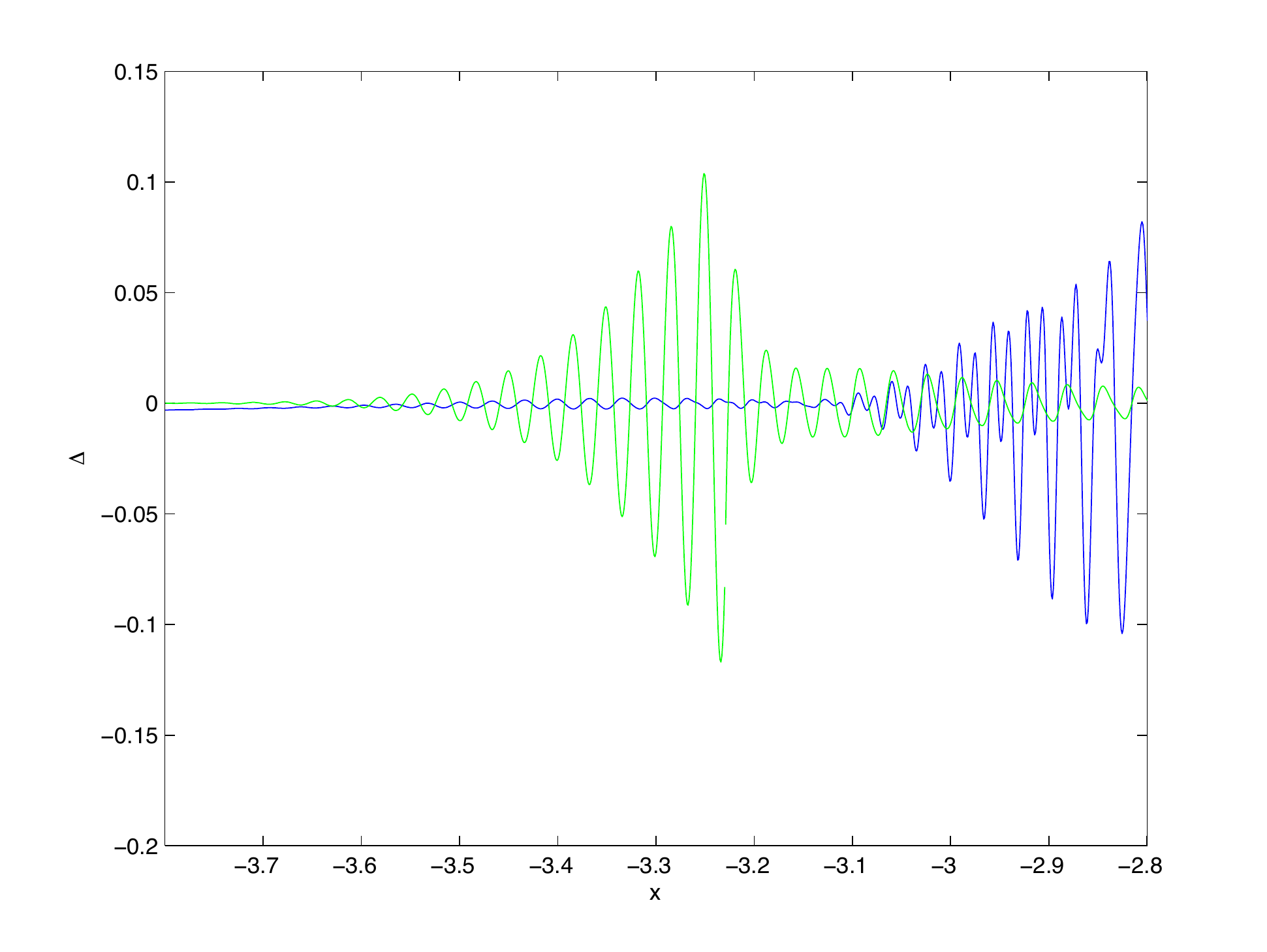}}     
  \subfloat[]{\label{leadzone}\includegraphics[width=0.5\textwidth]{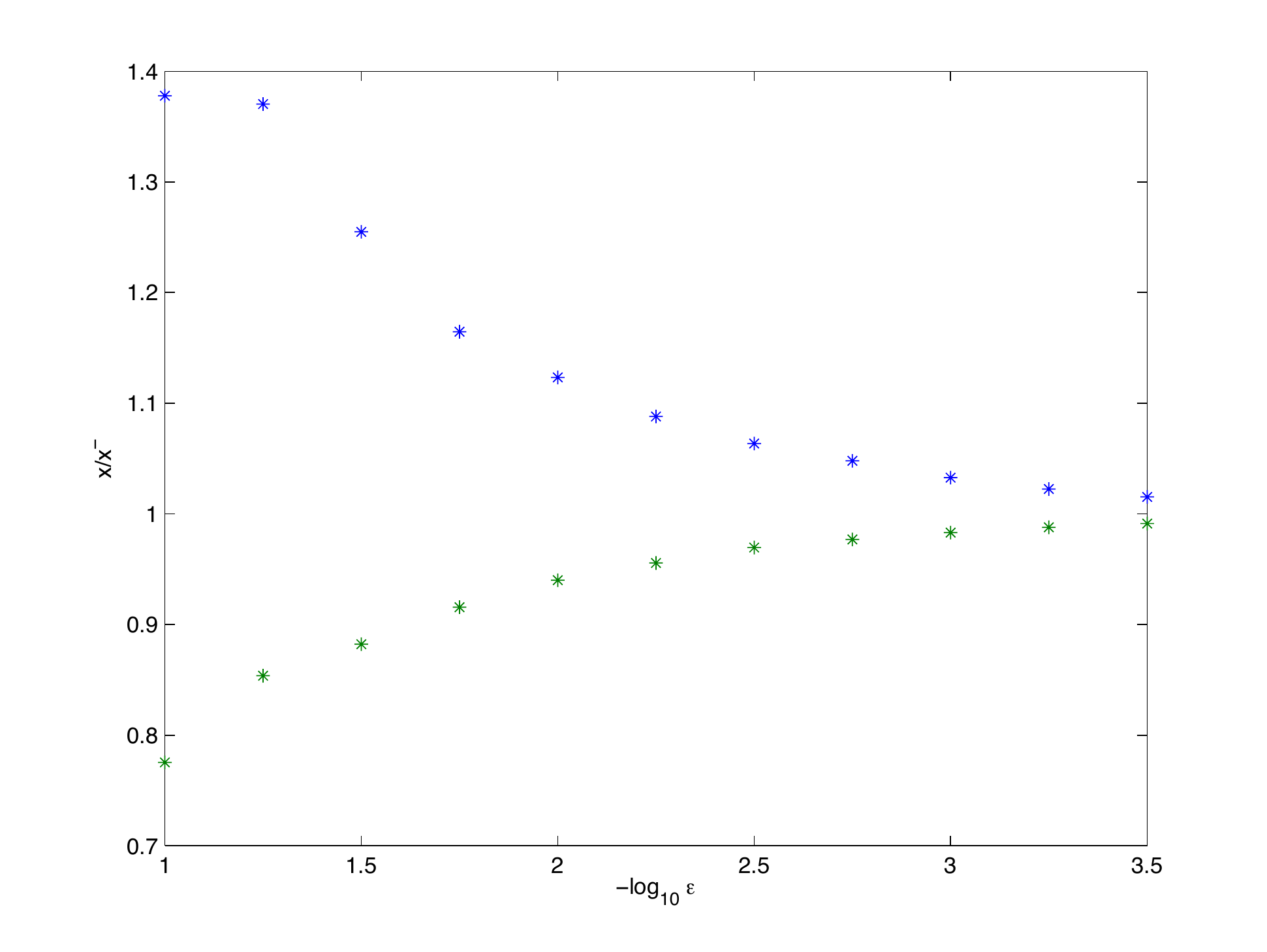}}
  \caption{In Fig.~(a) the
difference between the numerical solution to the KdV equation for
the initial datum $u_{0}=-\mbox{sech}^{2}x$ at $t=0.4$ for $\epsilon=10^{-2}$ and the
corresponding  ${\rm P}_{II}$ asymptotic solution  (\ref{expansionu}) in blue, and the difference
between KdV and Hopf and one-phase KdV solution in green.
In Fig.~(b) the edges
of the zone where the  ${\rm P}_{II}$ asymptotic solution  (\ref{expansionu}) provides a better
asymptotic description of KdV than the Hopf or the one-phase KdV
solution  in dependence of $\epsilon$.  }    
\end{figure}

As can be already seen from Fig.~\ref{kdv1e4t4lead3}, the multiscale
solution gives a better asymptotic description of KdV near the
leading edge of the Whitham zone than the Hopf and the one-phase KdV
solution. This is even more obvious in Fig.~\ref{kdv1e4t42d} where
the difference between KdV and the asymptotic solutions is shown.

\begin{figure}
  \centering
  \subfloat[]{\label{fig:gull}\includegraphics[width=0.5\textwidth]{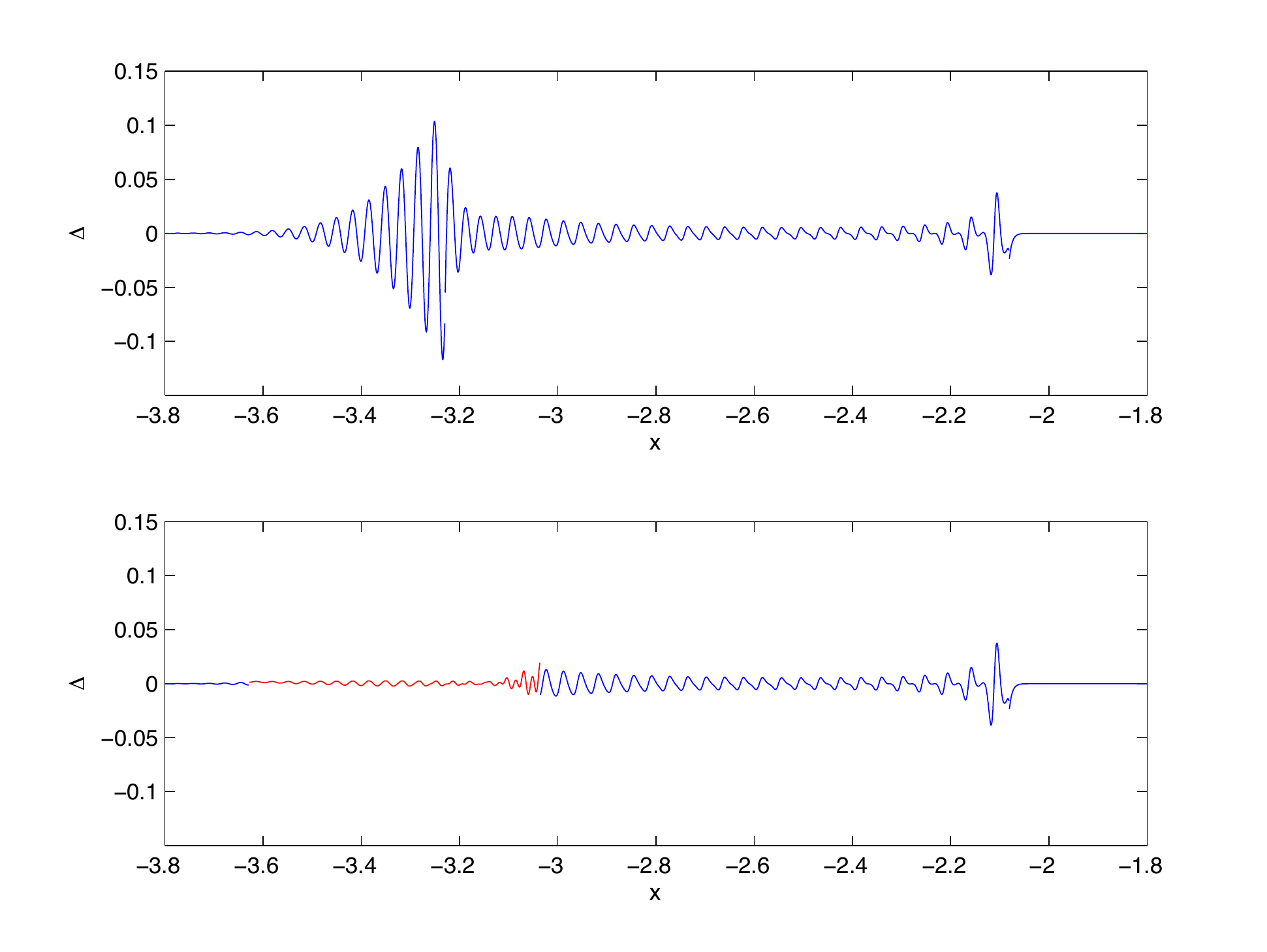}}                
  \subfloat[]{\label{fig:tiger}\includegraphics[width=0.5\textwidth]{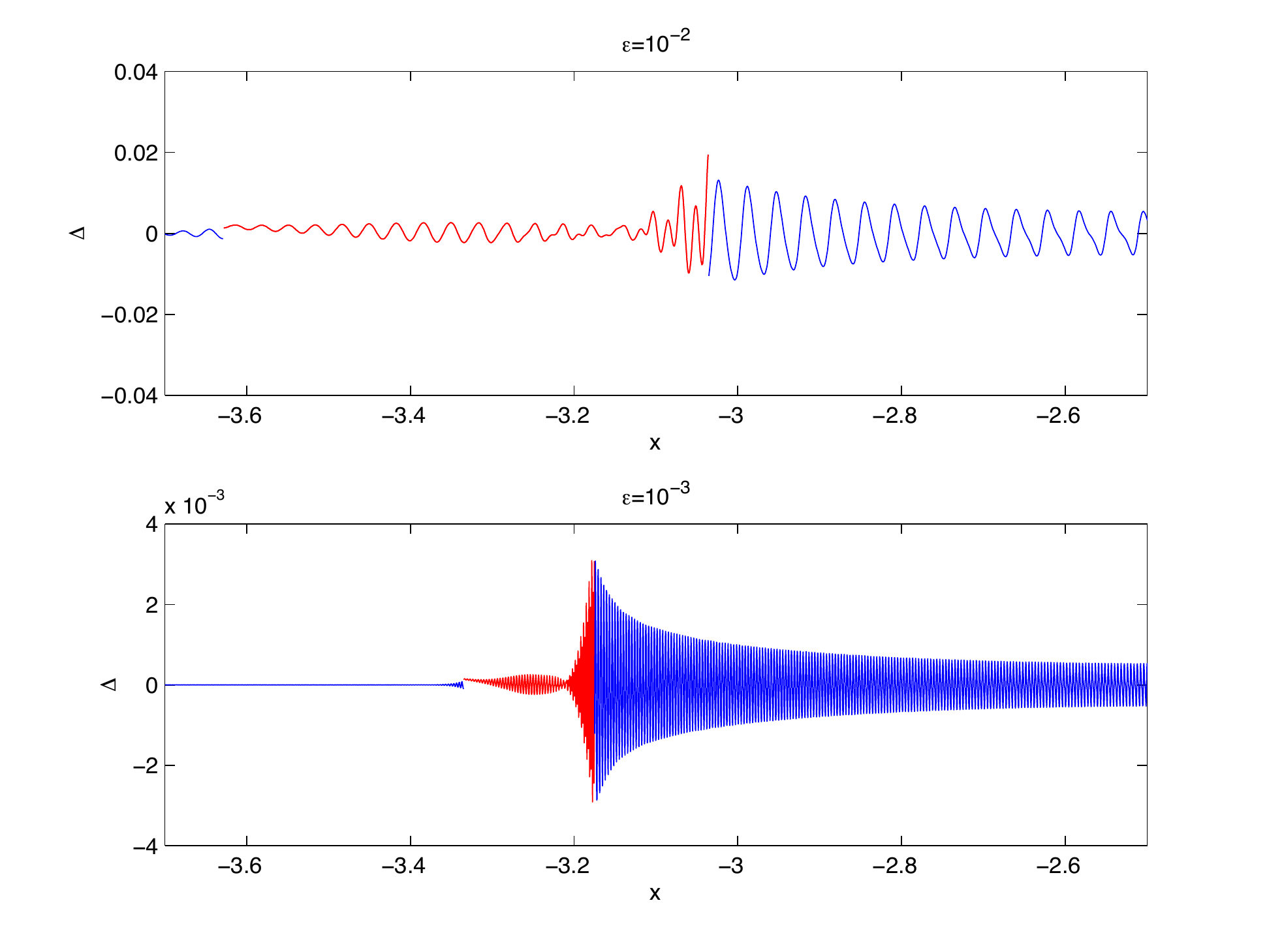}}
  \caption{
  In the upper part of Fig.~(a)  one can see the difference between the numerical
solution to the KdV equation for the initial datum
$u_{0}=-\mbox{sech}^{2}x$ and  $\epsilon=10^{-2}$ at
$t=0.4$ and the corresponding asymptotic solution in terms of Hopf
and one-phase KdV solutions. The lower part shows the same
difference, which is replaced close to the leading edge of the
Whitham zone by the difference between KdV solution and the ${\rm P}_{II}$ asymptotic solution (\ref{expansionu}) (shown in red where the error is smaller than the one
shown above). The figures in (b) show the same situation as in the 
lower part of (a) for two values of $\epsilon$. Notice the rescaling 
of the $\Delta$ axis with a factor $\epsilon$, the expected scaling 
of the error.
  }
   \label{kdv1e4t4matchl}
\end{figure}

This suggests to identify the regions where each of the asymptotic
solutions gives a better description of KdV than the other. The
results of this analysis can be seen in Fig.~\ref{kdv1e4t4matchl}.
This matching procedure clearly improves the KdV description near
the leading edge. We also show the difference between this matched
asymptotic solution and the KdV solution for two values of
$\epsilon$. Visibly the zone, where the solutions are matched,
decreases with $\epsilon$.

%\begin{figure}[htb!]
%  \includegraphics[width=0.6\textwidth]{kdv1e4t42d.pdf}
% \caption{The
%difference between the numerical solution to the KdV equation for
%the initial datum $u_{0}=-\mbox{sech}^{2}x$ and 
%$\epsilon=10^{-2}$ at $t=0.4$ for $\epsilon=10^{-2}$ and the
%corresponding  multiscale solution (\ref{expansionu}) in blue, and the difference
%between KdV and Hopf and elliptic solution in green.}
%   \label{kdv1e4t42d}
%\end{figure}
%\begin{figure}[htb!]
%  \includegraphics[width=0.6\textwidth]{kdv1e4t4matchl.pdf}
% \caption{In
%the upper part one can see the difference between the numerical
%solution to the KdV equation for the initial datum
%$u_{0}=-\mbox{sech}^{2}x$ and  $\epsilon=10^{-2}$ at
%$t=0.4$ and the corresponding asymptotic solution in terms of Hopf
%and elliptic solution. The lower figure shows the same
%difference, which is replaced close to the leading edge of the
%Whitham zone by the difference between KdV solution and multiscales
%solution (\ref{expansionu}) (shown in red where the error is smaller than the one
%shown above).}
%   \label{kdv1e4t4matchl}
%\end{figure}

There is a certain ambiguity in the precise definition of this
matching zone due to the oscillatory character of the solutions.
The limits of the matching zone for several values of
$\epsilon$ can be seen in Fig.~\ref{leadzone}.  Due to the lower
number of oscillations in the Hopf region, the matching zone
extends much further into this region than in the Whitham region.
There does not appear to be a clear scaling law for the  width of 
this zone. It can be already seen in Fig.~\ref{kdv1e4t4matchl} that 
the error at the matching does not scale with $\epsilon$.  In fact we find a scaling close to 
$\epsilon^{a}$ with $a\sim 2/3$ (in the Whitham zone we find $a=0.63$ 
and $\sigma_{a}=0.015$ with $r=0.9995$, and in the ${\rm P}_{II}$ zone $a=0.60$ 
and $\sigma_{a}=0.063$ with $r=0.99$). Thus it is not possible to 
obtain an error of order $\epsilon$ up to the trailing edge. It is 
clear that analytic connection formulae between the two 
asymptotic solutions must be established to obtain an error of order 
$\epsilon$ in the shown range.

%\begin{figure}[htb!]
%  \includegraphics[width=.5\textwidth]{leadzone.pdf}
% \caption{The edges
%of the zone where the multiscales solution (\ref{expansionu}) provides a better
%asymptotic description of KdV than the Hopf or the elliptic
%solution for the initial datum $u_{0}=-\mbox{sech}^{2}x$  at $t=0.4$ 
%in dependence of $\epsilon$.}
%   \label{leadzone}
%\end{figure}

\section{Trailing edge }
In this section we study numerically for times greater than the 
critical time the soliton asymptotic formula 
(\ref{expansion u}) that approximates as $\epsilon\rightarrow 0$ the solution of KdV near the trailing edge of the oscillatory zone.  We identify the
zone, where this asymptotic formula  gives a better description of 
KdV than the one-phase KdV (\ref{elliptic}) and Hopf  (\ref{Hopfasym}) solutions  and study the $\epsilon$-dependence of the errors. 

\begin{figure}[htb!]
  \includegraphics[width=0.7\textwidth]{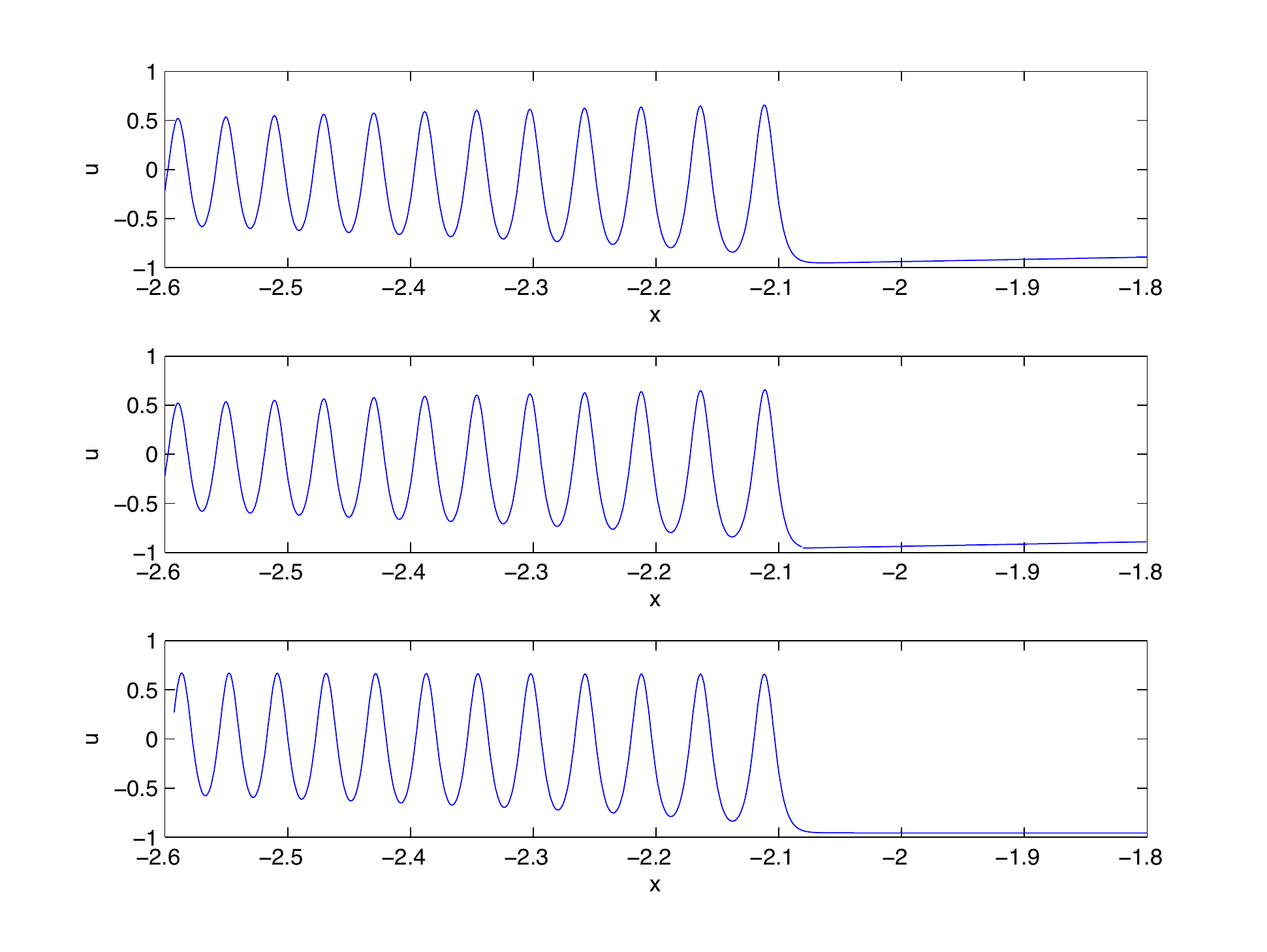}
 \caption{The
figure shows in the upper part the numerical solution to the KdV
equation for the initial datum $u_{0}=-\mbox{sech}^{2}x$ and 
$\epsilon=10^{-2}$ at $t=0.4$, in the middle the
corresponding asymptotic solution in terms of Hopf and one-phase KdV
solution, and in the lower part the multiscale solution 
(\ref{expansion u}).}
   \label{kdv1e4t4trail3}
\end{figure}

In Fig.~\ref{kdv1e4t4trail3} we show the KdV solution, the asymptotic
solution via Whitham and Hopf and the soliton asymptotics  near
the trailing edge of the Whitham zone. As before the
one-phase KdV solution gives a very good description in the interior
of the Whitham zone, whereas the
soliton asymptotic formula   gives as expected a better description near
the trailing edge.

\begin{figure}[htb!]
  \includegraphics[width=0.6\textwidth]{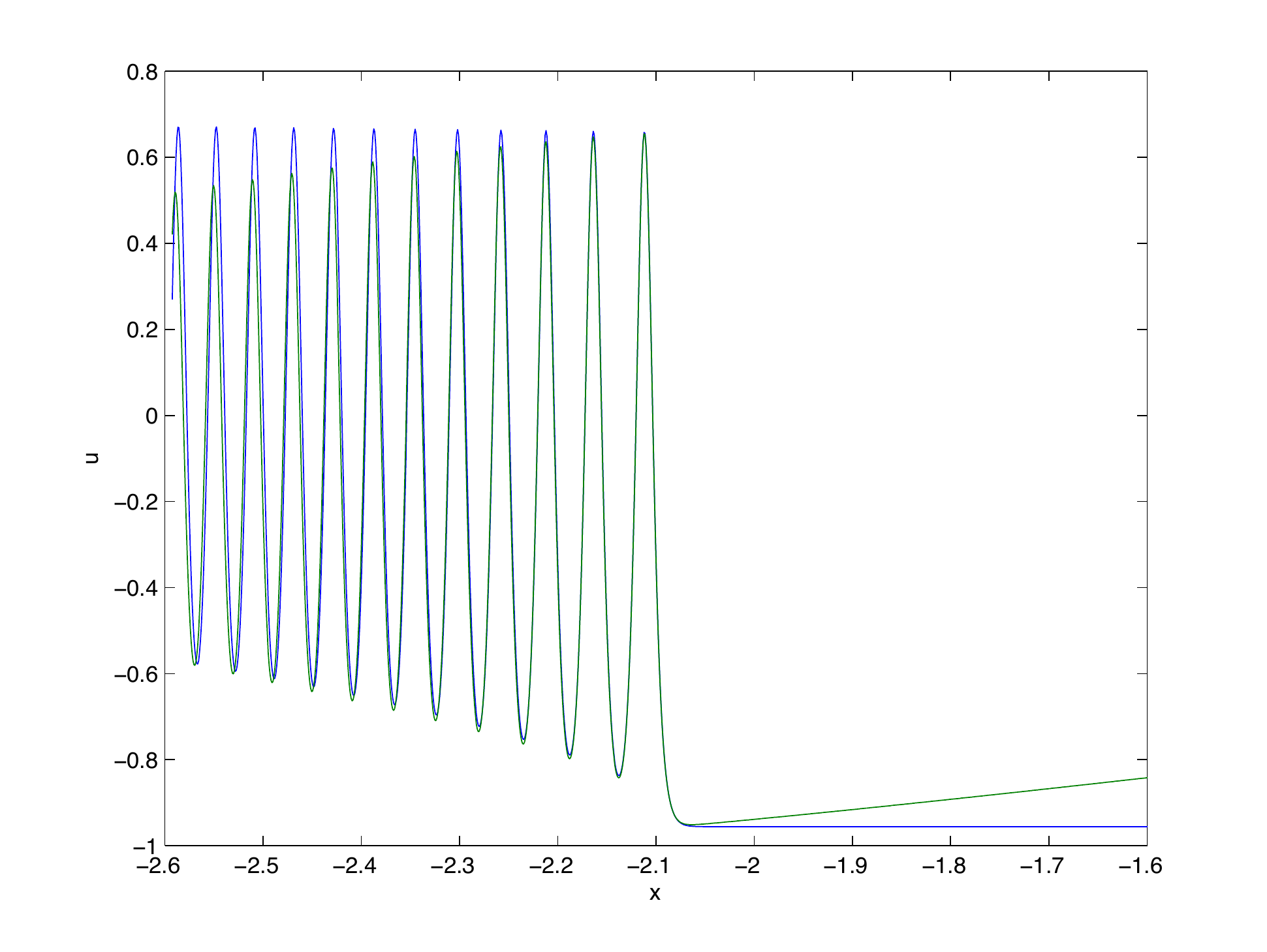}
 \caption{The
numerical solution to the KdV  equation for the initial datum
$u_{0}=-\mbox{sech}^{2}x$  and $\epsilon=10^{-2}$ at
$t=0.4$ in blue and the corresponding  multiscale solution 
(\ref{expansion u}) in
green.}
   \label{kdv1e4t4trail}
\end{figure}

In Fig.~\ref{kdv1e4t4trail} the KdV and the multiscale solution are
shown in one plot for $\epsilon=10^{-2}$. It can be seen that the
agreement very close to the boundary of the Whitham zone is once more 
so good that the difference of the solutions has to be studied. The solution only
gives locally an asymptotic description, and the quality of the 
approximation is not symmetric around the critical point.

\begin{figure}[htb!]
  \includegraphics[width=\textwidth]{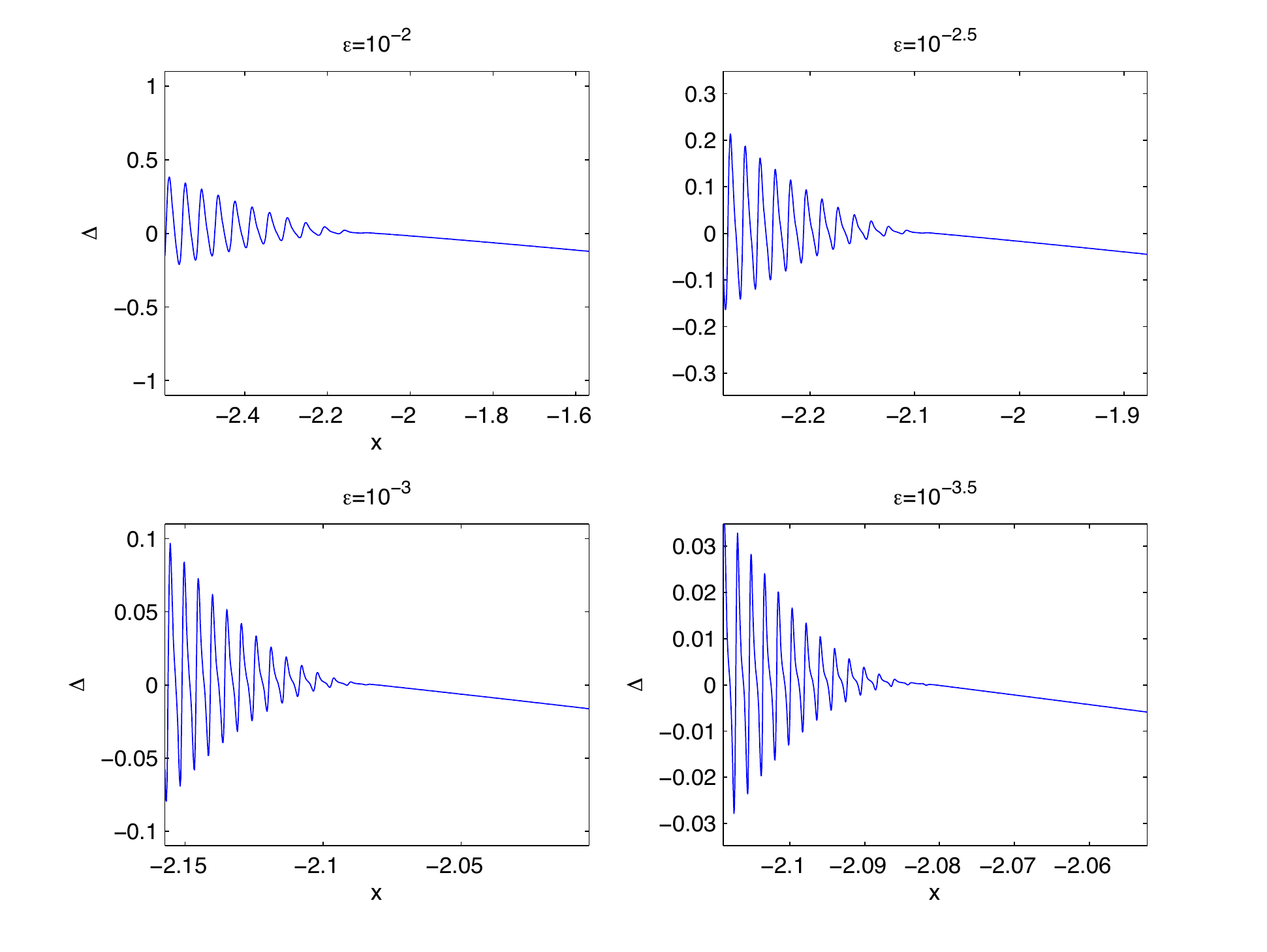}
 \caption{The difference between the numerical solution to the KdV
equation for the initial datum $u_{0}=-\mbox{sech}^{2}x$ at $t=0.4$ 
and the corresponding  soliton asymptotic  solution (\ref{expansion u}) for
four values of $\epsilon$.  Note the scaling of the $x$ and $\Delta$ axes with
a factor $\epsilon$ to take care of 
the expected scaling in $x$
and of the shown error next to the trailing edge of the Whitham zone.}
   \label{kdvtrail4e}
\end{figure}

The difference between KdV solution  and the soliton asymptotic  solution is shown for
several values of $\epsilon$ in Fig.~\ref{kdvtrail4e}. The scales in $x$ 
and $\Delta$ are both rescaled by a factor $\epsilon$ 
which is the
expected scaling behavior of the zone (numerically $\epsilon$ and 
$\epsilon\ln \epsilon$ are indistinguishable), where the  soliton asymptotic 
solution should be applicable, and of the expected error. It can be seen
that with these rescalings the error is of the same order for 
different values of $\epsilon$.  A linear
regression analysis for the logarithm of the difference $\Delta$
between KdV and multiscale solution in the interval 
$[x^{+}+\epsilon\ln\epsilon,x^{+}-\epsilon\ln\epsilon]$ gives a scaling
of the form $\Delta\propto \epsilon^{a}$ with $a=1.07$ with
standard deviation $\sigma_{a}=0.056$ and correlation coefficient 
$r=0.998$. The found scaling is thus compatible   
with $\epsilon\ln\epsilon$.

\begin{figure}
  \centering
  \subfloat[]{ \label{kdv1e4t4trail2d}\includegraphics[width=0.5\textwidth]{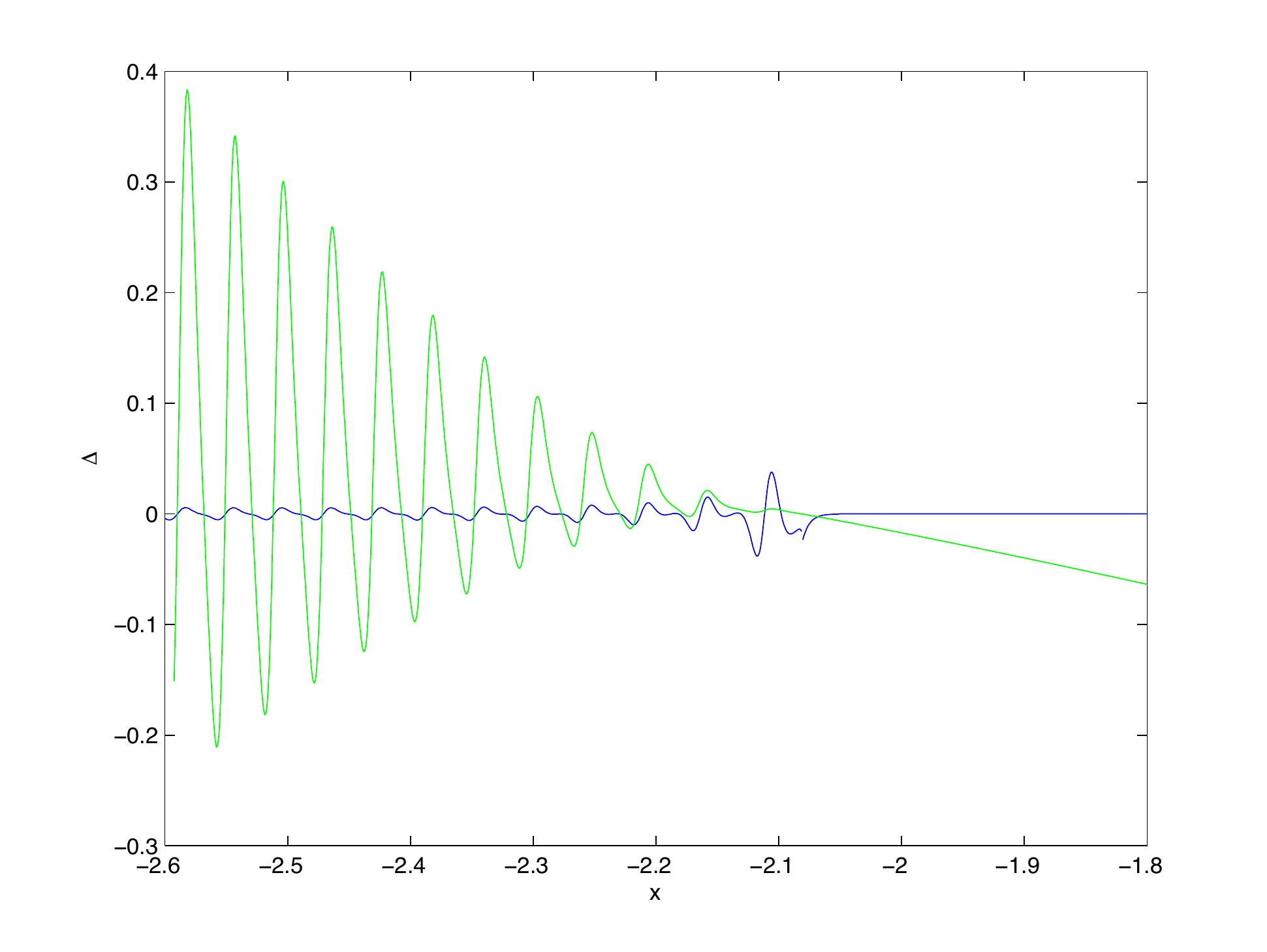}}                
  \subfloat[]{\label{trailzone}\includegraphics[width=0.5\textwidth]{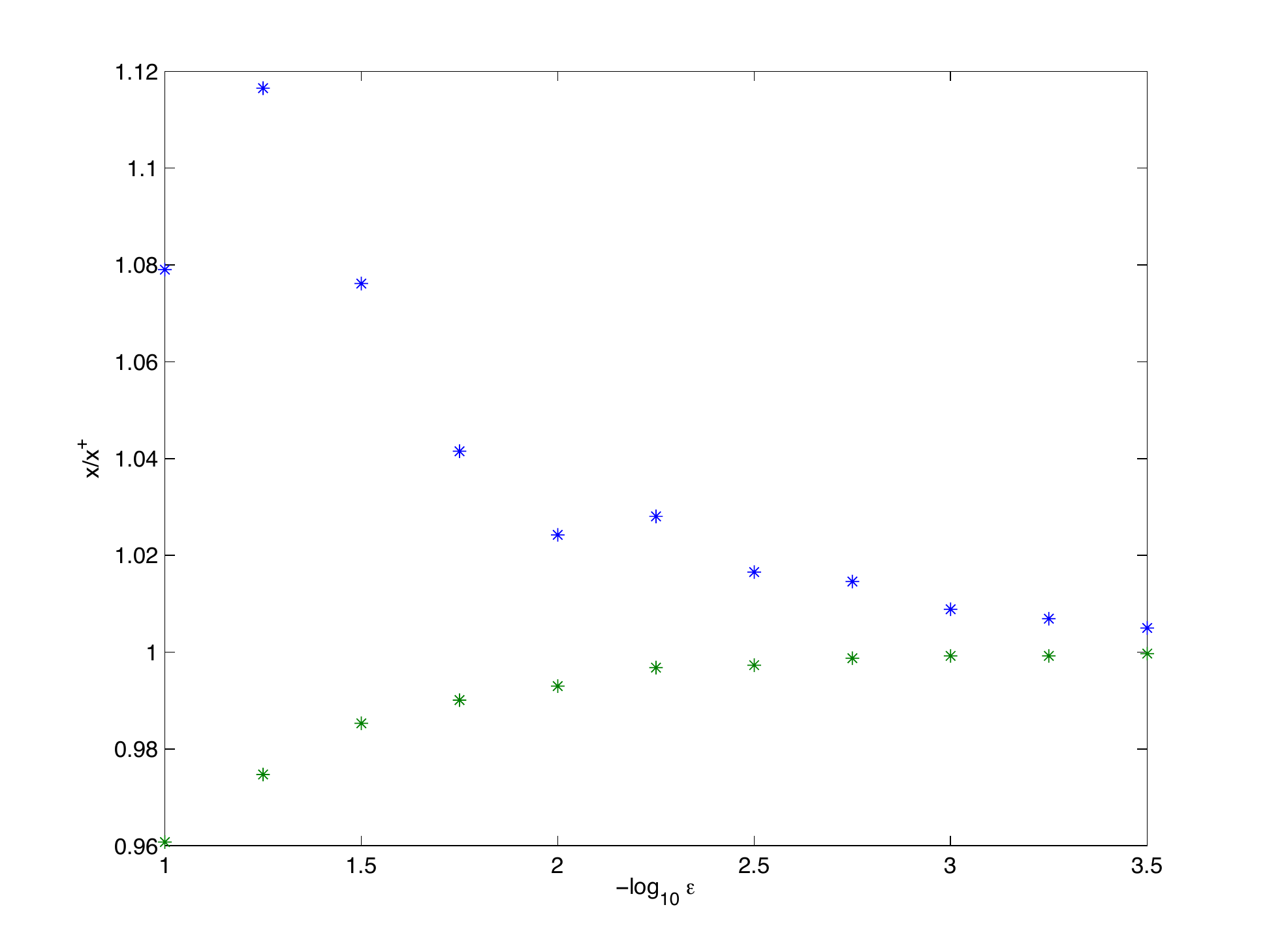}}
  \caption{In Fig.~(a), the
difference between the numerical solution to the KdV equation for
the initial datum $u_{0}=-\mbox{sech}^{2}x$ and 
$\epsilon=10^{-2}$ at $t=0.4$  and the
corresponding  soliton asymptotic (\ref{expansion u}) in green, and the difference
between KdV and Hopf and one-phase KdV solution in blue.
In Fig.~(b) the edges
of the zone where the soliton asymptotic  (\ref{expansion u}) provides a better
asymptotic description of KdV than the Hopf or the one-phase KdV
solution  in dependence of $\epsilon$.}
\end{figure}

It can be seen in Fig.~\ref{kdv1e4t4trail2d}, where
the difference between KdV and the asymptotic solutions is shown, 
that soliton asymptotic solution gives a much  a better description of KdV near the
trailing edge of the Whitham zone than the Hopf and the one-phase KdV solution.

%\begin{figure}[htb!]
%  \includegraphics[width=0.6\textwidth]{kdv1e4t4trail2d.pdf}
% \caption{The
%difference between the numerical solution to the KdV equation for
%the initial datum $u_{0}=-\mbox{sech}^{2}x$ and 
%$\epsilon=10^{-2}$ at $t=0.4$ for $\epsilon=10^{-2}$ and the
%corresponding  multiscales solution (\ref{expansion u}) in blue, and the difference
%between KdV and Hopf and elliptic solution in green.}
%   \label{kdv1e4t4trail2d}
%\end{figure}

\begin{figure}[htb!]
 \centering
  \subfloat[]{\label{kdv1e4t4matchr}\includegraphics[width=0.5\textwidth]{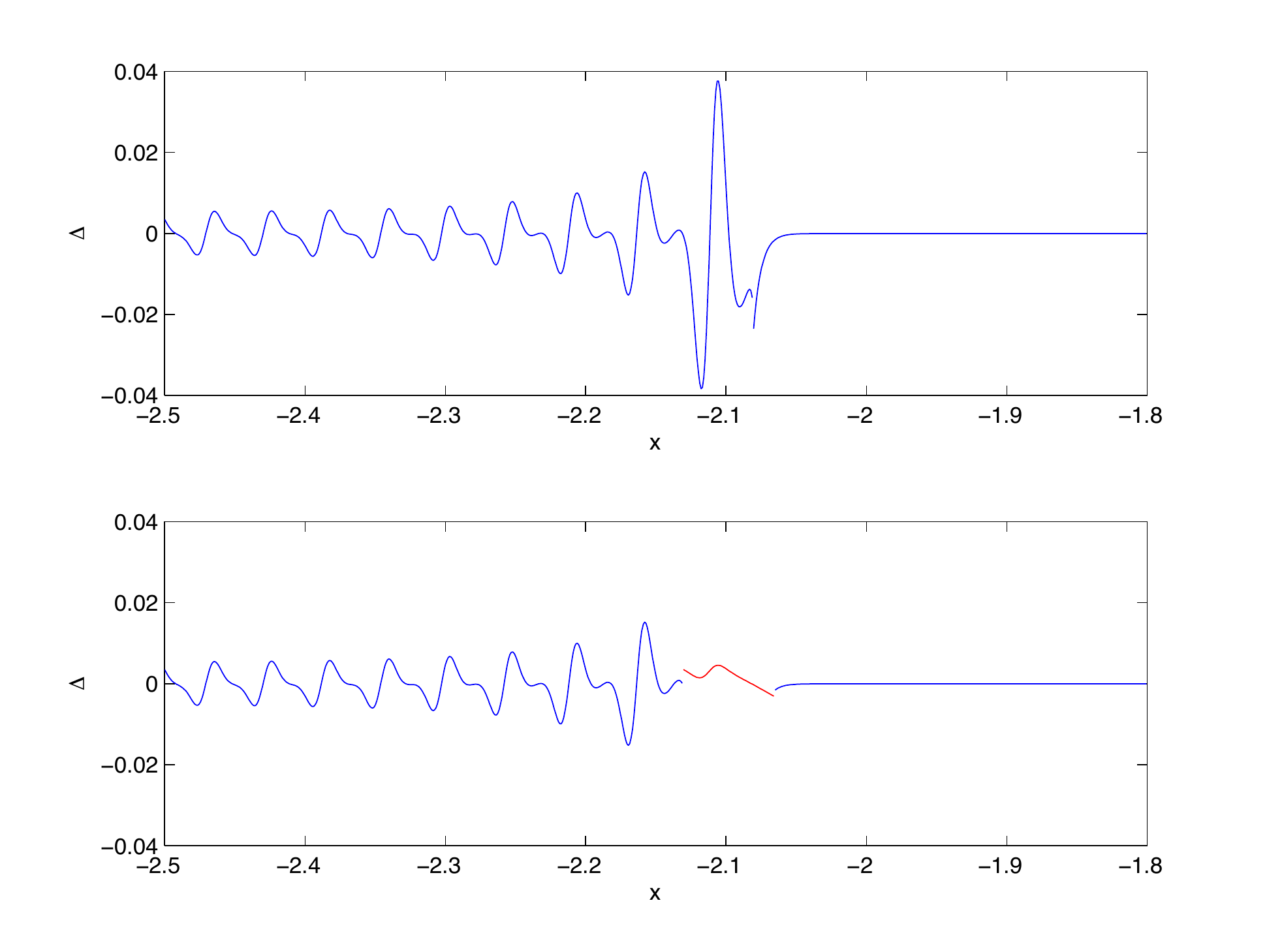}}                
  \subfloat[]{\label{kdv1e4t4matchr2e}\includegraphics[width=0.5\textwidth]{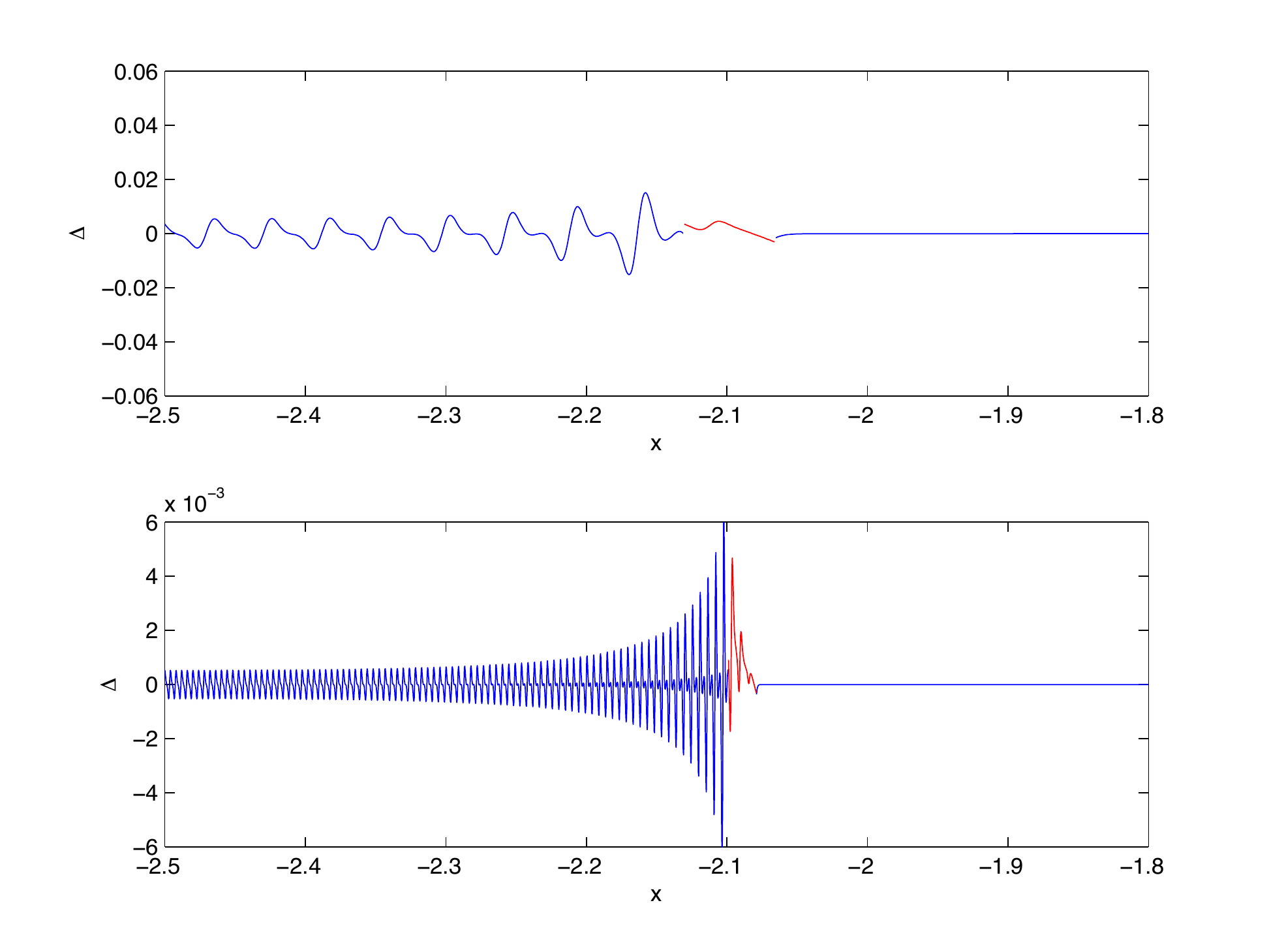}}
 \caption{In
the upper part of Fig.~(a) one can see the difference between the numerical
solution to the KdV equation for the initial datum
$u_{0}=-\mbox{sech}^{2}x$ and  $\epsilon=10^{-2}$ at
$t=0.4$ and the corresponding asymptotic solution in terms of Hopf
and one-phase KdV solution. The lower part shows the same
difference, which is replaced close to the trailing  edge of the
Whitham zone by the difference between KdV solution and the soliton asymptotic
 (\ref{expansion u}) (shown in red where the error is smaller than the one
shown above). In Fig.~(b) the same situation as in the lower part of 
(a) is shown for two values of $\epsilon$: $10^{-2},\;10^{-3}$.  The $\Delta$-axis is rescaled  by a factor $\epsilon$. }
\end{figure}

Again we can identify the regions where each of the asymptotic
solutions gives a better description of KdV than the other. The
results of this analysis can be seen in Fig.~\ref{kdv1e4t4matchr}.
This matching procedure clearly improves the KdV description near
the trailing edge. In Fig.~\ref{kdv1e4t4matchr2e} we see the difference between this matched
asymptotic solution and the KdV solution for two values of
$\epsilon$. Visibly the zone, where the solutions are matched,
decreases with $\epsilon$.

%\begin{figure}[htb!]
%  \includegraphics[width=0.7\textwidth]{kdv1e4t4matchr2e.pdf}
% \caption{The
%difference between the numerical solution to the KdV equation for
%the initial datum $u_{0}=-\mbox{sech}^{2}x$ at $t=0.4$
%and the corresponding asymptotic solution in terms of Hopf and
%elliptic solution in blue and KdV and multiscales solution (\ref{expansion u}) in red,
%where the latter error is smaller than the former, for two values
%of $\epsilon$. The $\Delta$-axis is rescaled by a factor $\epsilon$.}
%   \label{kdv1e4t4matchr2e}
%\end{figure}

Once more there is no precise definition of this
matching zone due to the oscillatory character of the solutions. We 
determine it as the point  where the curves of the differences intersect, or where 
they come closest, before one error dominates the other for all 
smaller 
respectively larger values of $x$.
The limits of the matching zone for several values of
$\epsilon$ can be seen in Fig.~\ref{trailzone}. 
There does not appear to be a clear scaling law for the  width of 
this zone. It can be already seen in Fig.~\ref{kdv1e4t4matchr2e} that 
the error in the matching zone does not scale with $\epsilon$ as close to 
the boundary of the Whitham zone. In fact we find a scaling close to 
$\epsilon^{1/2}$ in both cases, but the correlation is not very good. 
%Thus it is not possible to 
%obtain an error of order $\epsilon$ up to the trailing edge. 
As for  the leading edge, it is necessary to establish analytical connection 
formulae.

%\begin{figure}[htb!]
%  \includegraphics[width=.5\textwidth]{trailzone.pdf}
% \caption{The edges
%of the zone where the multiscales solution (\ref{expansion u}) provides a better
%asymptotic description of KdV than the Hopf or the elliptic
%solution for the initial datum $u_{0}=-\mbox{sech}^{2}x$  at $t=0.4$ 
%in dependence of $\epsilon$.}
%   \label{trailzone}
%\end{figure}
%
\section{Point of gradient catastrophe}
In this section we study  numerically the approximation   (\ref{ExpP12}) to the solution $u(x,t,\epsilon)$ of KdV  as $\epsilon\rightarrow 0$ near the point of gradient catastrophe $(x_c,t_c)$ for the  solution of
the Hopf equation. 
% It will be shown that it
%provides a better description of the asymptotic behavior near the
%critical points than the Hopf or the elliptic
%KdV solution. 
We identify the
zone, where  the ${\rm P}_{I}^2$ asymptotic formula   
(\ref{ExpP12})   gives a better asymptotic  description of KdV than 
the Hopf or the one-phase KdV solution
 and study the $\epsilon$-dependence of the errors. We 
qualitatively study for a time $t> t_{c}$ close to $t_{c}$ how the various 
multiscale approximations perform.

\subsection{Critical time}

\begin{figure}[htb!]
  \includegraphics[width=0.7\textwidth]{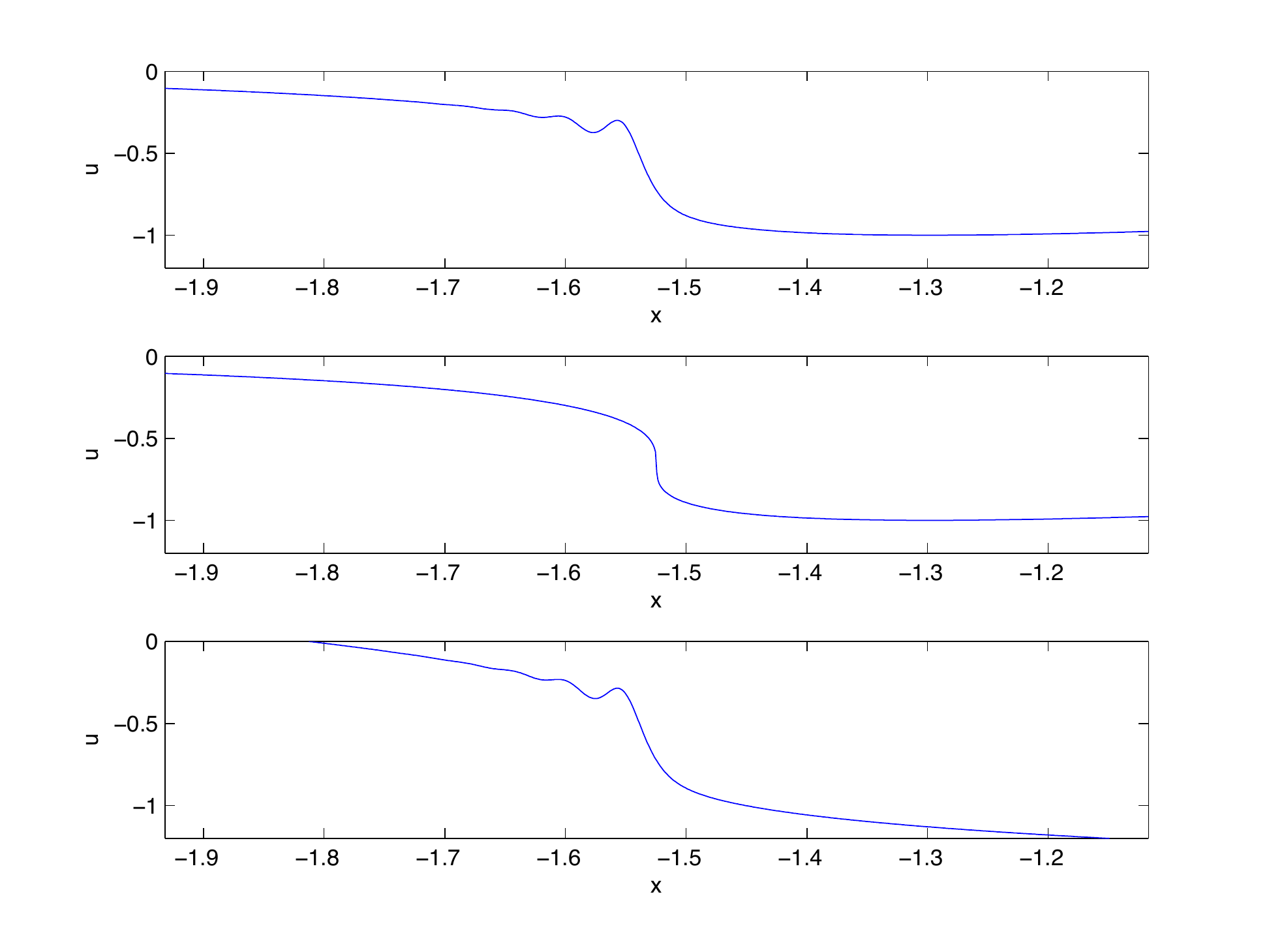}
 \caption{The
figure shows in the upper part the numerical solution to the KdV
equation for the initial datum $u_{0}=-\mbox{sech}^{2}x$ and 
$\epsilon=10^{-2}$ at $t=t_{c}$, in the middle the
corresponding Hopf
solution, and in the lower part the ${\rm P}_I^2$ asymptotics
(\ref{ExpP12}).}
   \label{kdv1e4tc3}
\end{figure}

For the initial datum $u_{0}(x)=-\mbox{sech}^{2}x$ the critical time 
is $t_{c}=\sqrt{3}/8\sim 0.2165$ and the critical point $x_{c} = 
-\sqrt{3}/2+\ln((\sqrt{3}-1)/\sqrt{2})\sim-1.5245$. In Fig.~\ref{kdv1e4tc3} we show the KdV solution, the Hopf
solution and the multiscale solution near
the critical point of the Hopf solution at the critical time. As before the
Hopf solution gives a very good description for $|x-x_{c}|\gg 0$, whereas the
${\rm P}_I^2$ asymptotic solution gives as expected a better description near
the critical point. The following figures are always symmetric with 
respect to $x_{c}$.

\begin{figure}[htb!]
  \includegraphics[width=0.6\textwidth]{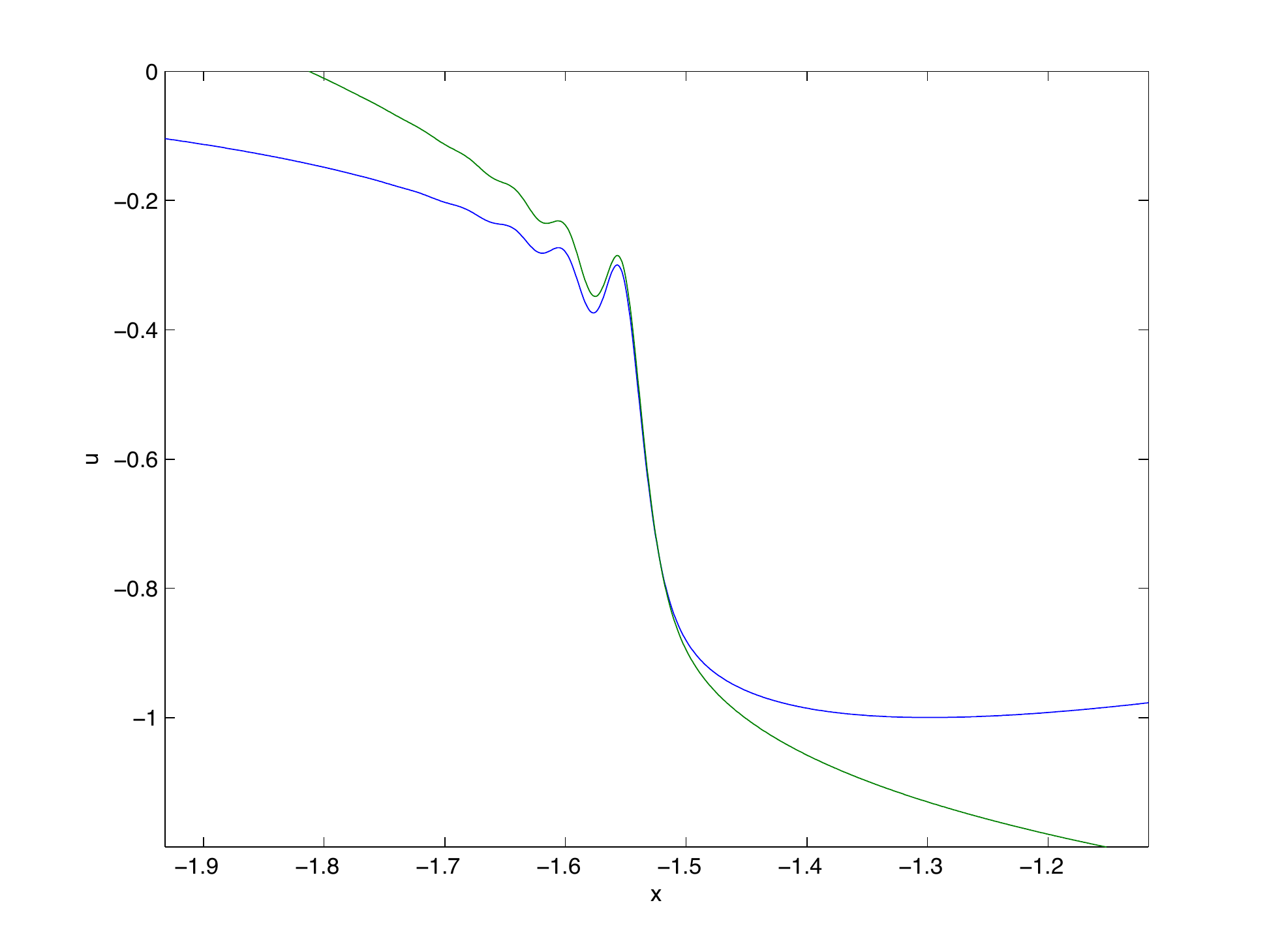}
 \caption{The
numerical solution to the KdV  equation for the initial datum
$u_{0}=-\mbox{sech}^{2}x$  and $\epsilon=10^{-2}$ at
$t=t_{c}$ in blue and the corresponding  ${\rm P}_I^2$ asymptotic solution  in
green.}
   \label{kdv1e4tc}
\end{figure}

In Fig.~\ref{kdv1e4tc} the KdV solution and the ${\rm P}_I^2$ asymptotic solution  are
shown in one plot for $\epsilon=10^{-2}$. It can be seen that the
agreement very close to the critical point of the Hopf solution is 
again  so good that the difference of the solutions has to be studied. The solution only
gives locally an asymptotic description.

\begin{figure}[htb!]
  \includegraphics[width=\textwidth]{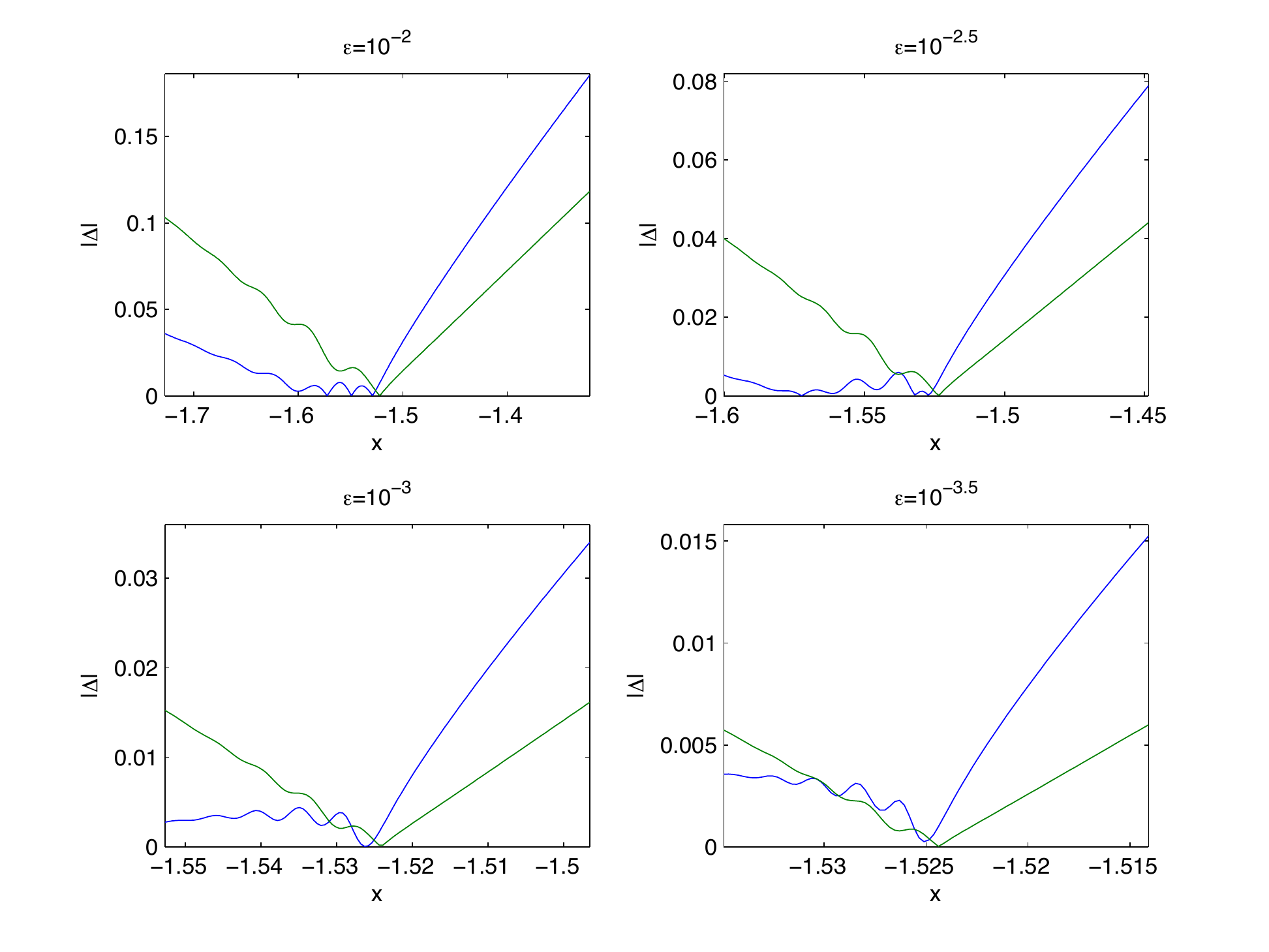}
 \caption{The difference between the numerical solution to the KdV
equation for the initial datum $u_{0}=-\mbox{sech}^{2}x$ at $t=t_{c}$ 
and the corresponding  multiscale solution for
four values of $\epsilon$, in blue the terms in (\ref{ExpP12}) up to 
order \( \epsilon^{2/7} \), in green up to order \( \epsilon^{4/7} \).  Note the scaling of the $x$  and $\Delta$ axes with a factor 
$\epsilon^{6/7}$ and $\epsilon^{5/7}$ respectively  to take care of 
the expected scaling in $x$
and of the shown error next to the critical point.}
   \label{kdv1e4tc4e}
\end{figure}

The difference between KdV solution  and  ${\rm P}_I^2$ asymptotic  solution is shown for
several values of $\epsilon$ in Fig.~\ref{kdv1e4tc4e}. The scales in 
$x$ are rescaled by a factor $\epsilon^{6/7}$, and the ones for 
$\Delta$ by a  factor $\epsilon^{5/7}$ 
respectively which is the
expected scaling behavior of the zone, where the  ${\rm P}_I^2$ asymptotic 
solution should be applicable, and of the expected error. It can be seen
that with these rescalings the error is of the same order for 
different values of $\epsilon$, at least close to the critical point. 
We also show in this figure the different behaviour of the terms in 
(\ref{ExpP12}) in order \( \epsilon^{2/7} \) and order \( 
\epsilon^{4/7} \). The former is not symmetric with respect to the critical point. In 
fact the approximation is better on the side where the oscillations appear. However if one 
studies positive initial data as in \cite{DubrovinGravaKlein2}, the 
oscillations are on the side of the critical point where the 
approximation in terms of the ${\rm P}_I^2$ solution is worse.

\begin{figure}[htb!]
\centering
  \subfloat[]{\label{kdv1e4tc2d}\includegraphics[width=0.5\textwidth]{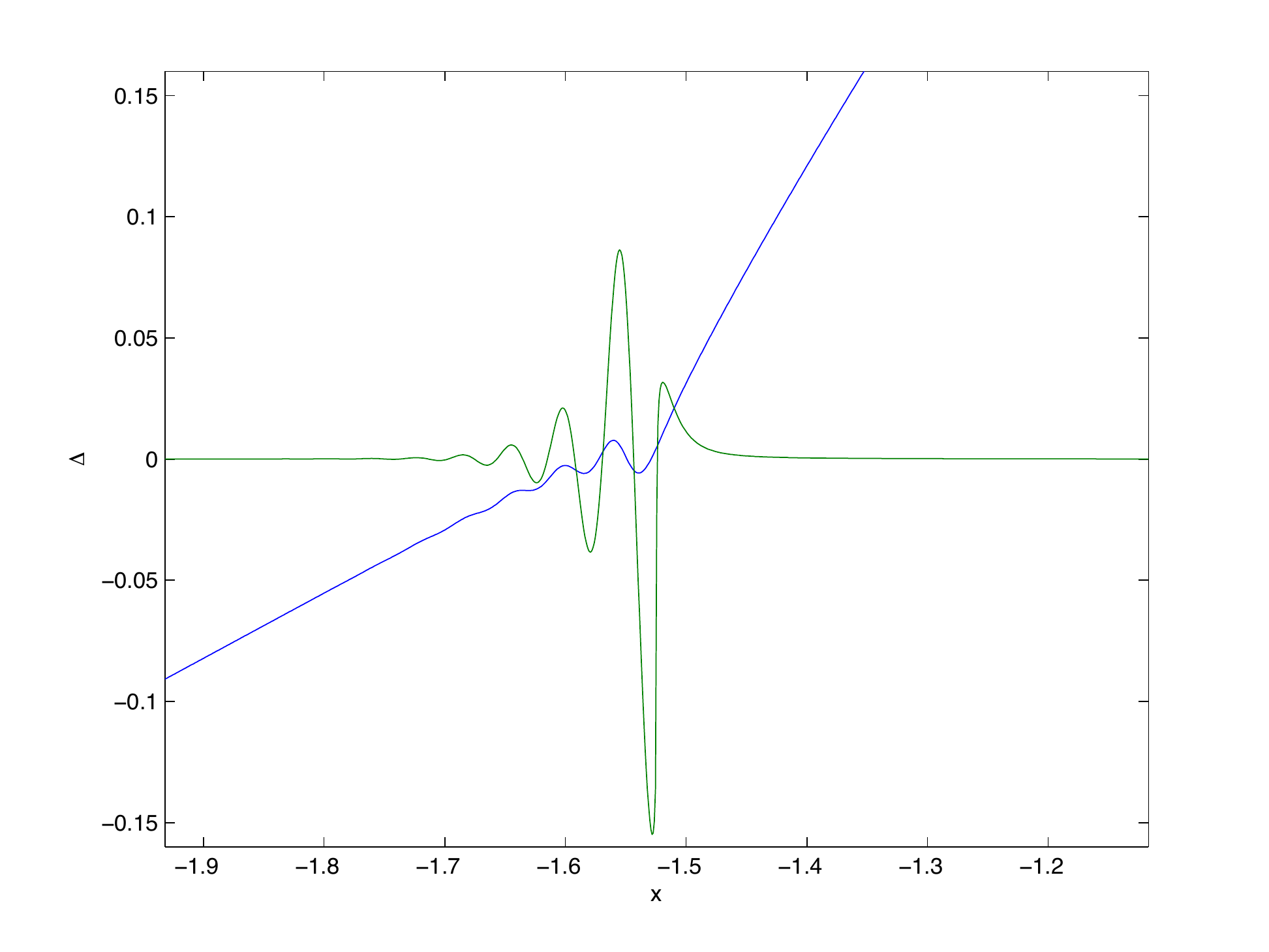}}                
  \subfloat[]{\label{critzone}\includegraphics[width=0.5\textwidth]{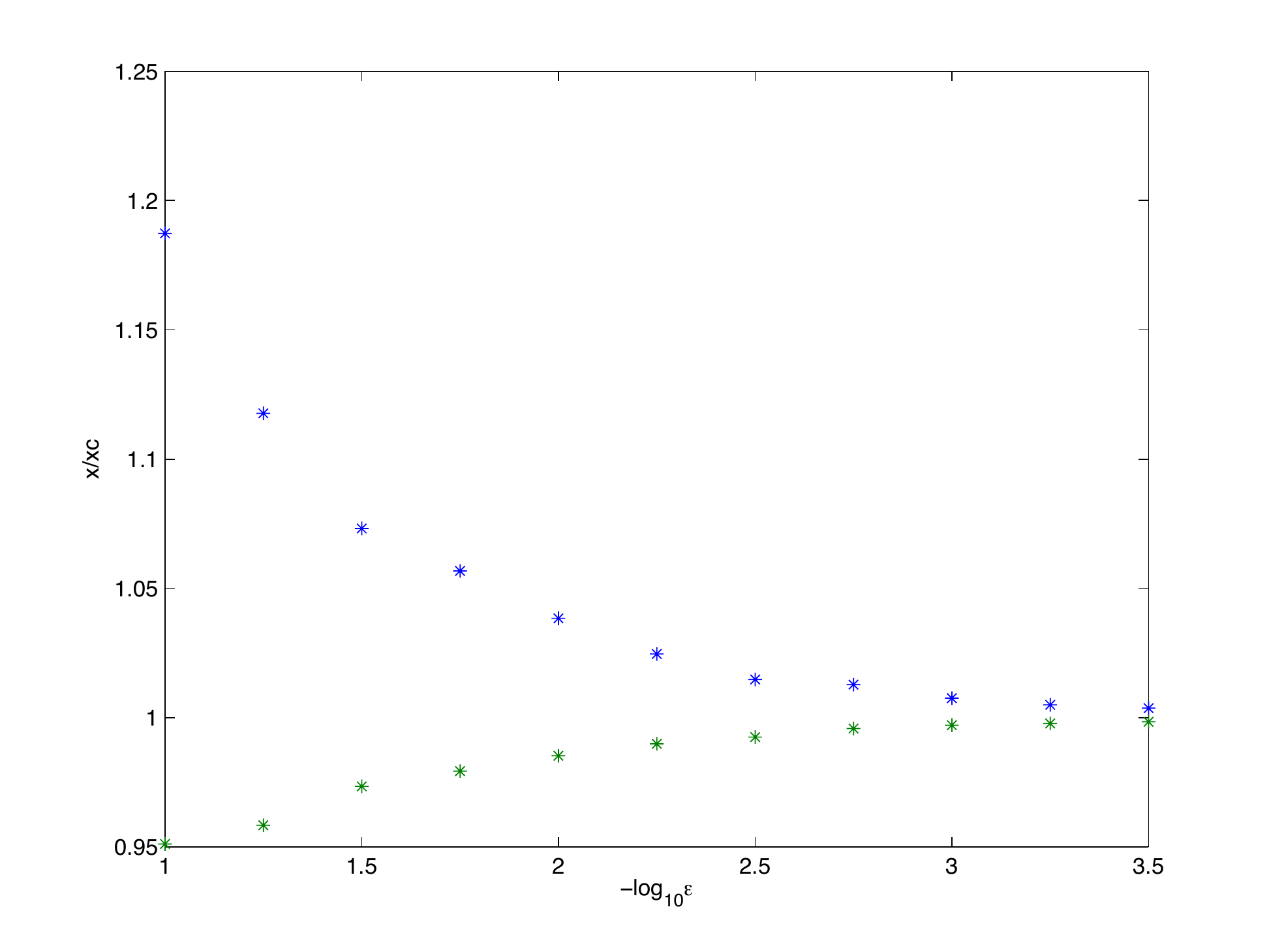}}
 \caption{ In Fig.~(a) the
difference between the numerical solution to the KdV equation for
the initial datum $u_{0}=-\mbox{sech}^{2}x$ and 
$\epsilon=10^{-2}$ at $t=t_{c}$  and the
corresponding  ${\rm P}_I^2$ asymptotic solution in blue, and the difference
between KdV and Hopf solution in green. In Fig.~(b) the edges
of the zone where the  ${\rm P}_I^2$ asymptotic solution provides a better
asymptotic description of KdV than the Hopf 
solution 
in dependence of $\epsilon$.}

\end{figure}

As is clear from Fig.~\ref{kdv1e4tc3}, the multiscale
solution gives a better asymptotic description of KdV near the
critical point than the Hopf. This is even more obvious in 
Fig.~\ref{kdv1e4tc2d} where
the difference between KdV and the Hopf solution is shown.

\begin{figure}[htb!]
 \subfloat[]{\label{kdv1e4tcmatch}\includegraphics[width=0.5\textwidth]{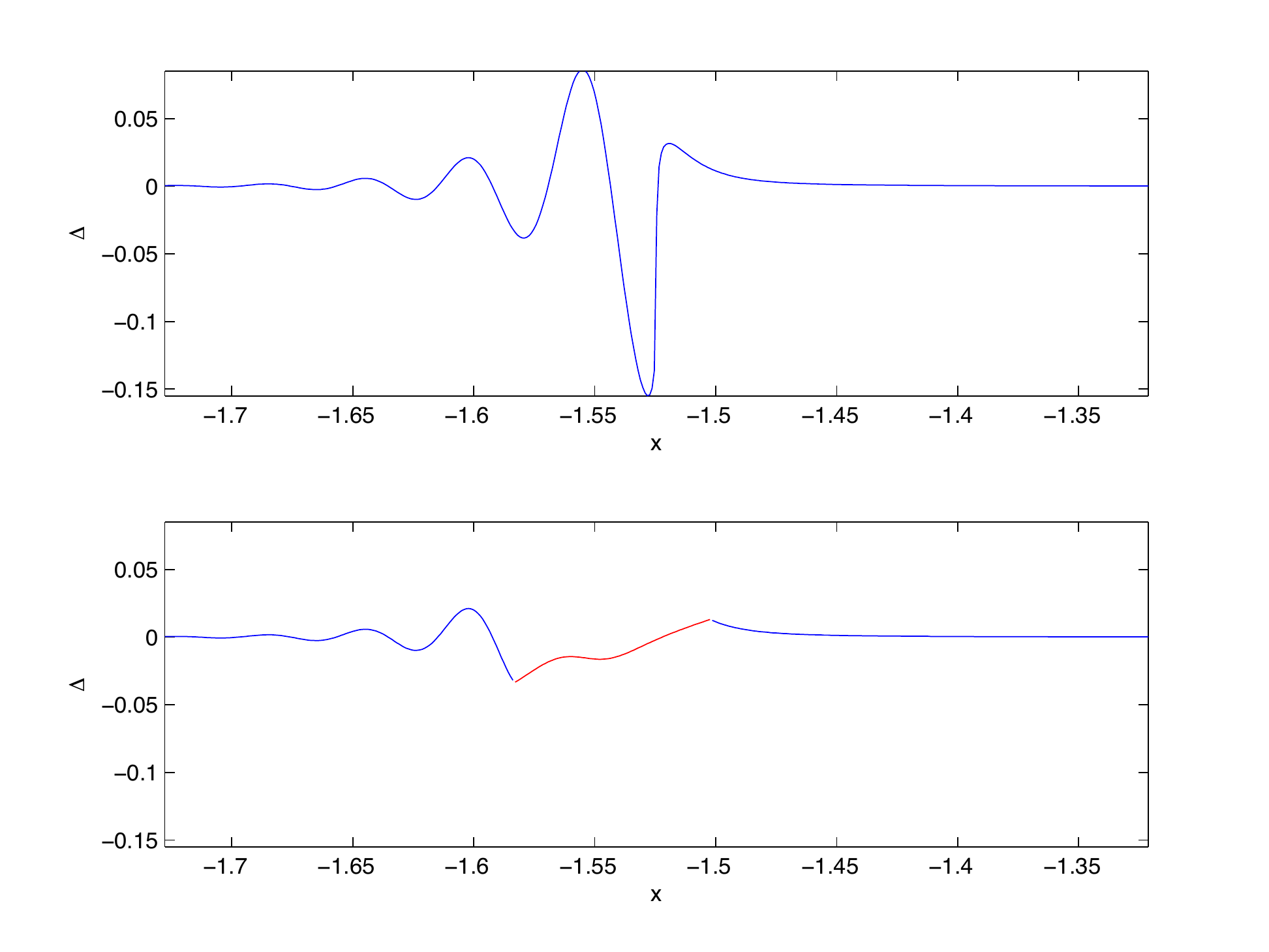}}                
  \subfloat[]{\label{kdv1e4tcmatch2e}\includegraphics[width=0.5\textwidth]{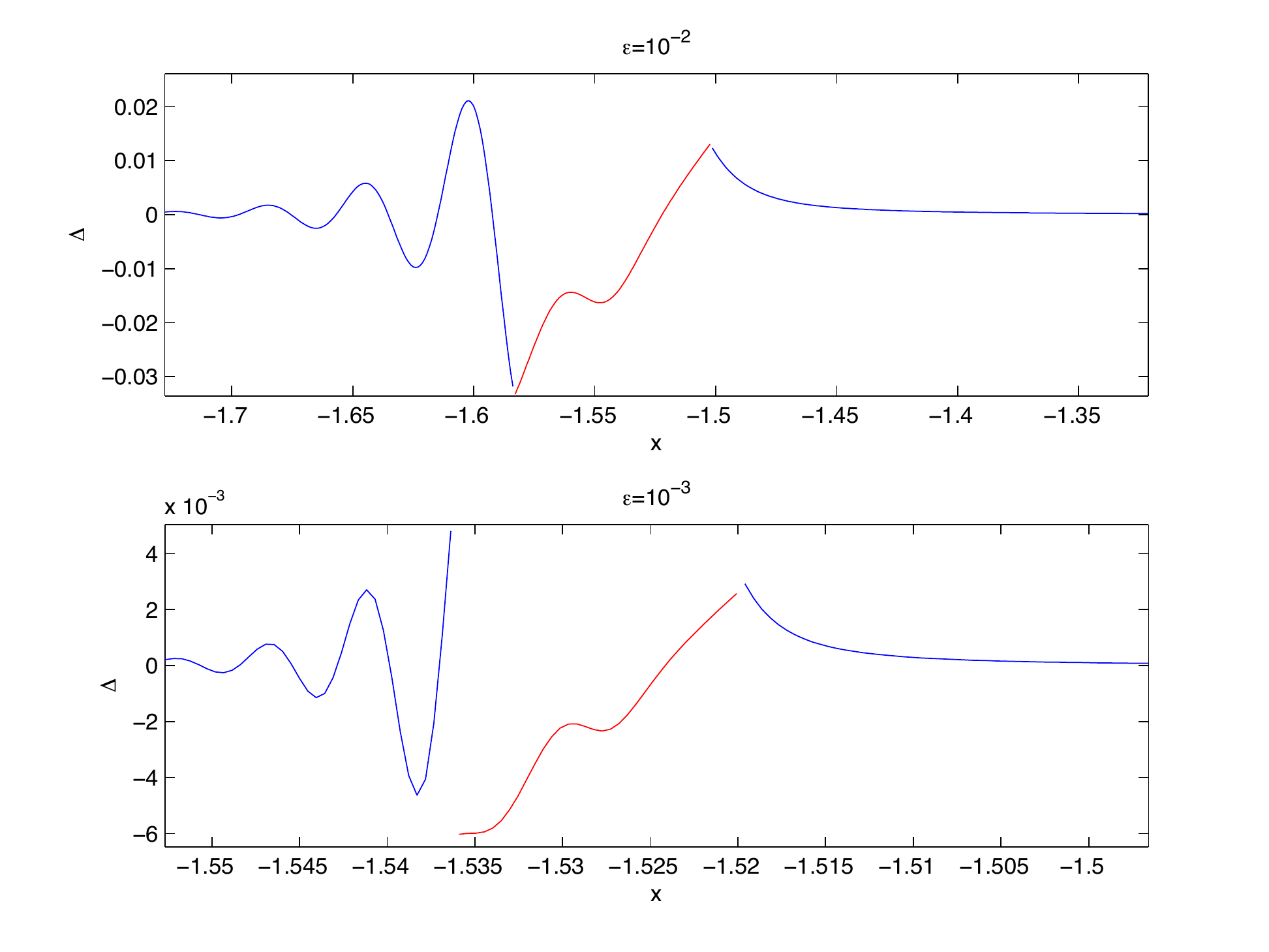}}
 \caption{In
the upper part of Fig.~(a) one can see the difference between the numerical
solution to the KdV equation for the initial datum
$u_{0}=-\mbox{sech}^{2}x$ and  $\epsilon=10^{-2}$ at
$t=t_{c}$ and the corresponding Hopf solution. The lower part shows the same
difference, which is replaced close to the critical point 
by the difference between KdV solution and  the ${\rm P}_I^2$ asymptotic solution (shown in red where the error is smaller than the one
shown above). Fig.~(b) shows the same situation as the lower figure 
in (a)  for two values
of $\epsilon$. The $\Delta$-axis is rescaled by a factor 
$\epsilon^{5/7}$, the $x$-axis by a factor $\epsilon^{6/7}$.}   
\end{figure}

Again we can identify the regions where each of the asymptotic
solutions gives a better description of KdV than the other. The
results of this analysis can be seen in Fig.~\ref{kdv1e4tcmatch}.
This matching procedure clearly improves the KdV description near
the critical point.
In Fig.~\ref{kdv1e4tcmatch2e} we see the difference between this matched
asymptotic solution and the KdV solution for two values of
$\epsilon$. Visibly the zone, where the solutions are matched,
decreases with $\epsilon$ (notice the rescaling of the axes with 
$\epsilon$).

%\begin{figure}[htb!]
%  \includegraphics[width=0.7\textwidth]{kdv1e4tcmatch2e.pdf}
% \caption{The
%difference between the numerical solution to the KdV equation for
%the initial datum $u_{0}=-\mbox{sech}^{2}x$ at $t=t_{c}$
%and the corresponding Hopf solution in blue and KdV and multiscales solution in red,
%where the latter error is smaller than the former, for two values
%of $\epsilon$. The $\Delta$-axis is rescaled by a factor 
%$\epsilon^{4/7}$, the $x$-axis by a factor $\epsilon^{6/7}$.}
%   \label{kdv1e4tcmatch2e}
%\end{figure}

Once more there is no precise definition of this
matching zone due to the oscillatory character of the solutions. We 
choose it as in the previous sections as given where the curves of the differences intersect, or where 
they come closest before one error dominates the other.
The limits of the matching zone for several values of
$\epsilon$ can be seen in Fig.~\ref{critzone}. 
There does not appear to be a clear scaling law for the  width of 
this zone. A linear
regression analysis for the logarithm of the difference $\Delta$
between KdV and multiscale solution in the matching zone gives a scaling
of the form $\Delta\propto \epsilon^{a}$  with $a=0.586$ 
($4/7\sim0.5714$) with
standard deviation $\sigma_{a}=0.06$ and correlation coefficient 
$r=0.99$  for the terms up to order \( \epsilon^{2/7} \) and with $a=0.62$ 
($5/7\sim0.7143$) with
standard deviation $\sigma_{a}=0.09$ and correlation coefficient 
$r=0.98$ for the terms up to order \( \epsilon^{4/7} \). The found scaling is thus compatible   
with the expected $\epsilon^{4/7}$ and \( \epsilon^{5/7} \) respectively.
The error in the Hopf zone at the limits of the matching zone are of 
the same order. As before it would be interesting to study the 
connection formulae between the Hopf and ${\rm P}_I^2$ zone.

%\begin{figure}[htb!]
%  \includegraphics[width=.5\textwidth]{critzone.pdf}
% \caption{The edges
%of the zone where the multiscales solutions provides a better
%asymptotic description of KdV than the Hopf 
%solution for the initial datum $u_{0}=-\mbox{sech}^{2}x$  at $t=t_{c}$ 
%in dependence of $\epsilon$.}
%   \label{critzone}
%\end{figure}
%
\subsection{Close to the critical time}
It is an interesting question to study how the various multiscale 
approximations perform for a time $t$ greater than the critical time 
with $t\sim t_{c}$, where all previously studied asymptotic formulae should 
be applicable. We consider just one value of $\epsilon$ 
($\epsilon=10^{-2}$) since the various multiscale expansions use 
different $\epsilon$-dependent rescalings of the time. We consider 
the time $t=0.23>t_{c}\sim0.2165$.

\begin{figure}[htb!]
  \includegraphics[width=0.7\textwidth]{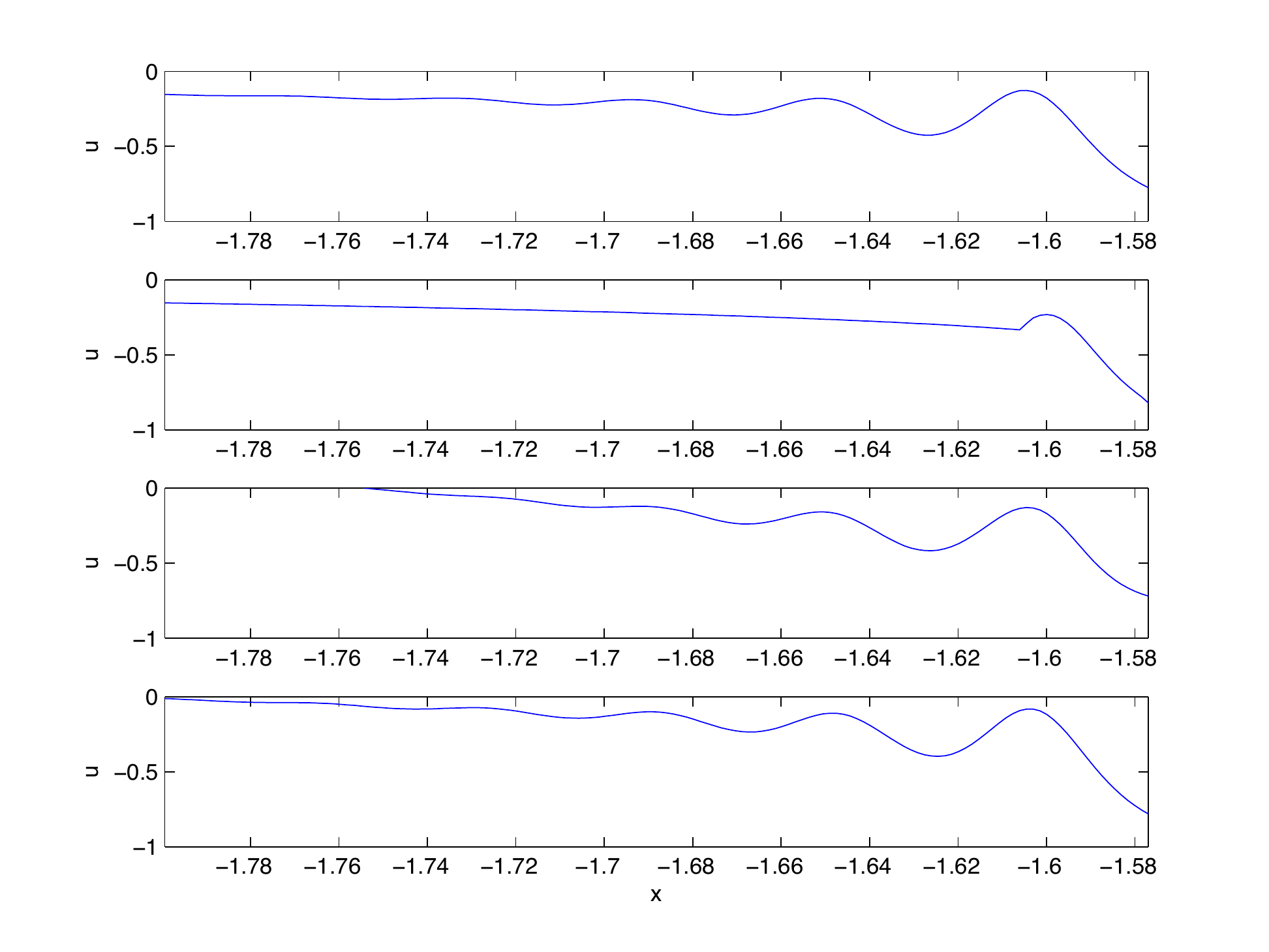}
 \caption{The
figure shows from top to bottom the numerical solution to the KdV
equation for the initial datum $u_{0}=-\mbox{sech}^{2}x$ and 
$\epsilon=10^{-2}$ at $t=0.23>t_{c}$, the asymptotic solution in 
terms of Hopf and one-phase KdV
solution,  the multiscale solution (\ref{expansionu}), and the multiscale solution 
(\ref{ExpP12}).}
   \label{kdv1e4_234}
\end{figure}

First we study the situation in the vicinity of the leading edge 
($x=-1.6051\ldots$). In Fig.~\ref{kdv1e4_234}, the KdV solution can be 
seen in this case as well as the asymptotic solution in terms of Hopf 
and one-phase KdV solution, the multiscale solution (\ref{expansionu}) 
near the leading edge and the multiscale solution (\ref{ExpP12}) 
close to the critical point.

\begin{figure}[htb!]
 \subfloat[]{\includegraphics[width=0.5\textwidth]{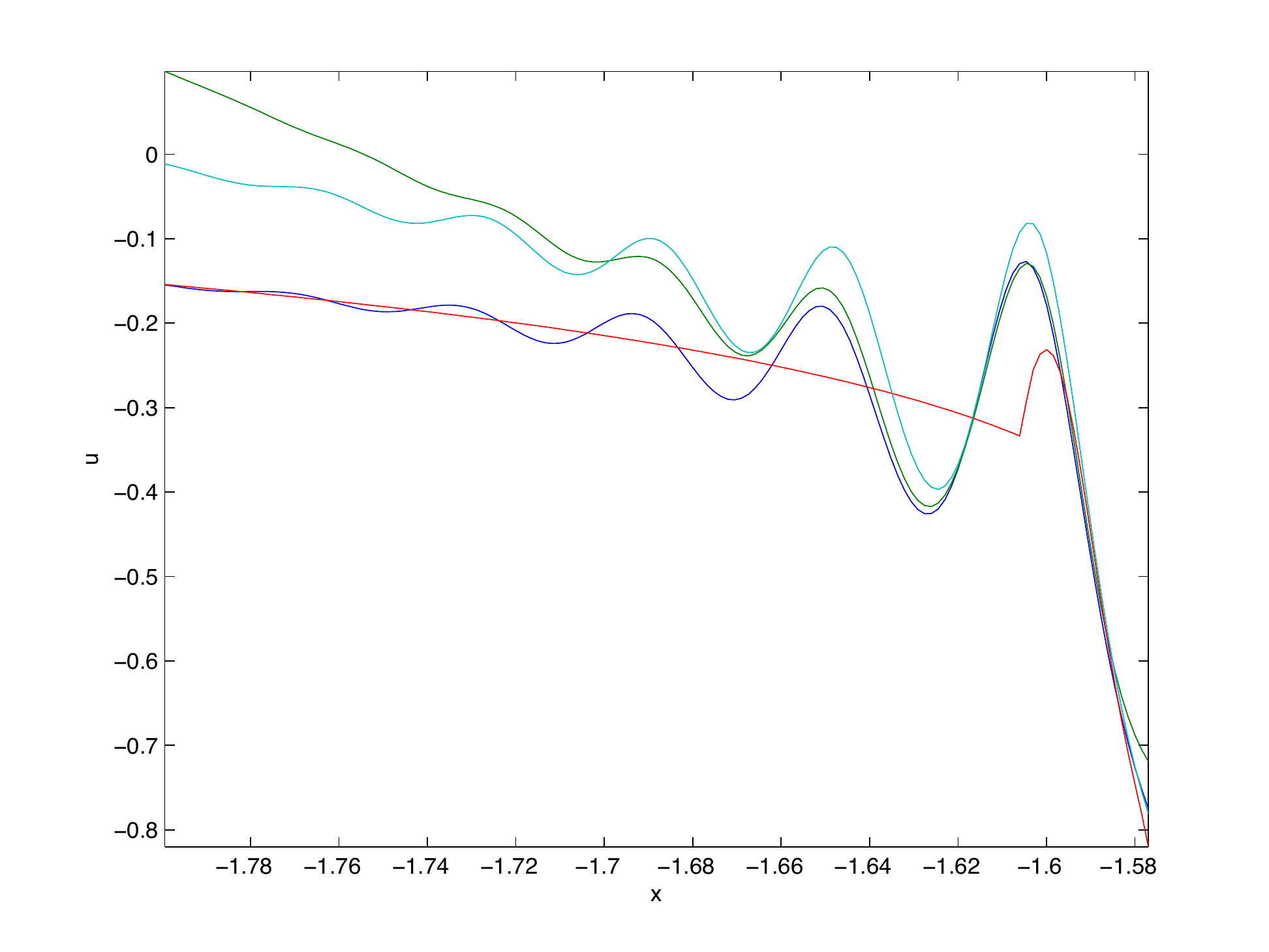}}                
  \subfloat[]{\label{kdv1e4_23delta}\includegraphics[width=0.5\textwidth]{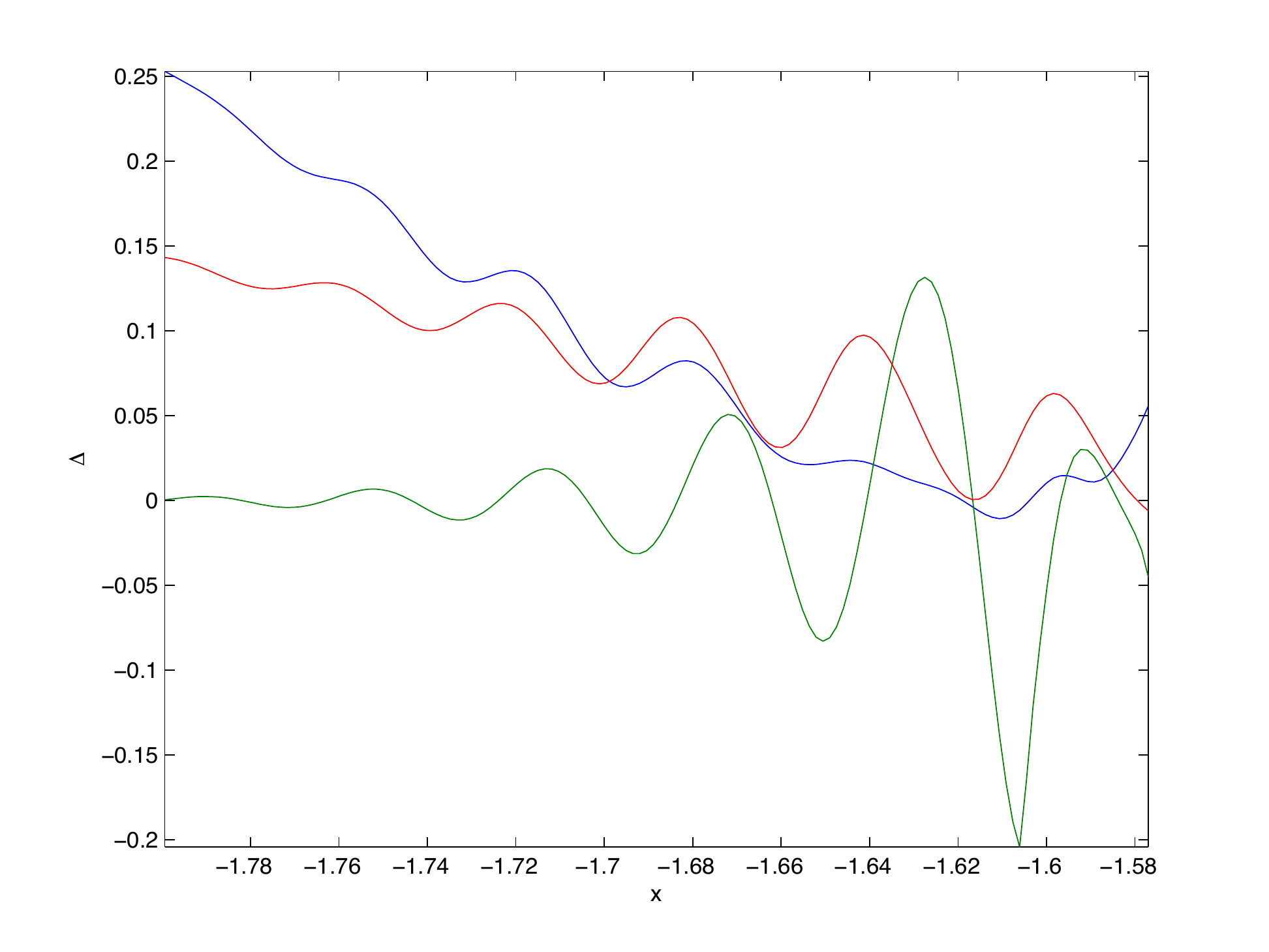}}
 \caption{Fig.~(a)  shows the numerical solution to the KdV
equation for the initial datum $u_{0}=-\mbox{sech}^{2}x$ and 
$\epsilon=10^{-2}$ at $t=0.23>t_{c}$ in blue, the asymptotic solution 
in terms of Hopf and one-phase KdV
solution in red,  the multiscale solution (\ref{expansionu}) in 
cyan, and the multiscale solution 
(\ref{ExpP12}) in green. Fig.~(b) shows the difference between the asymptotic solution in terms of Hopf and one-phase KdV
solution and the numerical solution to the KdV
equation  in green,  between the 
${\rm P}_{II}$ asymptotic  solution (\ref{expansionu}) and KdV in 
red, and the ${\rm P}_{I}^2$ asymptotic solution 
(\ref{ExpP12}) and KdV in blue.}
   \label{kdv1e4_234in1}
\end{figure}

In Fig.~\ref{kdv1e4_234in1} the same solutions can be seen in one 
figure.

It can be seen already from these figures or from the plot of the 
differences between the asymptotic solutions and the KdV solution in 
Fig.~\ref{kdv1e4_23delta} that the asymptotic solution in terms of 
Hopf and one-phase KdV solution performs worst close to the leading 
edge, and that the 
multiscale solution (\ref{ExpP12}) in terms of the ${\rm P}_I^2$ transcendent 
is most satisfactory. The multiscale solution (\ref{expansionu}) is 
only very close to the leading better than the ${\rm P}_I^2$ asymptotics. It 
captures qualitatively the oscillations in the Hopf region, but is 
not oscillating around the Hopf solution as KdV. As can be seen from 
(\ref{expansionu}) the reason for this is that the amplitude of this 
solution is divided by $u-v$ which tends to 0 at the critical point. 
The ${\rm P}_I^2$  asymptotics also quickly becomes out of phase in the Hopf 
region. 
Thus to obtain a satisfactory description of the small dispersion 
limit of KdV close to the critical point, one has to study 
connection formulae between the various asymptotic solutions. 

%\begin{figure}[htb!]
%  \includegraphics[width=0.6\textwidth]{kdv1e4_23delta.pdf}
% \caption{The
%figure shows the difference between the asymptotic solution in terms of Hopf and elliptic
%solution and the numerical solution to the KdV
%equation for the initial datum $u_{0}=-\mbox{sech}^{2}x$ and 
%$\epsilon=10^{-2}$ at $t=0.23>t_{c}$ in blue,  between the 
%${\rm P}_{II}$ asymptotic  solution (\ref{expansionu}) and KdV in 
%green, and the ${\rm P}_{I}^2$ asymptotic solution 
%(\ref{ExpP12}) and KdV in red.}
%   \label{kdv1e4_23delta}
%\end{figure}

\begin{figure}[htb!]
  \includegraphics[width=0.7\textwidth]{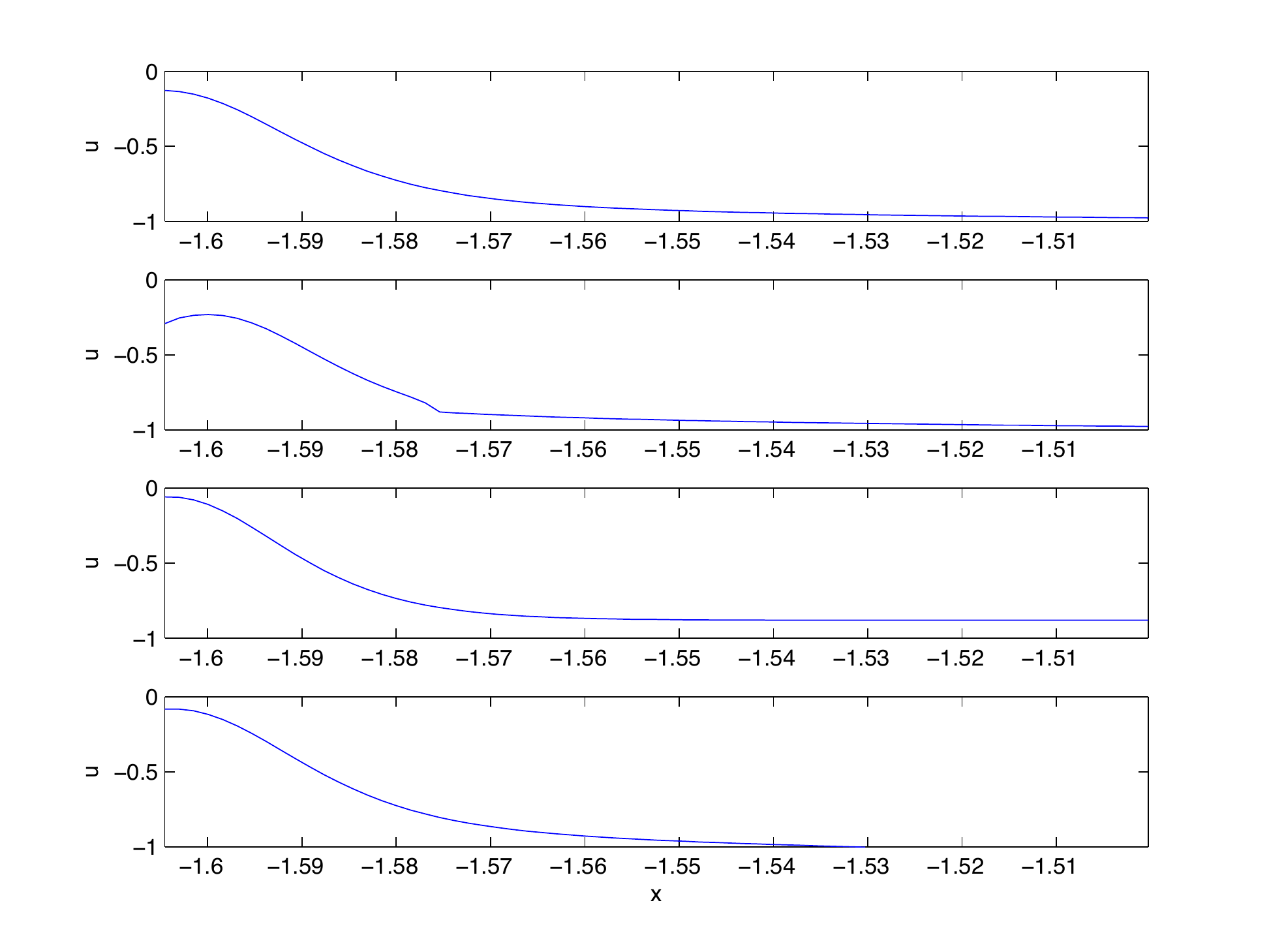}
 \caption{The
figure shows from top to bottom the numerical solution to the KdV
equation for the initial datum $u_{0}=-\mbox{sech}^{2}x$ and 
$\epsilon=10^{-2}$ at $t=0.23>t_{c}$, the asymptotic solution in terms of Hopf and one-phase KdV
solution,  the soliton asymptotic  solution (\ref{expansion u}), and the ${\rm P}_I^2$ asymptotic  solution 
(\ref{ExpP12}).}
   \label{kdv1e4_234trail}
\end{figure}

The situation near the trailing edge 
($x=-1.5757\ldots$) is similar. In Fig.~\ref{kdv1e4_234trail}, the KdV solution can be 
seen in this case as well as the asymptotic solution in terms of Hopf 
and one-phase KdV solution, the multiscale solution (\ref{expansion u}) 
near the leading edge and the multiscale solution (\ref{ExpP12}) 
close to the critical point.

\begin{figure}[htb!]
 \subfloat[]{\includegraphics[width=0.5\textwidth]{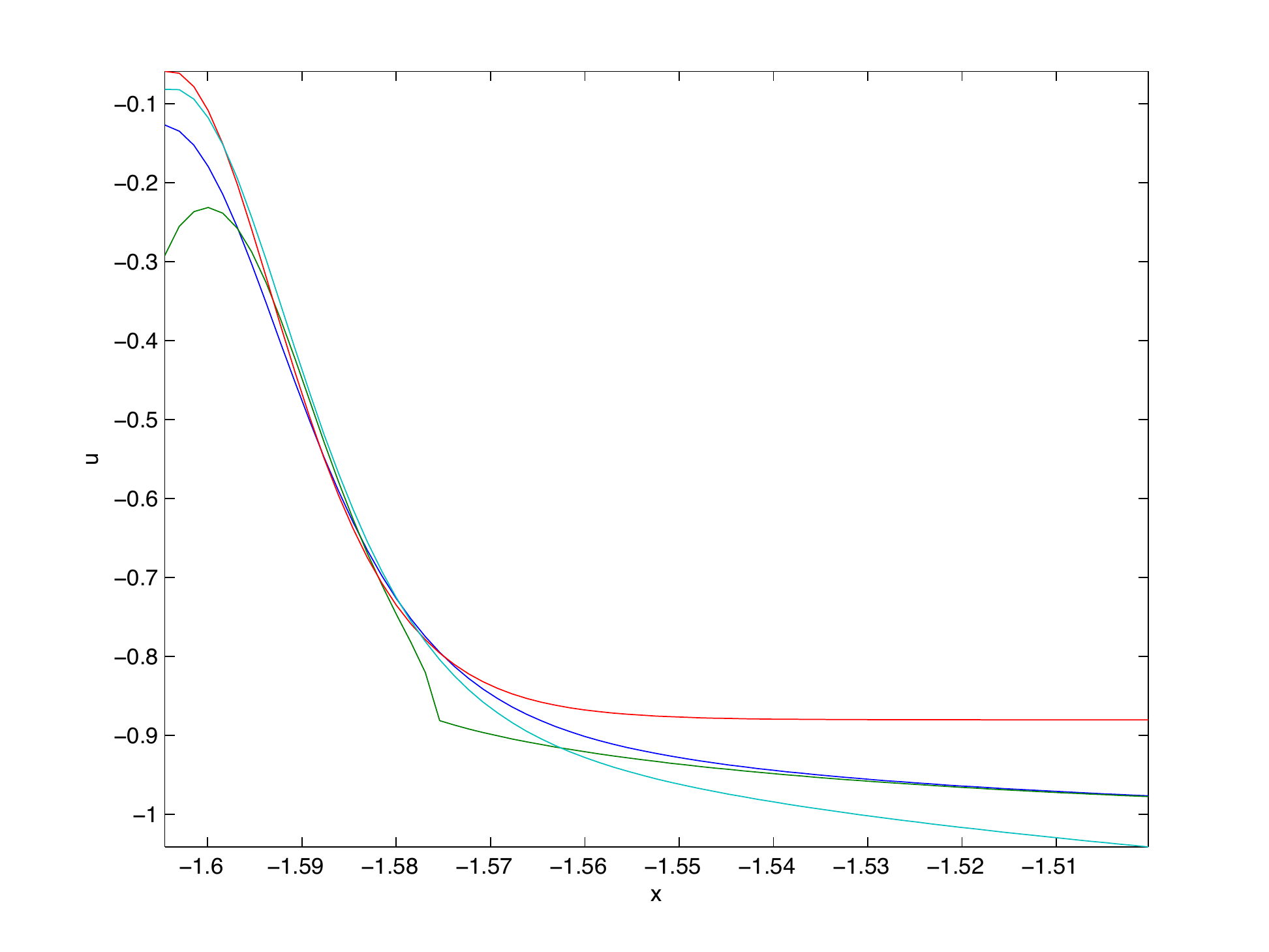}}                
  \subfloat[]{\label{kdv1e4_23deltatrail}\includegraphics[width=0.5\textwidth]{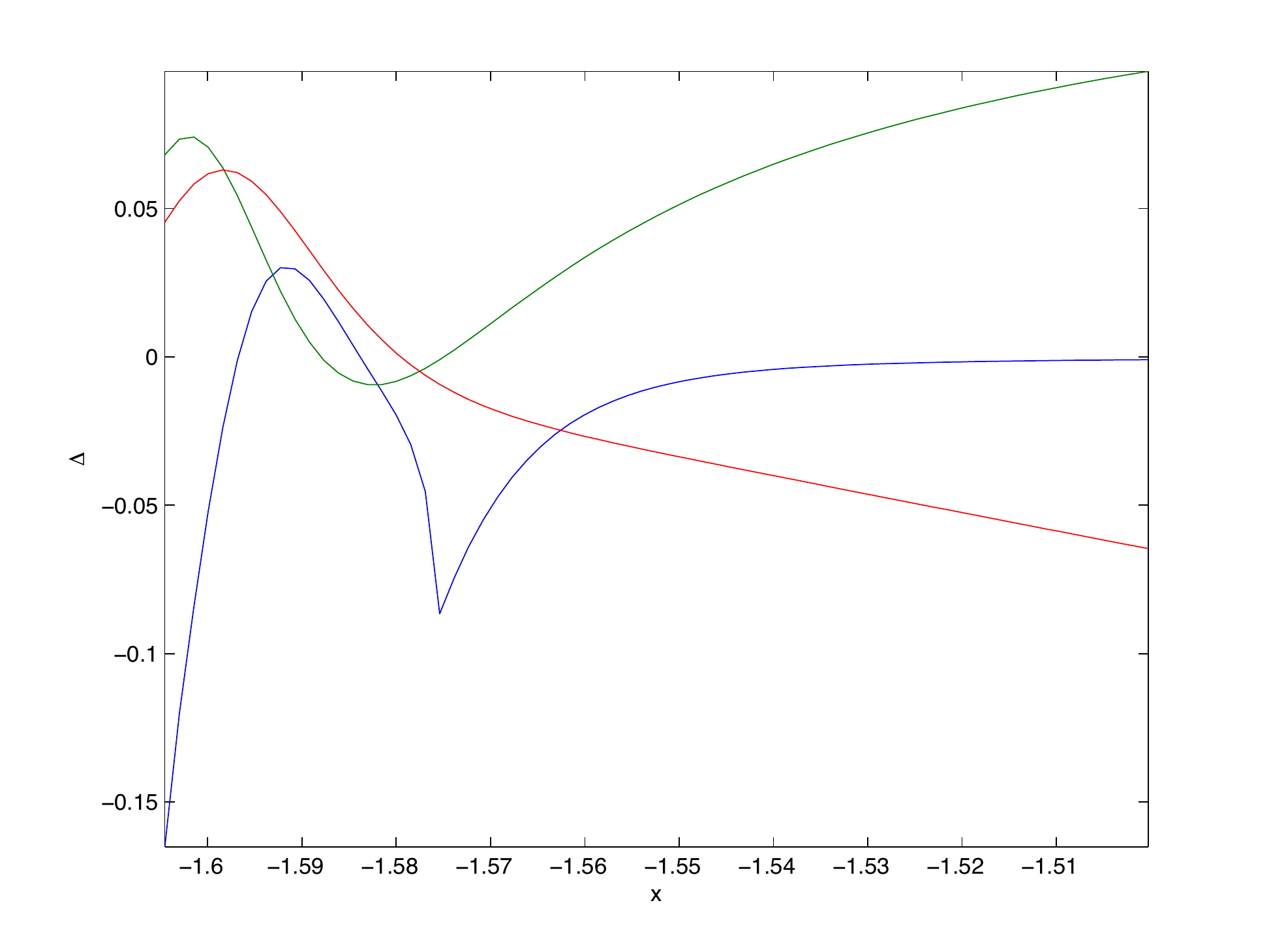}}
 \caption{Fig.~(a) shows the numerical solution to the KdV
equation for the initial datum $u_{0}=-\mbox{sech}^{2}x$ and 
$\epsilon=10^{-2}$ at $t=0.23>t_{c}$ in blue, the asymptotic solution in terms of Hopf and one-phase KdV
solution in green,  the multiscale solution (\ref{expansion u}) in 
red, and the multiscale solution 
(\ref{ExpP12}) in cyan. Fig.~(b) shows the difference between the asymptotic solution in terms of Hopf and one-phase KdV
solution and the numerical solution to the KdV in blue,  between the 
soliton asymptotic  solution (\ref{expansion u}) and KdV in 
green, and the ${\rm P}_I^2$ asymptotic  solution 
(\ref{ExpP12}) and KdV in red. }
   \label{kdv1e4_234in1trail}
\end{figure}
%\begin{figure}[htb!]
%  \includegraphics[width=0.6\textwidth]{kdv1e4_23deltatrail.pdf}
% \caption{The
%figure }
%   \label{kdv1e4_23deltatrail}
%\end{figure}

In Fig.~\ref{kdv1e4_234in1trail} the same solutions can be seen in one 
figure.
These figures as well as the plot of the 
differences between the asymptotic solutions and the KdV solution in 
Fig.~\ref{kdv1e4_23deltatrail} show that the asymptotic solution in terms of 
Hopf and one-phase KdV solution performs worst close to the trailing 
edge, and that the 
${\rm P}_I^2$ asymptotic  solution (\ref{ExpP12}) performs best. The 
soliton asymptotic  solution (\ref{expansionu}) is 
 better than the ${\rm P}_I^2$ asymptotics only very close to the 
 trailing edge. 
The ${\rm P}_I^2$  asymptotics also quickly becomes unsatisfactory in the Hopf 
region. 
Thus to obtain a better description of the small dispersion 
limit of KdV close to the critical point, one has to study 
connection formulae between the various asymptotic solutions. 

\section{Outlook}

The numerical results of the previous sections have shown that the 
proposed asymptotic descriptions lead in fact to an error of order 
$\epsilon$ in various regions of the $x,t$-plane where the respective 
formulae are supposed to hold. However the asymptotic description at the trailing edge of the oscillatory zone and the point of gradient catastrophe are characterized by errors of higher order.
 In order to obtain a complete analytic asymptotic description in the $(x,t)$ plane,  analytic connection formulae have to be established 
between the various asymptotic formulae.  For example, despite the 
asymptotic formulas (\ref{elliptic}) and (\ref{expansionu})  having an error of order $\epsilon$, the numerical results show that there is still a region near the leading edge at the boundary of the Whitham zone where  the error is bigger. This means that   a connection formula between the elliptic expansion (\ref{elliptic})
and the expansion (\ref{expansionu})  (where the terms of order  $\epsilon^{\frac{2}{3}}$ have been dropped),  is needed.   

In \cite{Claeys}  Claeys has derived connection formula  for the     ${\rm P}_I^2$  solution  $U(X,T)$  of  (\ref{PI2}) in different regions of the $(X,T)$ plane.
Using these relations, one can derive in a non rigorous way the corresponding connection formulas for the asymptotic solution of the KdV equation near the point of gradient catastrophe.
Indeed the ${\rm P}_{\rm I}^2$ solution
$U(X,T)$ describes a singular transition between a region of simple algebraic asymptotics and a region
of more complicated oscillatory asymptotics involving the Jacobi
elliptic $\theta$-function. One may thus expect that $U(X,T)$
itself also exhibits different types of asymptotics. Indeed the following result holds \cite{Claeys}: 
\begin{itemize} 
\item  $X\rightarrow \pm \infty$ and  $T\rightarrow -\infty$ or  $T\rightarrow +\infty$  in such a way $S=\dfrac{X}{T^{\frac{3}{2}}}$ remains bounded  away from the interval $[-12\sqrt{3}, \,\dfrac{4\sqrt{15}}{9}]$, then $U(X, T)$ has an algebraic asymptotics;
\item  if $ -12\sqrt{3}<S<\dfrac{4\sqrt{15}}{9}$, then $U(X,T)$ has 
an elliptic asymptotics;
\item for $S\rightarrow -12\sqrt{3}$,  $U(X,T)$ has a ${\rm P}_{II}$ asymptotics;
\item for $S\rightarrow \dfrac{4\sqrt{15}}{9}$, $U(X,T)$  has a soliton-like asymptotics.
\end{itemize}

Substituting the algebraic asymptotic \cite{Claeys} of the ${\rm P}_{\rm I}^2$ solution into the asymptotic expansion (\ref{ExpP12}), one obtains in a non rigorous way the connection formula
between the Hopf  asymptotic solution (\ref{Hopfasym})  and (\ref{ExpP12}). More precisely
in the limit when
\[
X=\dfrac{x-x_c-6u_c(t-t_c)}{(k)^{1/7}\epsilon^{\frac{6}{7}}},\quad T=\dfrac{t-t_c}{(k)^{3/7}\epsilon^{\frac{4}{7}} }
\]
goes to infinity in such a way that $s=\dfrac{X}{T^{\frac{3}{2}}}=\sqrt{k}\dfrac{x-x_c-6u_c(t-t_c)}{(t-t_c)^{\frac{3}{2}}}$ is outside the interval $(-12\sqrt{3}, \dfrac{4\sqrt{15}}{9})$, or 
$T\rightarrow -\infty$  then  the solution of the KdV equation is approximated by 
\[
u(x,t,\epsilon)=u_c+z(s)\left(\dfrac{t-t_c}{\sqrt{k}}\right)^{\frac{1}{2}}+O\left(\dfrac{\epsilon^{\frac{4}{7}}}{t-t_c}\right).
\]
where $z(s)$ solves the equation $s=6z-z^3$. The second term on the 
right hand side of the above expression coincides with the solution of the Hopf equation with initial data $f_L(u)=-ku^3$.

 \subsection{${\rm P}_I^2$ asymptotics and elliptic asymptotics}
The asymptotic expansions (\ref{elliptic}) and (\ref{ExpP12}) have a 
connection region that follows from  substituting into  
(\ref{ExpP12})  the elliptic asymptotics  for the ${\rm P}_I^2$ solution obtained in  \cite{Claeys}.
In the limit when
\[
X=\dfrac{x-x_c-6u_c(t-t_c)}{(k)^{1/7}\epsilon^{\frac{6}{7}}},\quad T=\dfrac{t-t_c}{(k)^{3/7}\epsilon^{\frac{4}{7}} }
\]
goes to infinity in such a way that $-12\sqrt{3}<s=\dfrac{X}{T^{\frac{3}{2}}}=\sqrt{k}\dfrac{x-x_c-6u_c(t-t_c)}{(t-t_c)^{\frac{3}{2}}}<\dfrac{4\sqrt{15}}{9}$, then  the solution of the KdV equation is approximated by 
\begin{equation}
\label{ellipticP12}
u(x,t,\epsilon)=u_c+\dfrac{\sqrt{t-t_c}}{\sqrt{k}}\left( b_1+b_2+b_3+2\alpha+ \dfrac{b_1-b_3}{2K^2(s)}(\log\vartheta)''(\tilde{\Omega}(s,t);\mathcal{T})\right)+O\left(\dfrac{\epsilon^{\frac{4}{7}}}{\sqrt{t-t_c}}\right),
\end{equation}
where $\alpha$,  $s$ and $\mathcal{T}$ are defined in (\ref{alpha}) (with the substitution $\beta_i\rightarrow b_i$) and  the argument of the Jacobi elliptic function $\vartheta$ is given by
\begin{equation}
\tilde{\Omega}
%=\dfrac{8(t-t_c)^{\frac{7}{4}} }{5\epsilon k^{\frac{3}{4}} }\int_{\beta_2}^{\beta_1}\sqrt{(\beta_1-\xi)(\xi-\beta_2)(\xi-\beta_3)}(\xi+\dfrac{1}{2}(\beta_1+\beta_2+\beta_3))d\xi\\
=\dfrac{(t-t_c)^{\frac{7}{4}}\sqrt{b_1-b_3}}{2\epsilon K(s) k^{\frac{3}{4}}}(s-2(b_1+b_2+b_3)-q).
\end{equation}
Here the quantities $b_i=b_i(s)$, with  $b_1(s)>b_2(s)>b_3(s)$, describe  the  Gurevich-Pitaevski \cite{GP} self-similar solution of   the Whitham equations (\ref{Whitham}) with cubic initial data, namely the function $q=q(b_1,b_2,b_3)$ in  (\ref{q0}) is such that $q(b,b,b)=b^3$ and the hodograph transform (\ref{hodograph}) takes the equivalent  form \cite{GT}
\begin{equation}
\label{selfsimilar}
\begin{split}
&6=-\sum_{i=1}^3 \dfrac{\partial}{\partial b_i}q=\dfrac{1}{5}\left[(b_1+b_2+b_3)^2+2(b_1^2+b_2^2+b_3^2)\right]\\
&s=\sum_{i=1}^3(2b_i-b_1+b_2+b_3)\dfrac{\partial}{\partial b_i}q+q=\dfrac{2}{15}\left[(b_1+b_2+b_3)^3-4(b_1^3+b_2^3+b_3^3)\right]\\
&\int_{b_3}^{b_2}\sqrt{(\xi-b_1)(\xi-b_2)(\xi-b_3)}(\xi+\dfrac{1}{2}(b_1+b_2+b_3))d\xi=0.
\end{split}
\end{equation}
The asymptotic expansion (\ref{ellipticP12}) should  give the connection formula  near the point of gradient catastrophe $(x_c,t_c)$  between the one-phase KdV asymptotics  (\ref{elliptic}) and ${\rm P}_I^2$  asymptotics (\ref{ExpP12}). It is straightforward to check that  formula  (\ref{elliptic}) reduces to (\ref{ellipticP12})  by expanding the initial data for the Whitham equations 
at the point of gradient catastrophe and keeping the first order correction (cubic term). Namely, if $\beta_i$, $i=1,2,3$ is the solution of the Whitham equation 
with the initial data (\ref{KdV}) and $b_i$ is the self-similar solution defined in (\ref{selfsimilar}), then
\[
\beta_i(x,t)= u_c+\sqrt{\dfrac{t-t_c}{k}}b_i(s)+O(t-t_c).
\]
With this formula (\ref{ellipticP12}) can be recovered from (\ref{elliptic}) by the above substitution.The error  in this limit is of order $t-t_c$. The connection formula (\ref{ellipticP12}) has already appeared in \cite{GS}.
%% \subsection{${\rm P}_I^2$ asymptotic and PII asymptotic}
%% The asymptotic expansions (\ref{expansionu}) and (\ref{ExpP12}) have a connection region that is provided by the ${\rm P}_I^2$ asymptotic for $S=-12\sqrt{3}$ \cite{Claeys}.
%% Indeed introducing the variable 
%% \[
%% \xi=\dfrac{X+12\sqrt{3}T^{\frac{3}{2}}}{c_0 T{\frac{1}{3}}}
%% \]
%% in the ${\rm P}_I^2$ solution and letting $T\rightarrow \infty$ such that 
%
%
\subsection{Connection between ${\rm P}_I^2$ asymptotic and  ${\rm P}_{II}$ asymptotic  or soliton asymptotic}
When $S=-12\sqrt{3}$  the ${\rm P}_I^2$ solution has  an asymptotic  expansion  that is provided by oscillations whose envelope is given by the Hasting McLeod solution of the Painlev\'e II equation   \cite{Claeys}.
Plugging this expansion into (\ref{ExpP12}), one obtains in a non rigorous way the connection formula between  (\ref{ExpP12}) and  the ${\rm P}_{II}$  asymptotic solution (\ref{expansionu}), where the terms of order $\epsilon^{\frac{2}{3}}$ have been dropped. Introducing the variable 
\[
\xi=-\dfrac{X+12\sqrt{3}T^{\frac{3}{2}}}{c_0c_1T^{\frac{1}{3}}}=-\dfrac{x-x_c-6u_c(t-t_c)+12\sqrt{3}k^{-\frac{1}{2}}(t-t_c)^{\frac{3}{2}}}{\epsilon^{\frac{2}{3}} c_0(t-t_c)^{\frac{1}{3}}},\quad c_0=\dfrac{2^{\frac{7}{6}}3^{\frac{1}{12}}}{5^{\frac{1}{6}}},\;\;c_1=\sqrt{\dfrac{5\sqrt{3}}{2}}
\]
in the ${\rm P}_I^2$ asymptotic solution (\ref{ExpP12}) and letting $T\rightarrow +\infty$ in such a way that  $X T^{-\frac{3}{2}}=12\sqrt{3}$ one obtains from \cite{Claeys}
\begin{equation}
%\begin{split}
\label{connectionPII}
u(x,t,\epsilon)=u_c+2\sqrt{3}\sqrt{\dfrac{t-t_c}{k}}-\dfrac{q(\xi)\left(\dfrac{\epsilon}{k}\right)^{\frac{1}{3}}}{c_0\left(\dfrac{t-t_c}{k}\right)^{\frac{1}{12}}}\cos\left(\dfrac{(t-t_c)^{\frac{7}{4}}}{\epsilon k^{\frac{3}{4}}}\omega\right)+O\left(\left(\dfrac{\epsilon^{\frac{4}{7}}}{t-t_c}\right)^{\frac{2}{3}}\right)
\end{equation}
where $q(\xi)$ is the Hasting-McLeod solution of Painlev\'e II,  and the phase $\omega$ is given by
\[
\omega=\dfrac{88}{7}c_1^3+2c_1^2c_0\xi\left(\dfrac{t-t_c}{k^{\frac{3}{7}}\epsilon^{\frac{4}{7}}}\right)^{-\frac{7}{6}}.
\]
This expansion coincides with (\ref{expansionu}) when the initial 
data at time $t=t_c$ is approximated by the cubic initial data  
$f_L(u)=-k(u-u_c)^3+O(u-u_c)^4$ with $k$ defined in (\ref{k}). Indeed 
in  this case the solution of system (\ref{leading})-(\ref{leading3}) takes the  form
\begin{equation*}
\begin{split}
&x^-(t)-x_c-6u_c(t-t_c)=12\sqrt{3}\dfrac{(t-t_c)^{\frac{3}{2}}}{\sqrt{k}}+O((t-t_c)^{\frac{5}{2}}),\\
&u(t)-u_c=2\sqrt{3}\sqrt{\dfrac{t-t_c}{k}}+O(t-t_c),\quad
v(t)-u_c=-\dfrac{\sqrt{3}}{2}\sqrt{\dfrac{t-t_c}{k}}+O(t-t_c).
\end{split}
\end{equation*}
Then plugging the above expressions of $x^-(t)$, $u(t)$ and $v(t)$  into (\ref{expansionu}) one arrives at 
(\ref{connectionPII}) (with a different error term though).

The  ${\rm P}_I^2$  solution has a connection region with the soliton 
asymptotics  (\ref{expansion u}) when  $S=\dfrac{4}{9}\sqrt{15}$ \cite{Claeys}. Substituting the  corresponding connection formula   in \cite{Claeys}  into (\ref{ExpP12})  one obtains in a non rigorous way the  connection formula for the asymptotic expansions   (\ref{ExpP12}) and (\ref{expansion u}). 
Indeed introducing the variable
\[
\xi=-\dfrac{8}{7}c_0^2\dfrac{X- \dfrac{4}{9}\sqrt{15}T^{\frac{3}{2}}}{T^{-\frac{1}{4}}\log T}=-\dfrac{8}{7}c_0^2\dfrac{x-x_c-6u_c(t-t_c)- \dfrac{4}{9}\sqrt{15k}(t-t_c)^{\frac{3}{2}}}{\epsilon(t-t_c)^{-\frac{1}{4}}\log \left(\dfrac{t-t_c}{\epsilon^{\frac{4}{7}}k^{\frac{3}{7}}}\right)},\quad c_0=\sqrt{\dfrac{7}{6}}(15)^{\frac{1}{4}},
\]
where $X$ and $T$ are defined as in (\ref{XT}), and letting $T\rightarrow +\infty$ in such a way that $X T^{-\frac{3}{2}}=\dfrac{4}{9}\sqrt{15}$ then 
\begin{equation}
\label{soliton}
\begin{split}
u(x,t,\epsilon)&=u_c-2\sqrt{\dfrac{5}{3}}\left(\dfrac{t-t_c}{k}\right)^{\frac{1}{2}}+2c_0^2\left(\dfrac{t-t_c}{k}\right)^{\frac{1}{2}}\sum_{j}\sech^2X_j+
\\
&O\left(\left(\dfrac{t-t_c}{\epsilon^{\frac{4}{7}}}\right)^{-\frac{5}{4}} \log^2\left(\dfrac{t-t_c}{\epsilon^{\frac{4}{7}}}\right)\right)
\end{split}
\end{equation}
where
\begin{equation}
\begin{split}
&X_j=-\frac{7}{8}(\frac12-\xi+j)\ln T-\ln(\sqrt{2\pi} h_j)-(j+\frac12)\log\left(\dfrac{16c_0^{\frac{5}{2}} }{ (15)^{\frac{1}{4}} }\right),\\
&h_j=\dfrac{2^{\frac{j}{2}} }{\pi^{\frac{1}{4}}\sqrt{j!}}.
%\quad \gamma=4(v-u)^{\frac{5}{4}}\sqrt{-\partial_v\theta(v;u)},
\end{split}
\end{equation}
This connection formula coincides with (\ref{expansion u})  when the 
initial data at $t=t_c$ is approximated by the cubic initial data  
$f_L(u)=-k(u-u_c)^3+O(u-u_c)^4$ where  $k$ is defined in (\ref{k}). 
In this case the solution of the system
of equation (\ref{trailing1})- (\ref{trailing3}) takes the form 
$x^+(t)-x_c-6u_c(t-t_c)= 
\dfrac{4\sqrt{15}}{9\sqrt{k}}(t-t_c)^{\frac{3}{2}}+O((t-t_c)^{\frac{5}{2}})$,  $u(t)-u_c=-\dfrac{2}{3}\sqrt{15}\sqrt{\dfrac{t-t_c}{k}} +O(t-t_c) $ and $v(t)-u_c=\dfrac{1}{2}\sqrt{15}\sqrt{\dfrac{t-t_c}{k}}+O(t-t_c) $. Plugging the above expressions of $x^+(t)$, $u(t)$ and $v(t)$ into  (\ref{expansion u}) one obtains the connection formula (\ref{soliton}).

The rigorous  derivation of the above connection formulas  and their  numerical  implementation will be investigated in a subsequent publication.

\end{document}